\documentclass[%
 reprint,
 amsmath,amssymb,
 aps,
pra,
]{revtex4-1}

\usepackage{graphicx}
\usepackage{dcolumn}
\usepackage{bm}

\usepackage{amsmath}
\usepackage{amssymb}
\usepackage{graphicx,xcolor}
\usepackage{subfigure}
\usepackage{adjustbox}
\usepackage{multirow}

\usepackage{epstopdf}
\newcommand{\dd}{{\rm{d}}}

\newcommand{\be}[1]{\begin{equation}\label{#1}}
\newcommand{\ee}{\end{equation}}
\newcommand{\ba}[1]{\begin{eqnarray}\label{#1}}
\newcommand{\ea}{\end{eqnarray}}
\newcommand{\rf}[1]{(\ref{#1})}
\newcommand{\nn}{\nonumber}

\begin{document}

\preprint{AIP/123-QED}

\title[Analysis of AMRI and TI via an extended Hain-L\"ust equation]{Analysis of azimuthal magnetorotational instability of rotating MHD\\ flows and Tayler instability via an extended Hain-L\"ust equation}

\author{R. Zou}
\affiliation{
Zhejiang Normal University, 688 Yingbin Road, Jinhua, Zhejiang, 321004, China}%


\author{J. Labarbe}%
\affiliation{%
Northumbria University, Newcastle upon Tyne, NE1 8ST, UK}

\author{Y. Fukumoto}%
\affiliation{%
Institute of Mathematics for Industry, Kyushu University, Fukuoka, 819-0395, Japan}

\author{O. N. Kirillov}
\affiliation{%
Northumbria University, Newcastle upon Tyne, NE1 8ST, UK}

\date{\today}

\begin{abstract}
\textcolor{black}{We consider a differentially rotating flow of an incompressible electrically conducting and viscous fluid subject to an external axial magnetic field and to an azimuthal magnetic field that is allowed to be generated by a combination of an axial electric current external to the fluid and electrical currents in the fluid itself. In this setting we derive an extended version of the celebrated Hain-L\"ust differential equation for the radial Lagrangian displacement that incorporates the effects of the axial and azimuthal magnetic fields, differential rotation, viscosity, and electrical resistivity. We apply the Wentzel-Kramers-Brillouin method to the extended Hain-L\"ust equation and derive a new comprehensive dispersion relation for the local stability analysis of the flow to three-dimensional disturbances. We confirm that in the limit of low magnetic Prandtl numbers, in which the ratio of the viscosity to the magnetic diffusivity is vanishing, the rotating flows with radial distributions of the angular velocity beyond the Liu limit, become unstable subject to a wide variety of the azimuthal magnetic fields, and so is the Keplerian flow.
In the analysis of the dispersion relation we find an evidence of a new long-wavelength instability which is caught also by the numerical solution of the boundary value problem for a magnetized Taylor-Couette flow.}

\end{abstract}

\maketitle


\section{Introduction}
\label{sec:Introduction}

\subsubsection*{Standard magnetorotational instability}

Due to the rediscovery of Velikhov's \cite{Vel59} and Chandrasekhar's \cite{Cha60} pioneering results by Balbus and Hawley \cite{BalHaw91}, the magnetorotational instability (MRI) has aroused strong interest in astrophysics as a promising mechanism for triggering turbulence in the flow of an accretion disk and for promoting outward transport of angular momentum, while the matter accretes to the center \cite{B2011, JB2013}. In magnetohydrodynamics (MHD) and plasma physics communities the MRI stimulated development of new experimental facilities for its detection in the magnetized Couette-Taylor flow of either liquid metal (sodium, gallium, and liquid eutectic alloy $GaInSn$) as in the Potsdam Rossendorf Magnetic Instability Experiment (PROMISE) or plasma as in the Madison plasma Couette flow experiment (MPCX) \cite{SGG2008, ELFB11,PhysRep2018,Stefani2019}.

Let us introduce the cylindrical coordinates $(r, \theta, z)$ with the $z$-axis along the axis of symmetry, along with $\bm{e}_r, \bm{e}_\theta$ and $\bm{e}_z$ being the unit vectors in the radial, azimuthal, and axial direction, respectively.
For an accretion disk, the Keplerian flow, a cylindrically symmetric flow with the profile $U_{\theta} \propto r^{-1/2}$ of rotational velocity, satisfies the force balance: $U^2_{\theta}(r)/r=\Omega^2(r)r=-\nabla\Phi;\ \Phi \propto 1/r$.
In general, a steady rotating flow with the angular velocity $\Omega(r) \bm{e}_z$, parallel to the $z$-axis can be considered as a base state.
To quantify the differential rotation the Rossby number is defined as $Ro=1/2\left(\dd \log \Omega/\dd \log r\right)=r\Omega'/(2\Omega)$, where the prime designates the derivative with respect to $r$ and $\Omega>0$ without loss of generality, see e.g. \cite{KirSte10, KirSte12}.

For a non-magnetized flow of an ideal incompressible fluid, Rayleigh's criterion states that the centrifugal instability with respect to axisymmetric disturbance occurs when the Rossby number, $Ro<-1$, which fails to include the Keplerian flow ($Ro=-3/4$).

According to \cite{Vel59,Cha60,BalHaw91}, a combined effect of fluid rotation and the imposed axial magnetic field is able to raise the critical Rossby number from $-1$ to 0 and destabilize the Rayleigh-stable flows (including the Keplerian one) of an incompressible fluid, for which the viscosity and the electric resistivity are neglected. The instability caused by the magnetic field that has only the axial component $\bm{B}=B_z\bm{e}_z$ is known as the Standard Magnetorotational Instability, or SMRI \cite{KirSte10}.

Already in \cite{Vel59,Cha60} a counterintuitive Velikhov-Chandrasekhar paradox for SMRI has been pointed out. In
the case of an ideal nonresistive flow, boundaries of the region of the magnetorotational instability are misplaced compared to the
Rayleigh boundaries of the region of the centrifugal instability,
and do not converge to those in the limit of a negligibly small
axial magnetic field \cite{KPS2011}. Willis and Barenghi established that the convergence is possible
in the presence of viscosity and resistivity \cite{WB2002}. Actually, the transition is parameterised  by the Lundquist number $S$, so that
the highly conducting fluids characterized by high values of $S$ have SMRI for $Ro<0$ and more resistive fluids with low Lundquist numbers are Rayleigh-unstable for $Ro<-1$, see the short-wavelength analysis in \cite{KS2011} and its recent confirmation by asymptotic and numerical methods in \cite{Deg18}.

\subsubsection*{Helical and azimuthal MRI and Tayler instability}

Given an axial magnetic field at some instant, the radial component is seeded by perturbing the axial field. Once the radial component arises, with the magnetic field frozen into a perfectly conducting accretion disk, the radial component is tilted by the differential rotation to produce azimuthal component, and the latter component is constantly stretched with time, resulting in establishing a strong azimuthal component \cite{Des04}. Three-dimensional numerical simulations demonstrated that the initial weak azimuthal magnetic field tends to be stretched out to become, at a later stage, dominant over the initial axial magnetic field \cite{BNST95,PapTer97}.

Instabilities induced by azimuthal magnetic fields have been studied already in \cite{BalHaw92a,TerPap96} for the accretion disks and in a more general setting for a differentially rotating flow of a perfectly conducting ideal fluid in \cite{FV95,OgiPri96}. A combined action of the azimuthal and the axial magnetic field, i.e. the helical field, on the stability of accretion disks in the ideal MHD setting was addressed in \cite{CurPud96}.

In a protoplanetary disk surrounding a young star, the ionization depends on the radiation from the X-rays and cosmic rays \cite{SalWar03}, and the temperature of the disk. The mid-plane of the accretion disk receives fewer radiation and the cold region of the disk is only weakly ionized. For the cold and less radiated parts of the protoplanetary and accretion disks as well as for the experiments with liquid metals, the effects of both the viscosity $\nu$ and the magnetic diffusivity $\eta$ are therefore not ignorable. Because of the low electric conductivity, the magnetic Prandtl number $Pm=\nu/\eta$ is very small \cite{RGSHS14} (e.g. $Pm\sim 10^{-5}$ for liquid sodium, $Pm\sim 10^{-6}$ for gallium and liquid eutectic alloy $GaInSn$). By contrast, in the hot parts of the accretion disks, because of the high electric conductivity, $Pm$ can become very large, see e.g. \cite{BalHen08} where $Pm$ ranges from $10^{-3}$ to $10^3$.

The case of $Pm=0$ is referred to as the inductionless limit \cite{Priede2007,Pri11,Priede2015,PhysRep2018}.
Viewing $Pm$ as a ratio of the magnetic and hydrodynamics Reynolds numbers, $Pm=Rm/Re$, one can deduce that $Pm\sim 10^{-5}$ and $Rm>1$ implies  $Re>10^5$ for the onset of SMRI that is governed by $Rm$ and $S$ and requires high values of these numbers for its excitation \cite{KSF14JFM}.
On the other hand, at $Re>10^5$  it is hard to keep the base flow of a liquid metal laminar in an experimental Couette-Taylor setup, which explains why SMRI is still not observed in an experiment \cite{B2011, JB2013}.

In 2005 Hollerbach and R\"udiger \cite{HolRud05} demonstrated that the simultaneous application of an axial and an azimuthal magnetic field
in the case of low $Pm$ can significantly reduce the critical value of the hydrodynamic Reynolds number at the onset of MRI in the Couette-Taylor flow. The predicted in \cite{HolRud05} axisymmetric helical MRI (HMRI) has been successfully detected in subsequent experiments on the PROMISE facility \cite{Rud10,SGGRSSH06,SGGHPRS09}.

The azimuthal MRI (AMRI), for which the magnetic field has only the azimuthal component $\bm{B}=B_\theta(r)\bm{e}_\theta$, was predicted to be non-axissymmetric and feasible for the parameters of the existing liquid-metal Couette-Taylor facilities in \cite{HolTee10}. It was detected by PROMISE in 2014 for the azimuthal magnetic field created by an axial current external to the liquid metal \cite{SGGGS14}.
We notice, however, that the domains of both AMRI and HMRI plotted in the $(Ha, Re)$-plane typically have a finite size along the $Re$-axis, which means that these instabilities can be inhibited at sufficiently large Reynolds numbers.

Both HMRI and AMRI observed in the liquid metal experiments were reported for the differential flows that are distant from the Keplerian one. Detection of HMRI and AMRI for the quasi-Keplerian Couette-Taylor flows is planned in the upcoming AMRI-TI liquid metal experiment in the frame of the DRESDYN project \cite{Stefani2019}. This advancement is based on a stability analysis initiated in \cite{KirSte13} and motivated by the work by Liu et al. \cite{LGHJ06} who, using a short wavelength approximation, identified critical steepnesses of the rotation profile, which prevent excitation of HMRI for $-0.828 \approx 2-2\sqrt{2} <Ro<2+2\sqrt{2}\approx 4.828$. These ``Liu limits'' were derived in the assumption that the radial profile of the azimuthal magnetic field is $B_\theta(r) \propto r^{-1}$ and $Pm$ is very low and thus excluded HMRI and AMRI of Keplerian flows, characterized by $Ro=-3/4$, in the liquid metal experiments where the azimuthal field is created by an isolated axial current (e.g. in PROMISE).

It is known, however, that the azimuthal magnetic field $B_\theta(r) \propto r$, corresponding to a homogeneous axial current density in a conducting fluid, may cause the kink-type Tayler instability (TI) \cite{Tay73,Tay80,RS2010}, even if the fluid is at rest, as it was observed in a recent liquid metal experiment \cite{Tayler2012}. By combining the field of an external to the fluid current with the currents through the fluid itself one can create azimuthal fields with the radial distributions that interpolate between $B_\theta(r)\propto r^{-1}$ and $B_\theta(r)\propto r$. This is the idea behind the design of the new AMRI-TI experimental setup \cite{Stefani2019}.

In view of these considerations, a helical magnetic field with the arbitrary radial dependence of the azimuthal component has been considered in \cite{KirSte13}. To characterize the magnetic shear, an appropriate \textit{magnetic Rossby number}, $Rb=r^2 (B_{\theta}/r)'/(2B_{\theta})$, has been defined \cite{KirSte13}. Then, $Rb=-1$ corresponds to $B_\theta(r)\propto r^{-1}$ and $Rb=0$ to $B_\theta(r)\propto r$. In the short wavelength approximation it was established that both the azimuthal and helical MRI are very sensitive to the parameter of the magnetic shear, $Rb$. In particular, it was discovered that, if the magnetic profile is made slightly shallower than $B_\theta\propto r^{-1}$, so as to satisfy the condition $Rb \geq -25/32$, the Keplerian flow invites both the AMRI and HMRI \cite{KirSte13, KSF14, KSF14JFM}. Later on, these results were confirmed numerically by solving a boundary value problem for the Couette-Taylor flow with the internal and external currents \cite{PhysRep2018,Hol15}. Numerous previous studies, e.g. \cite{FV95,OgiPri96,LGHJ06,Priede2007,Pri11,HolRud05,HolTee10,KirSte10,KSF12}, overlooked this important result because they were restricted to the current-free field $B_\theta\propto r^{-1}$ with $Rb=-1$.

\subsubsection*{The Hain-L\"ust equation and its extensions}

The HMRI and the AMRI were addressed for axisymmetric and non-axisymmetric perturbations in the short-wavelength regime by traditional Wentzel-Kramers-Brillouin (WKB) method \cite{KirSte10,SB2014} and within the geometrical optics \cite{FV95} approximation \cite{KSF12,KSF14,KirSte13,KSF14JFM,Kir15,Kir17}. The advantage of the latter is a possibility of a systematic derivation of asymptotic equations of different order controlled by a universal small parameter.

In the work \cite{HL1958} Hain and L\"ust derived an ordinary differential equation of Sturm-Liouville type \cite{HMS2017} for the radial Lagrangian displacement to determine the growth rates of MHD instabilities with respect to isothermal perturbations in a diffuse linear pinch. Since then the Hain-L\"ust equation (following form the Frieman-Rotenberg equation \cite{FR1960}) is widely used in the studies of local and global instabilities of cylindrical plasma equilibria \cite{GP04,GKP10,GKP2019}. In particular, it was established that the standard WKB analysis applied to the Hain-L\"ust equation produces the correct local dispersion relation compared to that following from the WKB analysis of the original system of first order MHD equations, see e.g. the discussion on page 103 in the book \cite{GKP10}.

Motivated by this advantage Zou and Fukumoto \cite{ZouFuk14} performed a rigorous derivation of the Hain-L\"ust equation for a differentially rotating ideal MHD fluid in a cylindrical configuration and subjected to an azimuthal magnetic field. After the substitution of the WKB form of the radial solution into the result they found a new local dispersion relation that contained the dispersion relation of the works \cite{FV95,OgiPri96} as a particular case in the limit of short axial wavelengths.

Compared to \cite{FV95,OgiPri96}, the dispersion relation by Zou and Fukumoto contained new terms affecting instabilities with respect to non-axissymmetric perturbations \cite{ZouFuk14}. By that reason, it is extremely interesting to apply this approximation scheme to the case of non-ideal MHD and derive a comprehensive local dispersion relation allowing for differential rotation, viscosity and resistivity and thus applicable to the studies of HMRI, AMRI and TI. This is the goal of the present paper. In it the extended Hain-L\"ust equation serves as a basis for the linear stability analysis of AMRI, HMRI and TI both in the limit of $Pm\to 0$ and in the case of general $Pm$ and $Rb$.

\subsubsection*{Overview of the article}

In Section \ref{sec:Formulation} we present the base state and the linearized MHD equations and derive the new version of the Hain-L\"ust differential equation for the incompressible fluid with allowance for differential rotation, viscosity and electrical resistivity.

In Section \ref{sec:Dispersion relation} we apply the Wentsel-Kramers-Brillouin (WKB) method to the extended Hain-L\"ust equation and obtain the comprehensive dispersion relation in the short radial wavelength limit.

In Section \ref{sec:Axisymmetric perturbations} we check that the new dispersion relation restores the known results for the SMRI and the HMRI, when restricted to axisymmetric disturbances.

In Section \ref{sec:Non-axisymmetric perturbations}, we derive the dispersion relation for the case of purely azimuthal magnetic field and arbitrary $Pm$. Then we focus on the non-axisymmetric AMRI at finite and vanishing magnetic Prandtl numbers. In the weak magnetic field limit we find that the Rayleigh criterion decides the instability. In the case of sufficiently strong azimuthal magnetic field we first deal exclusively with two extreme modes of $kr\rightarrow 0$ and $kr\rightarrow \infty$, being featured by the axial wavenumber $k$. For the Keplerian flow, the short axial-wavelength mode ($kr\rightarrow\infty$) is excitable for $Rb>-25/32$, in accordance with the earlier works \cite{KirSte13,KSF14JFM}.
We find that the long axial-wavelength mode ($kr\rightarrow 0$) is excitable for $Rb<-1/4$ even when the flow is non-rotating. These findings are supported by computation of the growth rates optimized over radial and axial wavelengths and by presenting the evolution of the stability diagrams in the $(Ro,Rb)$-plane as the radial wavenumber varies from small to large values.

In Section \ref{sec:Non-axisymmetric AMRI critical Re Ha}, we find that in the limit of $kr\rightarrow 0$, an upper limit of the value of $qr$, where $q$ is the radial wavenumber, is placed for the instability to occur. Then, we analyse numerically our WKB dispersion relation with a reasonable restriction on the radial wavenumber $q$. This results in the stability diagrams well compared with that of the global numerical analysis of the work \cite{RudHol07} and local analysis of the works \cite{KSF14JFM,Kir15} for various values of the magnetic Prandtl number.

Finally, in Section \ref{sec:BVP} we complement the local stability analysis with the global stability analysis of the original MHD system equipped with boundary conditions that we solve by the pseudo-spectral method \cite{H2000,Deg11,Deg17} to validate the theory.



\section{Extending the Hain-L\"ust Equation}
\label{sec:Formulation}

We consider the linear stability of a cylindrically symmetric rotating flow, of an \textit{incompressible} viscous fluid with finite electric conductivity, to three-dimensional disturbances. 
\textcolor{black}{The basic state is a rotating flow in equilibrium with the velocity field $\bm{U}=\bm{U}(r)$, characterized by the angular velocity $\Omega(r)$, in the steady magnetic field $\bm{B}=\bm{B}(r)$, of the same symmetry, with the azimuthal and the axial components $r\mu(r)$ and $B_z(r)$, respectively:
\begin{equation}
\bm{U}=r\Omega(r)\bm{e}_{\theta},\ \bm{B}=r\mu(r)\bm{e}_{\theta}+B_z\bm{e}_z.
\label{eqn:base-state}
\end{equation} The constant axial component of the magnetic field can be assumed to be externally imposed whereas the azimuthal component can be thought of as created by axial electric currents both external to the fluid and running through the fluid itself \cite{Priede2015,Hol15}.}

The velocity $\bm{u}$, the magnetic field $\bm{b}$ and the total pressure $p$ are partitioned into the basic flow, and the disturbance as
\begin{equation}
\bm{u}=\bm{U}+\tilde{\bm{u}}, \ \bm{b}=\bm{B}+\tilde{\bm{b}}, \ p=P+\tilde{p}.
\label{eqn:base-state}
\end{equation}
The Navier-Stokes and the induction equations linearized in the disturbance $(\tilde{\bm{u}},\tilde{\bm{b}}, \tilde{p})$ are
\begin{eqnarray}
&&\frac{\partial \tilde{\bm{u}} }{\partial t}+(\tilde{\bm{u}}\cdot \nabla)\bm{U}+(\bm{U}\cdot\nabla)\tilde{\bm{u}}
=-\frac{1}{\rho}\nabla \tilde{p}
 \notag\\
&&
\hspace*{10mm}
+\frac{1}{\rho\mu_0}(\bm{B}\cdot \nabla)\tilde{\bm{b}}+
\frac{1}{\rho\mu_0}(\tilde{\bm{b}}\cdot \nabla)\bm{B}+\nu\nabla^2\tilde{\bm{u}},
\label{eqn:Euler-disturb}
    \\
&&\frac{\partial \tilde{\bm{b}} }{\partial t}=\nabla\times(\bm{U} \times \tilde{\bm{b}})+\nabla\times(\tilde{\bm{u}} \times \bm{B})+\eta\nabla^2\tilde{\bm{b}},
\label{eqn:induc-disturb}
    \\
&&\nabla\cdot \tilde{\bm{u}}=0,
\label{eqn:solnoid-disturb}
   \\
&&\nabla\cdot \tilde{\bm{b}}=0,
\label{eqn:div-b=0-disturb}
\end{eqnarray}
where $\mu_0,\ \nu$ and $\eta$ represent the magnetic permeability, the kinematic viscosity and the magnetic diffusivity, respectively. We assume that $\mu_0,\ \nu,\ \eta$ are all constant \cite{KSF14}.

Owing to the steadiness and to the symmetries with respect to translation along and rotation about the $z$-axis, we pose the disturbances in the normal-mode form \be{nf}\tilde{\bm{u}},\tilde{\bm{b}},\tilde{p} \propto\exp[\lambda t+i(m\theta+k z)].\ee The azimuthal wavenumber $m$ takes an integer value, the axial wavenumber $k$ is taken to be a real number, and $\lambda$ is the eigenvalue to be calculated.
Substituting \rf{nf} into (\ref{eqn:Euler-disturb})--(\ref{eqn:div-b=0-disturb}) yields a coupled system of 8 ordinary differential equations for functions of $r$.

With a view to incorporate only the leading-order effect of short-wave radial disturbances under the assumption  of $\nu$ and $\eta$ being small, we may simply replace $-\nabla^2$ with $|\bm{k}|^2=k^2+q^2+m^2/r^2$, where $q(r)$ is the radial wavenumber.
\textcolor{black}{Indeed, if the disturbance is thought to be $$\propto\exp[\lambda t+i(m\theta+k z)]c(r)\exp\left\{i\int{q(r)\dd r}\right\}$$
and $L$ is the characteristic length, then $c'(r)\approx c(r)/L$ and $c''(r)\approx c(r)/L^2$, and $q'(r)\approx q(r)/L$. For $q(r)L\gg1$, $c(r)q^2(r)$ becomes the leading order term and we can write
\be{am}-\nabla^2\approx q^2(r)+k^2+m^2/r^2\ee
in the dissipation terms. This procedure amounts to discarding terms in the short wavelength regime, and should be justified \textit{a posteri\'ori}.}

Within the assumptions made, we write the resulting equations in the matrix form for the vector-function
$$\bm{\xi}=(\tilde{u_r},\
\tilde{u_\theta},\
\tilde{u_z},\
\tilde{b_r},\
\tilde{b_\theta},\
\tilde{b_z},\
\tilde{p})$$
as
\begin{equation}
\sf{M}\bm {\xi=0}
\label{eqn:disturb-eq}
\end{equation}
with the matrix operator
 \begin{eqnarray}
&\sf{M}=&
 \notag\\
&
 \displaystyle
\left(\begin{footnotesize}\begin{array}{cccccccc}
 \displaystyle\tilde{\lambda}_\nu  &\displaystyle-2\Omega &\displaystyle0 &\displaystyle-\frac{iF}{\rho\mu_0}&\displaystyle\frac{2\mu}{\rho\mu_0}&\displaystyle0&\displaystyle\frac{1}{\rho}\frac{d}{d r}
        \\
 \displaystyle\frac{1}{r}\frac{d}{dr}(r^2\Omega)& \displaystyle\tilde{\lambda}_\nu&\displaystyle0&\displaystyle-\frac{2\mu+r\frac{d\mu}{dr}}{\rho\mu_0}&\displaystyle-\frac{iF}{\rho\mu_0}&\displaystyle 0& \displaystyle\frac{1}{r\rho} im
          \\
 \displaystyle0&\displaystyle0 &\displaystyle\tilde{\lambda}_\nu &\displaystyle0&\displaystyle0&\displaystyle-\frac{iF}{\rho\mu_0}& \displaystyle\frac{1}{\rho} ik        \\
 \displaystyle-iF&0&0&\displaystyle\tilde{\lambda}_\eta&0&0&0
                  \\
 \displaystyle r\frac{d\mu}{dr}&\displaystyle-iF&0&\displaystyle-r\frac{d\Omega}{dr}&\displaystyle\tilde{\lambda}_\eta&0&0\\
 0&0&\displaystyle-iF&0&0&\displaystyle\tilde{\lambda}_\eta&0\\
 \displaystyle\frac{1}{r}+\frac{d}{dr}&\displaystyle\frac{im}{r}&ik&0&0&0&0\\
0&0&0&\displaystyle\frac{1}{r}+\frac{d}{dr}&\displaystyle\frac{im}{r} &ik &0
\end{array}\end{footnotesize}
\right),& \notag\\
&&
\label{eqn:M}
\end{eqnarray}
where $F=m\mu+B_zk$ \cite{GKP2019},
\ba{}
\tilde{\lambda}_\nu=\lambda+im\Omega+\omega_\nu,\quad \tilde{\lambda}_\eta=\lambda+im\Omega+\omega_\eta
\ea
and {\color{black}$\omega_\nu=|\bm{k}|^2\nu$, $\omega_\eta=|\bm{k}|^2\eta$ \cite{KSF14JFM}. }

\textcolor{black}{The assumption \rf{am} allows us to reduce the system \rf{eqn:disturb-eq} to a single ordinary differential equation of \textit{second order}, governing the radial Lagrangian displacement of a fluid particle, --- an equivalent to the famous Hain-L\"ust equation \cite{HL1958}, which is a Sturm-Liouville equation with coefficients depending rationally on the eigenvalue parameter $\lambda$ \cite{HMS2017}. Note that without \rf{am} the resulting differential equation would be of order higher than 2.}

For the ideal MHD, the magnetic field is frozen into the fluid and the Lagrangian variable helps to construct the iso-magnetovortical \cite{VMI1999} perturbations, with respect to which the stability analysis is typically made. 
This is no longer true for the non-ideal case. We find that, with $\nu$ and $\eta$ included, the following `quasi' radial displacement $\xi_r=u_r/\tilde{\lambda}_{\eta}$, connected with the radial component $\tilde{u_r}$, is advantageous for simplifying the resulting equation. This differs from the radial Lagrangian displacement by the $\omega_\eta$ term in $\tilde{\lambda}_{\eta}$.

Therefore, we introduce a new dependent variable $\chi=-ru_r/\tilde{\lambda}_\eta$, with the minus sign chosen for convenience, and the following notation
\begin{eqnarray}
&\Lambda= \displaystyle\tilde{\lambda}_\nu+\frac{F^2}{\tilde{\lambda}_{\eta}\rho\mu_0},&
\label{eqn:def-Lambda}
    \\
&h^2=k^2+\frac{m^2}{r^2}.&
\label{eqn:def-h}
\end{eqnarray}

With this, as we show in detail in the Appendix~\ref{sec:Derivation of HL equation}, the system (\ref{eqn:disturb-eq}) collapses into a single second-order ordinary differential equation for $\chi(r)$
\begin{eqnarray}
&&\frac{\dd }{\dd r}\left(f\frac{\dd \chi}{\dd r}\right)+s\frac{\dd\chi}{\dd r}-g\chi=0,
\label{eqn:Hain-Lust-eq}
\end{eqnarray}
where
\begin{eqnarray}
f&=&\frac{\tilde{\lambda}_\eta \Lambda}{h^2r},\notag \\
s&=&\frac{{\color{black}im (\tilde{\lambda}_\nu-\tilde{\lambda}_\eta)}}{h^2r}\Omega',
             \notag \\
g&=&\frac{\dd }{\dd r}\left\{\frac{im\tilde{\lambda}_\eta}{h^2r^2}\left[\left(1-\frac{\tilde{\lambda}_\nu}{\tilde{\lambda}_\eta}\right)r\Omega'+2\left(\Omega-\frac{iF\mu}{\rho\mu_0\tilde{\lambda}_\eta}\right)\right]\right\}
             \notag \\
&&+ \frac{E\tilde{\lambda}_\eta}{\Lambda r}-\left(\Omega-\frac{iF\mu}{\rho\mu_0\tilde{\lambda}_\eta}\right)
\notag \\
&&
\times \frac{2m^2\tilde{\lambda}_\eta}{\Lambda h^2r^3}\left[\left(1-\frac{\tilde{\lambda}_\nu}{\tilde{\lambda}_\eta}\right)r\Omega'+2\left(\Omega-\frac{iF\mu}{\rho\mu_0\tilde{\lambda}_\eta}\right)\right],
%
\label{eqn:fsg}
\end{eqnarray}
and the prime denotes the derivative with respect to $r$. The expression for the coefficient $E$ is given by the formula (\ref{eqn:E-def}) in the Appendix~\ref{sec:Derivation of HL equation}.

\textcolor{black}{Equation (\ref{eqn:Hain-Lust-eq}) with the coefficients (\ref{eqn:fsg}) is thought of as a new version of the Hain-L\"ust equation  \cite{HL1958,HMS2017} for the incompressible fluid extended with allowance for the effect of differential rotation, viscous dissipation, and magnetic diffusion. To the best of our knowledge in this generality  it has not been previously reported in the literature. With $\nu=0$ and $\eta=0$, it reduces to the extended Hain-L\"ust equation for the ideal incompressible MHD flow in differential rotation \cite{ZouFuk14}. If, additionally, $\Omega=0$, it exactly coincides with the classical Hain-L\"ust equation for the non-rotating ideal incompressible MHD fluid in cylindrical configuration \cite{GP04,GKP10,GKP2019}.}

\section{Dispersion relation in short radial wavelength approximation}
\label{sec:Dispersion relation}

Following \cite{GP04,GKP10,GKP2019} we apply the WKB approximation  to (\ref{eqn:Hain-Lust-eq}) by introducing the ansatz $\chi(r)=c(r)\exp\{i\int{q(r)\dd r}\}$ and assuming that the \textit{radial} wavelength is very short, i.e. $ q(r)L\gg 1$, where $L$ is the length scale for the radial inhomogeneity. This results in the algebraic dispersion relation
\begin{eqnarray}
&&(h^2+q^2) \tilde{\lambda}_\eta^2\Lambda^2
+4k^2\left(\Omega\tilde{\lambda}_\eta-\frac{iF\mu}{\rho\mu_0}\right)
\notag\\
&&\times\left[\Omega Ro(\omega_\eta-\omega_\nu)
+\left(\Omega\tilde{\lambda}_\eta-\frac{iF\mu}{\rho\mu_0}\right)\right]
     \notag\\
&&
+ 4\Lambda h^2\tilde{\lambda}_\eta\Bigg[\left(\Omega^2Ro-\frac{\mu^2}{\rho\mu_0}Rb\right)\notag\\
&&+{\color{black}\frac{imr}{4} \frac{\dd}{\dd r}\left(\frac{2(\Omega\tilde{\lambda}_\eta-\frac{i\mu F}{\rho\mu_0})+(\omega_\eta-\omega_\nu)r\Omega'}{h^2r^2}\right)}\Bigg]
  =0,\nn\\
\label{eqn:dispers-rel0}
\end{eqnarray}
where we have introduced the Rossby number $Ro$
and the magnetic Rossby number $Rb$ by \cite{KSF12, KirSte13, KSF14JFM}
\begin{equation}
Ro=\frac{1}{2}\frac{r}{\Omega}\Omega',\quad
Rb=\frac{1}{2}\frac{r}{\mu}\mu'.
\label{eqn:def-Rossby-num}
\end{equation}

In the ideal case when $\omega_{\nu}=0$ and $\omega_{\eta}=0$ the dispersion relation (\ref{eqn:dispers-rel0}) reduces to that of the work \cite{ZouFuk14} that, in its turn reduces to the ideal dispersion relation derived by Ogilvie and Pringle \cite{OgiPri96} and Friedlander and Vishik \cite{FV95} as well as to the ideal versions of the dispersion relation of the works \cite{SB2014,KirSte13,KSF14JFM} in the limit of large axial wavenumbers, $k\to \infty$.

Applying the WKB approximation to the extended Hain-L\"ust equation for the radial Lagrangian displacement rather than to the coupled system of the ordinary differential equations (\ref{eqn:disturb-eq}) we obtain an additional term $\frac{imr}{4} \frac{\dd}{\dd r}\left(\frac{2(\Omega\tilde{\lambda}_\eta-\frac{i\mu F}{\rho\mu_0})+(\omega_\eta-\omega_\nu)r\Omega'}{h^2r^2}\right)$ in the resulting dispersion relation \rf{eqn:dispers-rel0}.
We notice that the axisymmetric mode $(m=0)$ remains intact since this term is irrelevant. However,
it can improve the prediction accuracy in the case of non-axisymmetric perturbations with long axial wavelength.

For our purpose of stability analysis, it is expedient to define two kinds of Alfv\'en frequency $\omega_A$ and $\omega_{A\theta}$, along with their ratio $\beta$ representing the helical geometry of the magnetic field, by \cite{KSF14JFM}
\begin{equation}
\omega_A=\frac{kB_z}{\sqrt{\rho\mu_0}},\quad
\omega_{A\theta}=\frac{\mu}{\sqrt{\rho\mu_0}},\quad
\beta=\frac{\omega_{A\theta}}{\omega_A}.
\label{eqn:def-non-dim-num}
\end{equation}
In addition, we introduce three dimensionless parameters, namely, the magnetic Prandtl number $Pm$, the Reynolds number $Re$ and the Hartmann number $Ha$ by \cite{KSF14JFM}
\begin{equation}
Pm=\frac{\omega_\nu}{\omega_\eta},\quad
Re=\frac{\Omega}{\omega_\nu},\quad
Ha=\frac{\omega_A}{\sqrt{\omega_\nu\omega_\eta}}.
\label{eqn:def-non-dim-num1}
\end{equation}
The dispersion relation for non-dimensional variables, with the derivative term in (\ref{eqn:dispers-rel0}) being expanded out, leads to
\begin{eqnarray}
&&(\Lambda_1\Lambda_2+\widehat{Ha}^2)^2
\notag\\
&&+4\frac{\widehat{h}^2(\Lambda_1\Lambda_2+\widehat{Ha}^2)}{\widehat{h}^2+\widehat{q}^2 }(Re^2PmRo-\beta^2Ha^2Rb)
 \notag\\
&&
+ \frac{4im(\Lambda_1\Lambda_2+\widehat{Ha}^2)}{\widehat{h}^2+\widehat{q}^2}\bigg[ReRo\sqrt{Pm}(\Lambda_2+imRe\sqrt{Pm})
 \notag\\
&&
-i(2m\beta+1)\beta Ha^2Rb+(i\widehat{Ha}\beta Ha-Re\sqrt{Pm}\Lambda_2)\frac{\widehat{k}^2}{\widehat{h}^2}
\notag\\
&&+{\color{black}RoRe(1-Pm)\left(Ro-\frac{\widehat{k}^2}{\widehat{h}^2}\right)}\bigg]
 \notag\\
&&
+4\alpha^2\bigg[(Re\Lambda_2\sqrt{Pm}-i\widehat{Ha}\beta Ha)\Big(Re\Lambda_2\sqrt{Pm}-i\widehat{Ha}\beta Ha
\notag\\
&&+RoRe(1-Pm)\Big)\bigg]
 =0,
\label{eqn:dispers-rel}
\end{eqnarray}
where
\begin{eqnarray}
\Lambda_1&=&\frac{\lambda}{\Omega}Re\sqrt{Pm}+imRe\sqrt{Pm}+\sqrt{Pm},
\notag\\
\Lambda_2&=&\frac{\lambda}{\Omega}Re\sqrt{Pm}+imRe\sqrt{Pm}+\frac{1}{\sqrt{Pm}},
\notag\\
\widehat{Ha}&=&Ha(1+m\beta),
\notag\\
\widehat{k}&=&kr,\quad \widehat{q}=qr,\quad \widehat{h}=hr,\quad
\alpha^2=\frac{\widehat{k}^2}{\widehat{h}^2+\widehat{q}^2}.
\label{eqn:dispers-rel-para}
\end{eqnarray}

In the rest of the paper, this form of the dispersion relation plays the decisive role for determining the instability criteria and for calculating the growth rates.

\section{Axisymmetric perturbations}
\label{sec:Axisymmetric perturbations}

To begin with, we confirm that (\ref{eqn:dispers-rel}) and (\ref{eqn:dispers-rel-para}) reproduce the known results in the axisymmetric case.

\textcolor{black}{\subsection{Standard MRI in the ideal MHD and beyond}}
\label{sec:Axisymmetric perturbations}

\textcolor{black}{For axisymmetric perturbations $(m=0)$ and purely axial magnetic field ($\beta=0$) the dispersion relation (\ref{eqn:dispers-rel}) simplifies as follows
\ba{mb0}
&&Pm^2\frac{\lambda^4}{\Omega^4}+2\frac{(Pm+1)Pm}{Re}\frac{\lambda^3}{\Omega^3}\nn\\
&+&\left(4Pm^2\alpha^2(Ro+1)+\frac{2(Ha^2+1)Pm+(Pm+1)^2}{Re^2}\right)\frac{\lambda^2}{\Omega^2}\nn\\
&+&2\left(\frac{4Pm\alpha^2(Ro+1)}{Re}+\frac{(Ha^2+1)(Pm+1)}{Re^3}\right)\frac{\lambda}{\Omega}\nn\\
&+&4\alpha^2\frac{Ha^2PmRo+Ro+1}{Re^2}+\frac{(Ha^2+1)^2}{Re^4}.
\ea
Expressing $Ha$, $Re$, and $Pm$ in terms of the Alfv\'en, viscous, and resistive frequencies according to \rf{eqn:def-non-dim-num1}
and then setting $\omega_{\nu}=0$ and $\omega_{\eta}=0$ we arrive at the well-known dispersion relation of the standard MRI of the ideal MHD \cite{BalHaw91, KirSte12, ZouFuk14}
\be{smridr}
\lambda^4+2(2\alpha^2\Omega^2 (Ro+1)+\omega_A^2)\lambda^2+4\alpha^2\Omega^2 Ro \omega_A^2+\omega_A^4=0,
\ee
from which $Ro<0$ follows as a necessary condition for the standard magnetorotational instability, established first in \cite{Vel59,Cha60}, and $Ro<-1$ as a criterion for the Rayleigh centrifugal instability in the absence of the magnetic field. The Velikhov-Chandrasekhar paradox is that the exact criterion for SMRI produced by \rf{smridr}
$$
Ro<-\frac{\omega_A^2}{4\alpha^2\Omega^2}
$$
does not tend to the Rayleigh criterion as $\omega_A \rightarrow 0$, \cite{KirSte12,KPS2011}.}

\textcolor{black}{Willis and Barenghi \cite{WB2002} realized, using numerical computation, that viscosity and resistivity are necessary to connect the two criteria.
To show this, we require negativity of the free term in the dispersion relation \rf{mb0} which yields the generalized criterion for SMRI of the non-ideal MHD \cite{KirSte10,KS2011}
$$
Ro<-\frac{1+\frac{1}{4\alpha^2}\left(\frac{Ha^2}{Re}+\frac{1}{Re}\right)^2}{Ha^2Pm+1}.
$$
Introducing the magnetic Reynolds number $Rm=Pm Re$ and the Lundquist number $S^2=Ha^2 Pm$
we re-write it as
$$
Ro<-\frac{1+\frac{1}{4\alpha^2}\left(\frac{S^2}{Rm}+\frac{1}{Re}\right)^2}{S^2+1},
$$
which in the limit of $Re\rightarrow \infty$ and $Rm\rightarrow \infty$ reduces to the condition \cite{KS2011}
\be{vcp}
Ro<-\frac{1}{S^2+1},
\ee
recently confirmed by the asymptotic and numerical analysis of Deguchi \cite{Deg18}.
At $S=0$ the inequality \rf{vcp} yields the Rayleigh criterion $Ro<-1$ whereas for $S \rightarrow \infty$ it restores the Velikhov-Chandrasekhar
condition $Ro<0$.}

\textcolor{black}{\subsection{Helical MRI in the limit $Pm \rightarrow 0$}}

Now we revisit the axisymmetric ($m=0$) HMRI occurring in the presence of both azimuthal and axial components of the magnetic field $\bm{B}=r\mu(r)\bm{e}_{\theta}+B_z \bm{e}_{z}$.

\textcolor{black}{It is well known that \cite{PhysRep2018}: ``AMRI, HMRI and TI survive also at low magnetic Prandtl numbers. One finds for their lines of neutral stability convergence in the (Ha/Re) coordinate plane for decreasing magnetic Prandtl number $Pm \rightarrow 0$, which can also be obtained with the inductionless approximation of the MHD equations for $Pm = 0$.''}

By that reason we can consider  (\ref{eqn:dispers-rel}) in the limit of $Pm \rightarrow 0$ and solve it for the eigenvalue as \cite{KSF14JFM}
\begin{eqnarray}
\frac{\lambda}{\Omega}&=&-\frac{1}{Re} +\frac{ Ha^2}{Re} (2 \alpha^2 \beta^2 Rb-1 )
        \notag\\
&&\pm
\frac{ 2\alpha}{Re}\left[ \beta^2Ha^4 ( 1+ \alpha^2 \beta^2 Rb^2) -
     Re^2 (1 + Ro) \right.
		 \notag\\
&&\left.+ i \beta Ha^2 Re (2 + Ro)\right]^{1/2}.
\label{eqn:HMRI-growthrate}
\end{eqnarray}
At large values of $Re$, (\ref{eqn:HMRI-growthrate}) is expanded as
\begin{eqnarray}
\frac{\lambda}{\Omega}&\approx&\pm  2i \alpha\sqrt{1 + Ro} +\bigg[-1 + Ha^2\Big(2 \alpha^2 \beta^2  Rb-1
 \notag\\
& &
\pm \frac{
    (2  +Ro)\alpha \beta }{
  \sqrt{1 + Ro}}\Big)\bigg] \frac{1}{Re}, \  \ (Ro\neq -1),
	                    \notag\\
	\frac{\lambda}{\Omega}&\approx&\pm 2\alpha  Ha\sqrt{i\beta}
  \frac{1}{\sqrt{Re} } + (-1 - Ha^2 + 2 \alpha^2 \beta^2 Ha^2 Rb)\frac{1}{Re}, \notag\\
& &\ \ (Ro=-1).
\label{eqn:HMRI-growthrate-largeRe}
\end{eqnarray}
From the zeroth-order term in (\ref{eqn:HMRI-growthrate-largeRe}), $Ro<-1$ is sufficient for instability and so is $Ro=-1$ unless $Ha=0$ or $\beta=0$.
The remaining task is classification for the case of $Ro>-1$.
Equation (\ref{eqn:HMRI-growthrate-largeRe}) tells that the growth rate, if it is positive, increases with $|Ha|$.

For $1\ll Ha\ll Re$ and $Ro\neq-1$, (\ref{eqn:HMRI-growthrate-largeRe}) reads for the growth rates \cite{KSF14JFM}
\begin{eqnarray}
\frac{\Re(\lambda)}{\Omega}&=&\left(2 \alpha^2 \beta^2  Rb-1\pm \alpha \beta\frac{
    Ro+2}{
  \sqrt{1 + Ro}}\right)N{-}\frac{1}{Re},
\label{eqn:HMRI-growthrate_R}
\end{eqnarray}
where $N={Ha^2}/{Re}$ is known as the Elsasser number \cite{KSF14JFM} and $\Re(\,)$ designates the real part. The coefficient at $N$ is a quadratic equation
with respect to $\alpha\beta$. Its discriminant is
$$D=8Rb+\frac{(Ro+2)^2}{Ro+1}.$$
Therefore, for $Rb<0$ the coefficient at $N$ can be positive, if $D>0$, which yields \cite{KirSte13,KSF14JFM}
\begin{eqnarray}
Rb>-\frac{1}{8}\frac{(Ro+2)^2}{Ro+1}
\label{eqn:HMRI-RbRo}
\end{eqnarray}
as a necessary condition for instability.

   \begin{figure*}
  \begin{center}
{\includegraphics*[scale=0.4]{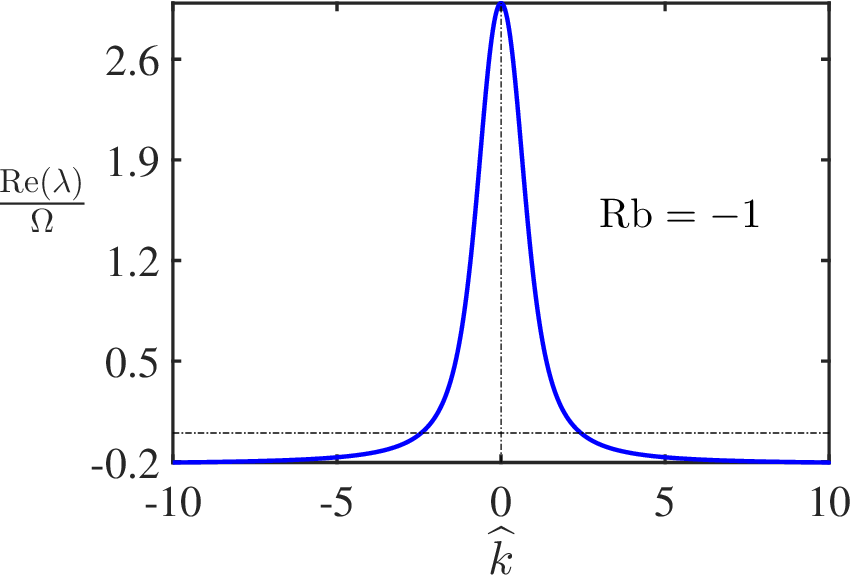}~
\includegraphics*[scale=0.4]{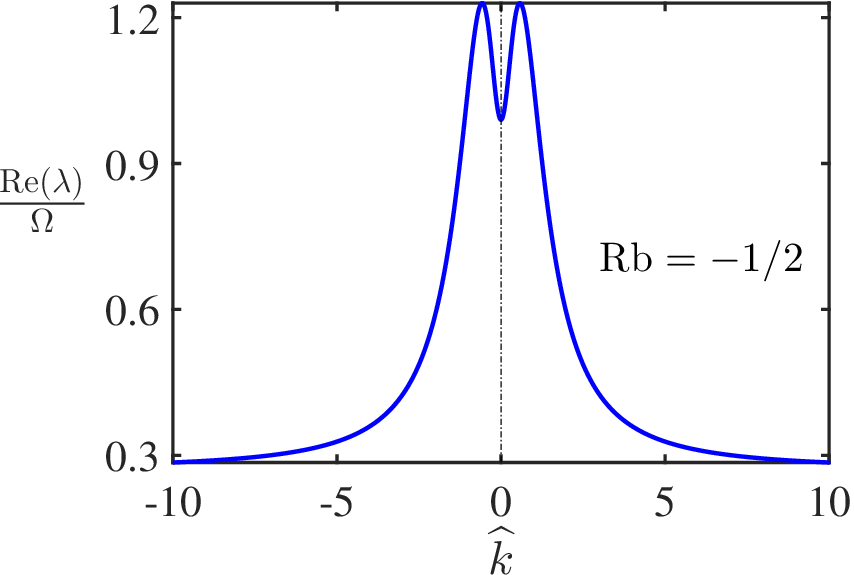}}
{\includegraphics*[scale=0.4]{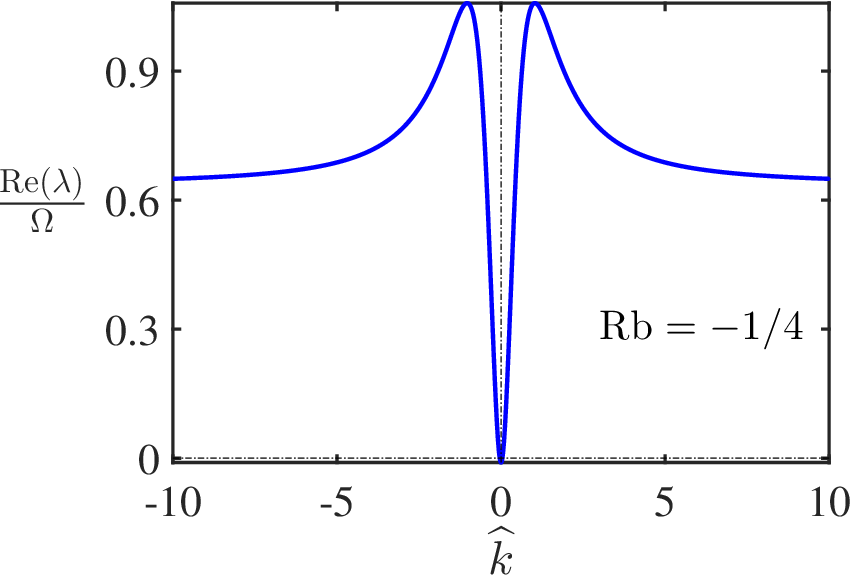}~
\includegraphics*[scale=0.4]{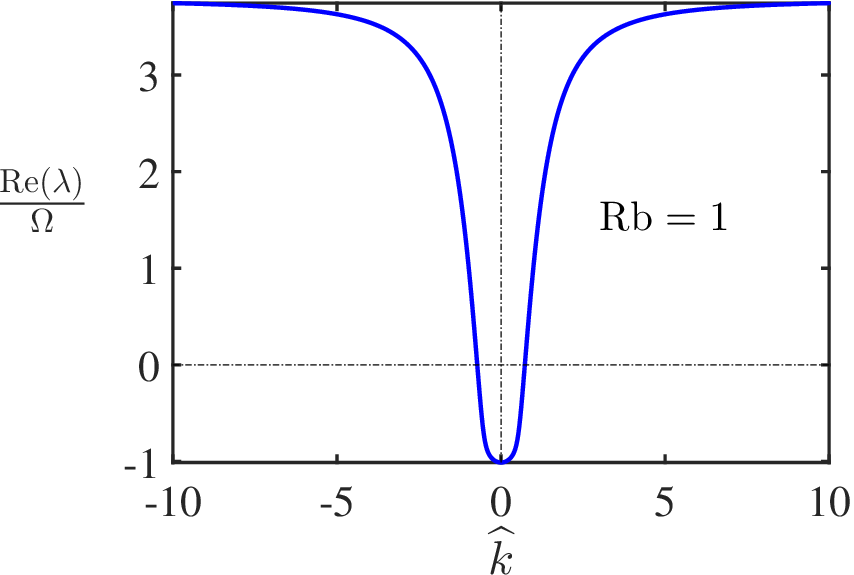}}
 \caption{The growth rate $\mathrm{Re}(\lambda)/\Omega$ given by \rf{eqn:AMRI-disp} versus the dimensionless axial wavenumber $\widehat{k}=kr$  for $\mathrm{Re}=100,\  \mathrm{Ha}_\theta=10,\ m=1,\ \widehat{q}=0,\ \mathrm{Ro}=-3/4$. From upper left to lower right, $\mathrm{Rb}$ is varied from $-1$ to $1$. As $\mathrm{Rb}$ increases, the value of $\widehat{k}$ corresponding to the maximum growth rate increases from $\widehat{k}=0$ to finite but nonzero value and ultimately this $\widehat{k}\rightarrow\infty$.}
\label{fig:k 0 to infty}
\end{center}
\end{figure*}

Note that when $Rb<0$ and $Re \to \infty$ the maximum of the growth rate, as a function of $\alpha\beta$, turns out to be
$$\frac{\Re(\lambda)}{\Omega}=-\frac{DN}{8Rb}$$
and is attained at
$$
\alpha\beta=\mp\frac{Ro+2}{4Rb\sqrt{Ro+1}}.$$
Correspondingly, {\color{black}when $Rb\le -1/2$ the instability occurs in the region \cite{KirSte13,KSF14JFM}
\begin{eqnarray}
Ro&\in&\left[-1,2 \left( -\sqrt{2} \sqrt{ 2 Rb^2+Rb }-1-2 Rb\right)\right]
                \notag\\
&&\cup\left[2 \left( \sqrt{2} \sqrt{ 2 Rb^2+Rb }-1-2 Rb\right),+\infty\right],
\label{eqn:instability HMRI-m=0, rb<0}
\end{eqnarray}
and when $-1/2<Rb<0$ the instability occurs in the region
\begin{eqnarray}
Ro&\in&\left[-1,+\infty\right].
\label{eqn:instability HMRI-m=0, 0.5<rb<0}
\end{eqnarray}
}
In particular, for $Rb=-1$ the critical Rossby numbers are $Ro_c=2 (1 \pm \sqrt{2})$ at $\alpha \beta=\pm1/\sqrt{2}$, and thus the upper and lower Liu limits are recovered  \cite{LGHJ06, KirSte10, KirSte12}.

\vspace{0.5cm}

\section{Non-axisymmetric perturbations}

\label{sec:Non-axisymmetric perturbations}

Hereafter we limit ourselves to the magnetic field that has only the azimuthal component $\bm{B}=r\mu(r)\bm{e}_{\theta}$. Let
\begin{equation}
\label{eqn:Ha_thetha}
{Ha}_\theta=\frac{\omega_{A\theta}}{\sqrt{\omega_\nu\omega_\eta}}
\end{equation}
be the azimuthal Hartmann number.

We first substitute $\beta =\mathit{Ha}_\theta/Ha$ into (\ref{eqn:dispers-rel}) and then take the limit $Ha \to 0$. As a result, we get the dimensionless dispersion relation of AMRI for arbitrary $Pm$
\begin{eqnarray}
&&(\Lambda_1\Lambda_2+m^2\mathit{Ha}_\theta^2)^2
\notag\\
&&+4\frac{\widehat{h}^2(\Lambda_1\Lambda_2+m^2\mathit{Ha}_\theta^2)}{\widehat{h}^2+\widehat{q}^2 }(Re^2PmRo-\mathit{Ha}_\theta^2Rb)
 \notag\\
&&
+ \frac{4im(\Lambda_1\Lambda_2+m^2\mathit{Ha}_\theta^2)}{\widehat{h}^2+\widehat{q}^2}\bigg[ReRo\sqrt{Pm}(\Lambda_2+imRe\sqrt{Pm})
 \notag\\
&&
-2im\mathit{Ha}_\theta^2Rb+(im\mathit{Ha}_\theta^2-Re\sqrt{Pm}\Lambda_2)\frac{\widehat{k}^2}{\widehat{h}^2}
\notag\\
&&{\color{black}+RoRe(1-Pm)\left(Ro-\frac{\widehat{k}^2}{\widehat{h}^2}\right)\bigg]}
 \notag\\
&&
+4\alpha^2\bigg[(Re\Lambda_2\sqrt{Pm}-im\mathit{Ha}_\theta^2)\Big(Re\Lambda_2\sqrt{Pm}-im\mathit{Ha}_\theta^2
\notag\\
&&+RoRe(1-Pm)\Big)\bigg]
 =0.
\label{eqn:dispers-rel-inducless-azim}
\end{eqnarray}

Taking the limit of $Pm\rightarrow  0$ in (\ref{eqn:dispers-rel-inducless-azim}), we find
\begin{eqnarray}
&&\widehat{\lambda}^2 +\frac{
  4  \widehat{\lambda} }{\widehat{h}^2+\widehat{q}^2}\bigg\{ \mathit{Ha}_\theta^2\left( 2m^2 Rb-\widehat{h}^2Rb -\frac{\widehat{k}^2 m^2}{\widehat{h}^2}\right)\notag\\	
	&&{\color{black}+imRe(Ro+1)( Ro-\frac{\widehat{k}^2}{\widehat{h}^2}) }\bigg\} \notag\\	
	&&+
  4 \alpha^2(Re-im\mathit{Ha}_\theta^2) (Re-im\mathit{Ha}_\theta^2+ReRo)
              =0,
\label{eqn:AMRI-disp}
\end{eqnarray}
where
$$\widehat{\lambda}=1 + \mathit{Ha}_\theta^2 m^2 + \frac{\lambda Re}{\Omega} + i m Re. $$

   \begin{figure*}[tbhp]
  \begin{center}
{\includegraphics*[scale=0.4]{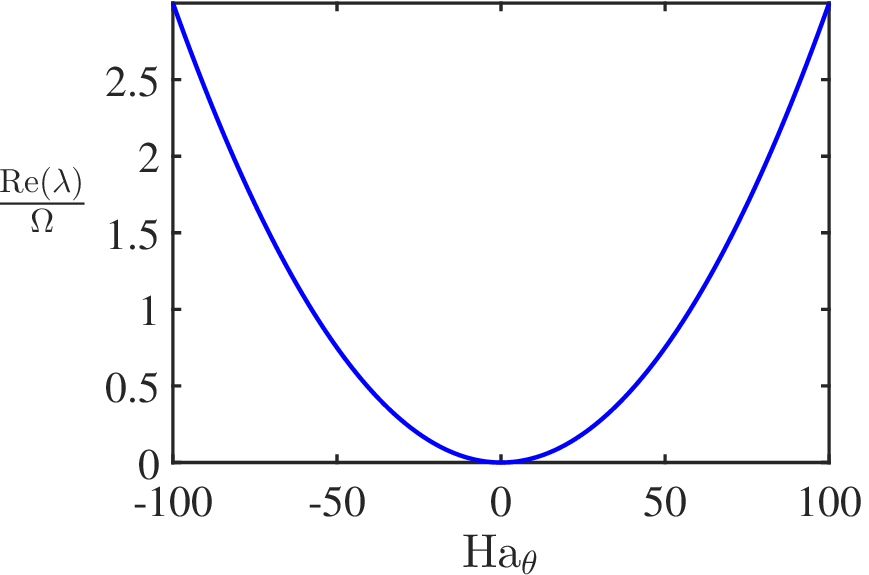}~
\includegraphics*[scale=0.4]{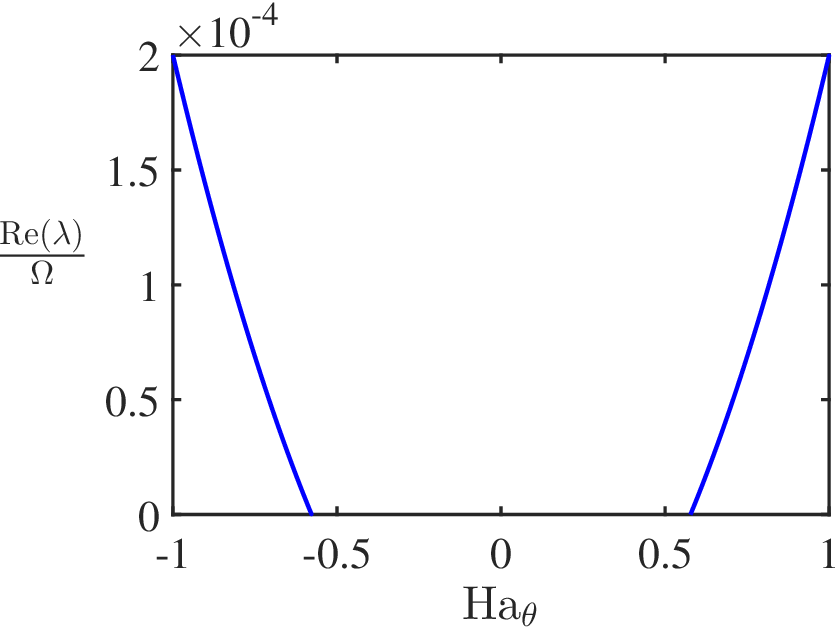}}
 \caption{The growth rate $\mathrm{Re}(\lambda_2)$ of \rf{eqn:AMRI-growthrate-k0} versus $\mathrm{Ha}_\theta$ when $\mathrm{Re}=10^4$, $m=1$, $\widehat{k}=\widehat{q}=0$, $\mathrm{Ro}=-3/4$ and $\mathrm{Rb}=-1$. The right panel is the close-up view of the left one near $\mathrm{Ha}_\theta=0$, demonstrating a certain strength of magnetic field needed for instability.
}
\label{fig:hatheta}
\end{center}
\end{figure*}

\subsection{Weak field}
\label{sec:weak field}

To examine the instability when magnetic field is weak, we express the solution of (\ref{eqn:AMRI-disp}) in powers of small parameter $\mathit{Ha}_\theta$. Then its leading-order term reads
\begin{eqnarray}
\frac{\lambda}{\Omega}&=&-\frac{1}{Re}-i m\left[1+\frac{2{\color{black}(1+Ro)}}{\widehat{h}^2+ \widehat{q}^2} \left(Ro -\frac{\widehat{k}^2}{\widehat{h}^2}\right)
    \right]
              \notag\\	
&\pm&2 \sqrt{-\alpha^2(1 + Ro)-\frac{m^2{\color{black}(1+Ro)^2}}{(\widehat{h}^2+ \widehat{q}^2)^2} \left(Ro-\frac{\widehat{k}^2}{\widehat{h}^2 }\right)^2}
		\notag\\	
&+&O(\mathit{Ha}_\theta).
\label{eqn:AMRI-samll-ha}
\end{eqnarray}
The radicand should be positive in total for instability. The first term in the radicand $-\alpha^2(1 + Ro)$ becomes positive for $Ro<-1$ and the second one is definitely non-positive. This non-positive term has the effect of decreasing the growth rate.
In particular, setting $m=0$ in (\ref{eqn:AMRI-samll-ha}) yields
\begin{eqnarray}
\frac{\lambda}{\Omega}&=&\pm 2\alpha i\sqrt{1+Ro}-\frac{1}{Re}.
\label{eqn:AMRI-samll-hare}
\end{eqnarray}
From (\ref{eqn:AMRI-samll-hare}) it follows that instability requires  \cite{KSF14JFM}
$$Ro<Ro_c=-1-\frac{1}{4\alpha^2 Re^2}.$$
Compared with the ideal hydrodynamics, for which the critical Rossby number is $Ro_c=-1$, the critical Rossby number is lowered by $1/(4\alpha^2Re^2)$ and the maximum growth rate is decreased by $1/Re$ due to viscosity.

When $Ro>-1$, to which the Keplerian flow ($Ro=-3/4$) belongs, the nonaxisymmetric as well as the axisymmetric modes decay as $\lambda/\Omega\approx -1/Re$.

\subsection{Strong field}
\label{sec:strong field}
We turn to the case of a strong magnetic field. The Reynolds number is assumed to be large. The axial wavenumber $\widehat{k}$ is an important parameter for determining the maximum growth rate and the instability region.

FIG.~\ref{fig:k 0 to infty} shows the growth rate given by equation \rf{eqn:AMRI-disp} as a function of $\widehat{k}$ for different values of $Rb$. We fix $m=1$, $Ro=-3/4$ and $\widehat{q}=0$, because numerically the modes of $\widehat{q}=0$ exhibit the fastest growth. We observe that at around $Rb=-1/4$, there is some finite $\widehat{k}$ at which the growth rate takes the maximum value. When $Rb$ is smaller than $-1/4$, the fast growth rate gives way to the $\widehat{k}=0$ mode at $Rb=-1$. When $Rb$ is increased above $-1/4$ by a certain amount, the maximum growth rate is attained in the limit of $\widehat{k} \to \infty$.

\subsubsection{The limit $\widehat{k}\rightarrow 0$}

The observations described above suggest us to examine closer the limit of $\widehat{k}\rightarrow 0$, which means letting $\alpha\rightarrow 0$ and $\widehat{h}\rightarrow m$ in (\ref{eqn:AMRI-disp}). In this limit the roots of (\ref{eqn:AMRI-disp}) at ${\rm Re} \gg 1$ take the form
\begin{eqnarray}
\frac{\lambda_{1}}{\Omega}&=&-i m -\left(1 + \mathit{Ha}_\theta^2 m^2\right)\frac{1}{Re},
              \notag\\	
\frac{\lambda_{2}}{\Omega}&=&-i m\left(1 + \frac{4Ro{\color{black}(1+Ro)}}{m^2+\widehat{q}^2}\right)
\notag\\	
&&
-\left[1 +\mathit{Ha}_\theta^2 m^2\left(1 +  \frac{4Rb}{m^2+\widehat{q}^2}\right)\right]\frac{1}{Re}.
\label{eqn:AMRI-growthrate-k0}
\end{eqnarray}

A glance at (\ref{eqn:AMRI-growthrate-k0}) shows that the axisymmetric mode ($m=0$) is excluded from the unstable ones and the growth rate $\Re(\lambda_1)$ is always negative.
The growth rate $\Re(\lambda_2)$ is positive provided that
\begin{eqnarray}
&&Rb<-\frac{1}{4}(m^2+\widehat{q}^2)\ \text{and} \ \mathit{Ha}_\theta^2>\frac{1}{m^2 \left(\frac{4|Rb|}{m^2+\widehat{q}^2}-1\right)}.
\notag\\
&&
\label{eqn:AMRI-growthrate-k0-instab}
\end{eqnarray}

FIG.~\ref{fig:hatheta} displays the growth rate $\Re(\lambda_2)$ as a function of $\mathit{Ha}_\theta$ when $Re=10^4$, $m=1$, $\widehat{k}=\widehat{q}=0$, $Ro=-3/4$ and $Rb=-1$. The left panel shows that the growth rate increases with $\mathit{Ha}_\theta$; the right panel is the close-up view near the origin. We recognize that the small but nonzero value  $|\mathit{Ha}_\theta|=1/\sqrt{3}\approx 0.5774$ is necessary for the onset of instability.

Note that rather than $Ro$, it is now $Rb$ that is tied with the instability and the negative value of $\dd \mu/\dd r$ is required. The maximum growth rate is attained at $\widehat{q}=0$.

When $Rb=-1$, the $m=\pm 1$ modes are the only possible modes for instability.

When $m=\pm 1$ is fixed, $Rb<-1/4$ is necessary for the instability of the $\widehat{k}=0$ mode.

It is remarkable that the instability exists, beyond the restriction of the Liu limit, for arbitrary Rossby number $Ro$.
However we should be cautious about this result, because the modes of $\widehat{q}=0$ lie outside the regime of validity of the radial WKB approximation. Later in the article, we argue about the limitation on $\widehat{q}$.

   \begin{figure*}[tbhp]
  \begin{center}
{\includegraphics*[scale=0.4]{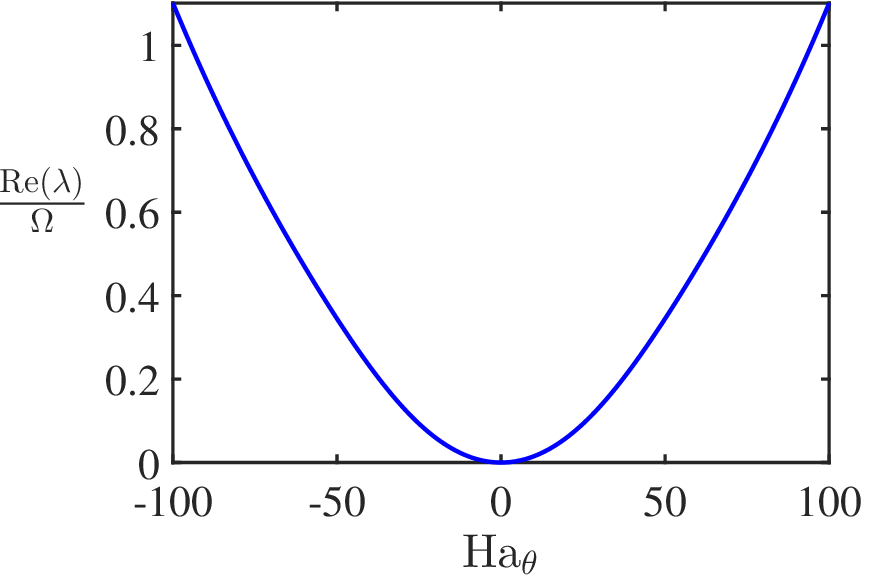}~
\includegraphics*[scale=0.4]{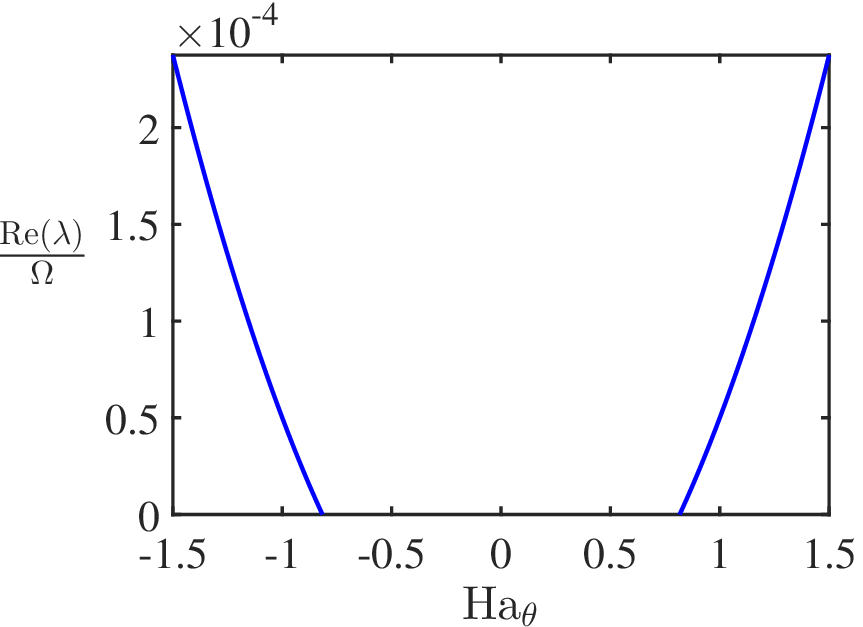}}
 \caption{The growth rate $\mathrm{Re}(\lambda)$ to $\mathrm{Ha}_\theta$ when $\mathrm{Re}=10^4$, $m=1$, $\widehat{k}\rightarrow \infty$, $\alpha=1$, $\mathrm{Ro}=-3/4$ and $\mathrm{Rb}=0$ according to \rf{eqn:AMRI-growthrate-kinfty-1}. The left panel shows that large $\mathrm{Ha}_\theta$ increase the growth rate and the right panel is the amplification of the left one when $\mathrm{Ha}_\theta$ is small, which demonstrates that a certain strength of magnetic field is needed for instability.
}
\label{fig:hatheta2}
\end{center}
\end{figure*}

\subsubsection{The limit $\widehat{k}\rightarrow \infty$}

{\color{black}In the limit $\widehat{k}\rightarrow\infty$, where $\widehat{q}^2$ is replaced by $\widehat{k}^2/\alpha^2-\widehat{k}^2-m^2$ ($0 \le \alpha \le 1$), the roots of (\ref{eqn:AMRI-disp}) take the form \cite{KSF14,KSF14JFM}
\ba{eqn:AMRI-growthrate-kinfty-1}
\frac{\lambda_{1,2}}{\Omega}&=&N_A(2\alpha^2 Rb-m^2)-im-\frac{1}{Re}
              \nn \\
  &\pm& 2\alpha \Big\{N_A^2 (m^2 {+} \alpha^2 Rb^2)
	   +
   im N_A (2 + Ro) -1 {-} Ro\Big\}^\frac{1}{2},\nn\\
\ea
where $N_A={\mathit{Ha}_\theta^2}/{Re}$ is the Elsasser number for the azimuthal magnetic field \cite{KSF14,KSF14JFM}.}

\textcolor{black}{By expanding the eigenvalues (\ref{eqn:AMRI-growthrate-kinfty-1}) to first order in $1/Re$ we get \cite{KSF14JFM,Pri11}
\ba{eqn:AMRI-growthrate-kinfty}
\frac{\lambda_{1,2}}{\Omega}&=&-i m \pm 2 \alpha \sqrt{-(1 + Ro)}
             \nn \\
  &+& N_A \left(2 \alpha^2 Rb-m^2\pm
     \frac{ \alpha m  (2 + Ro)}{\sqrt{1 + Ro}}\right)-\frac{1}{Re}.
		\nn \\
\ea
}
\textcolor{black}{When $Ro<-1$, the instability occurs with the growth rate $\Re(\lambda)/\Omega\approx 2 \alpha \sqrt{-(1 + Ro)}$. This mode pertains to the classical Rayleigh instability since no magnetic field is required.}

\textcolor{black}{When $Ro>-1$, the instability criterion becomes
 \begin{eqnarray}
&&-n^2+
     |n|\frac{2 + Ro}{\sqrt{1 + Ro}}+2 Rb>0,
              \notag\\	
&&\text{and} \ \alpha^2\mathit{Ha}_\theta^2>\frac{\sqrt{1 + Ro}}{(2  Rb-n^2)\sqrt{1 + Ro}+
   |n|  (2 + Ro)},\notag\\
   &&
\label{eqn:AMRI-growthrate-kinfty-instab1}
\end{eqnarray}
where $n=m/\alpha$. If we choose that, e.g. $Rb=0$, $m=1$ and $\alpha=1$, then $|\mathit{Ha}_\theta|\approx 0.8165$ is the onset of AMRI at the Keplerian $Ro=-3/4$ as shown in FIG.~\ref{fig:hatheta2}.}

The left-hand side of the first of the inequalities (\ref{eqn:AMRI-growthrate-kinfty-instab1})
is a quadratic polynomial with respect to the real-valued number $m$. Hence, the discriminant of this polynomial
$$
D=\frac{(2+Ro)^2}{1+Ro}+8Rb>0
$$
in order that the polynomial can take positive values. This yields the familiar \cite{KirSte13,KSF14JFM} necessary condition for instability \rf{eqn:HMRI-RbRo}. For instance, for Keplerian flow $Ro=-3/4$ in \rf{eqn:HMRI-RbRo} the inequality $Rb>-25/32$ is necessary for instability \cite{KirSte13,KSF14JFM}.

{\color{black}
The first inequality in (\ref{eqn:AMRI-growthrate-kinfty-instab1}), when $Rb\le n^2/2-n$, is written for $Ro$ as
\begin{eqnarray}
&&-1<Ro<-2+\frac{n^2-2Rb-\sqrt{(n^2-2Rb)^2-4n^2}}{2n^2(n^2-2Rb)^{-1}}\nn\\
&&{\rm or} \,\, Ro>-2+\frac{n^2-2Rb+\sqrt{(n^2-2Rb)^2-4n^2}}{2n^2(n^2-2Rb)^{-1}},
\label{eqn:AMRI-growthrate-kinfty-instab}
\end{eqnarray}
and when $Rb>n^2/2-n$, as
\begin{eqnarray}
&&Ro>-1.
\label{eqn:AMRI-growthrate-kinfty-instab1}
\end{eqnarray}
When $n=\pm\sqrt{-2Rb}$,} the domain \rf{eqn:AMRI-growthrate-kinfty-instab} reduces to \rf{eqn:instability HMRI-m=0, rb<0}, which, at $Rb=-1$ takes the form
$$
-1<Ro<2-2\sqrt{2}\quad {\rm or} \quad Ro>2+2\sqrt{2},
$$
where $2-2\sqrt{2}$ and $2+2\sqrt{2}$ are the lower and the upper Liu limits, respectively  \cite{LGHJ06}.

\subsubsection{Growth rate optimized by $\widehat{k}$ and $\widehat{q}$}

In the long wavelength limit of $\widehat{k}\rightarrow 0$, $Rb<-1/4$ is necessary for the instability of $m=1$ mode as shown by (\ref{eqn:AMRI-growthrate-k0-instab}), while in the short wavelength limit of $\widehat{k} \rightarrow \infty$, the condition $Rb>-\frac{1}{8}\frac{(Ro+2)^2}{Ro+1}$ given by \rf{eqn:HMRI-RbRo} is necessary for the instability. Since the latter one overlaps with the former one, we conclude that for each value of $Rb$ there exist wavenumbers $\widehat{k}$ and $\widehat{q}$ such that the mode with $m=1$ is unstable.

\begin{figure}[tbhp]
  \begin{center}
\includegraphics*[scale=0.4]{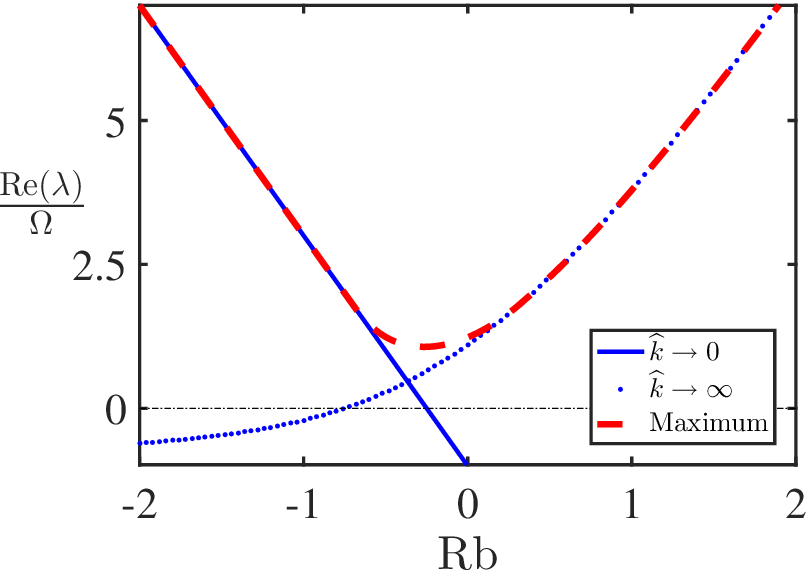}
\includegraphics*[scale=0.4]{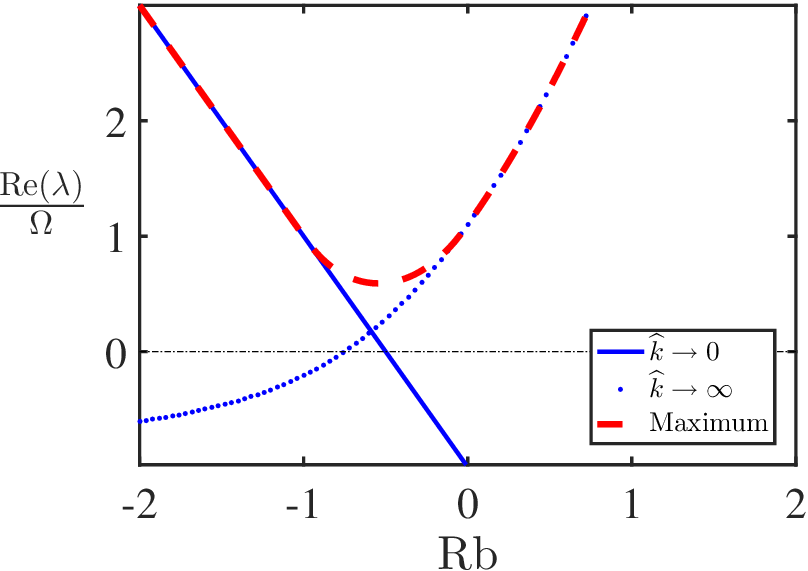}
 \caption{ The growth rate to magnetic Rossby number $\mathrm{Rb}$ for $\mathrm{Re}=10^4$, $\mathrm{Ha}_\theta=100$, $m=1$, $\mathrm{Ro}=-3/4$ and $\widehat{q}=0$ (upper panel) or $\widehat{q}=1$ (lower panel) according to \rf{eqn:AMRI-disp}. The solid line is $\widehat{k}=0$ mode; the dotted one is the $\widehat{k}\rightarrow\infty$ mode and the dashed line stands for the growth rate maximized over $\widehat{k}$, whose left part tends to the $\widehat{k}=0$ mode and the right part tends to the $\widehat{k}=\infty,\ \alpha=1$ mode.}
\label{fig:crossover3}
\end{center}
\end{figure}

Either the mode of $\widehat{k}\rightarrow 0$ or $\widehat{k}\rightarrow \infty$ dominate in large range of $Rb$, and the maximum growth rate is attained at a finite value of $\widehat{k}$ for every particular value of the magnetic Rossby number, $Rb$, as illustrated in FIG.~\ref{fig:crossover3}. In this figure the optimized with respect to $\widehat{k}$ growth rate is plotted against $Rb$ for $Re=10^4$, $\mathit{Ha}_\theta=100$, $m=1$ and $Ro=-3/4$ and $\widehat{q}=0$ (upper panel) and $\widehat{q}=1$ (lower panel). We observe the crossover of the $\widehat{k}=0$ mode and the $\widehat{k}=\infty$ mode. The range of large negative values of $Rb$ is dominated by the $\widehat{k}=0$ mode and the one of large positive values of $Rb$ is dominated by the $\widehat{k}\rightarrow \infty$ mode.

\begin{figure*}[tbhp]
\begin{center}
{\includegraphics*[scale=0.35]{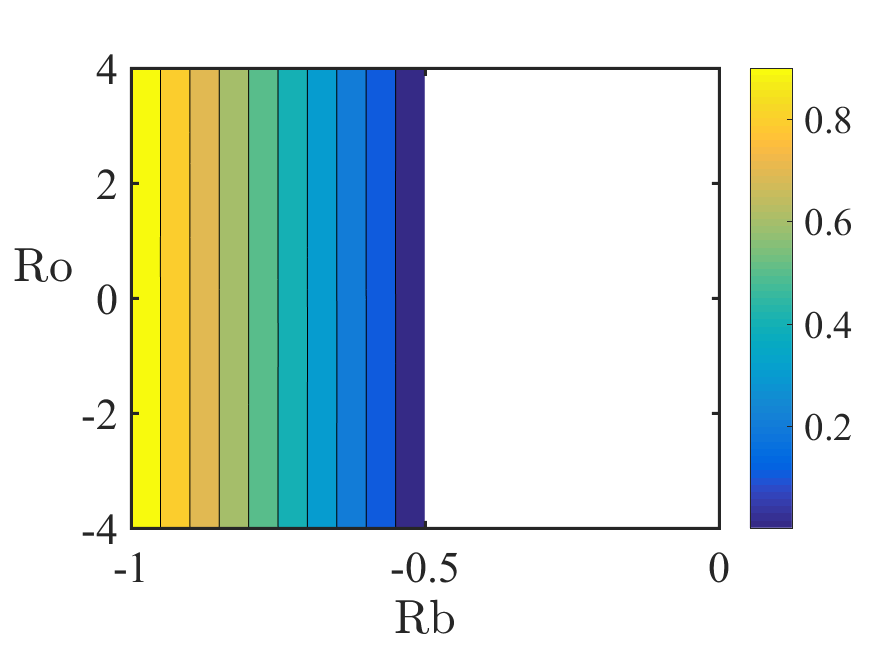}~
\includegraphics*[scale=0.35]{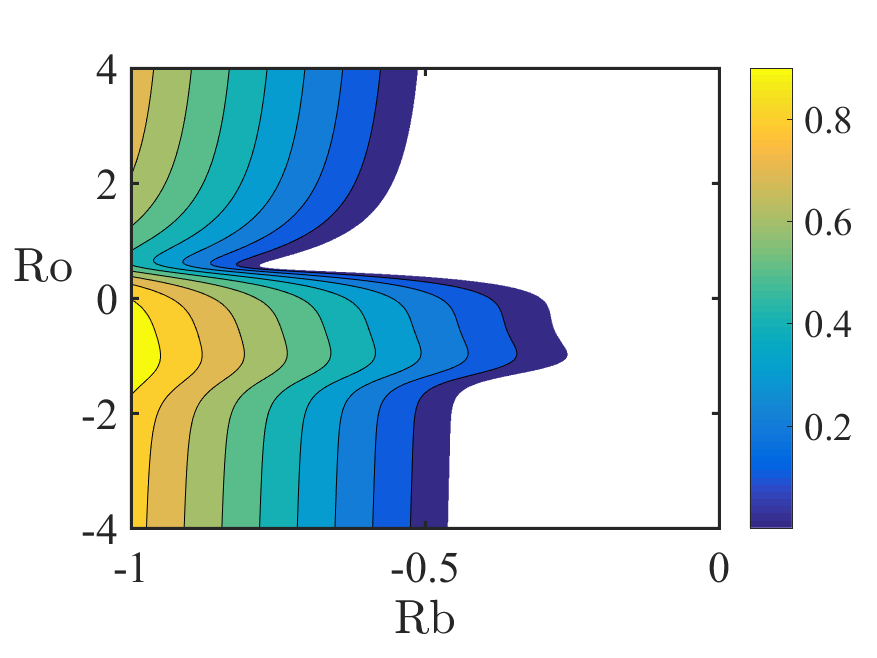}~
\includegraphics*[scale=0.35]{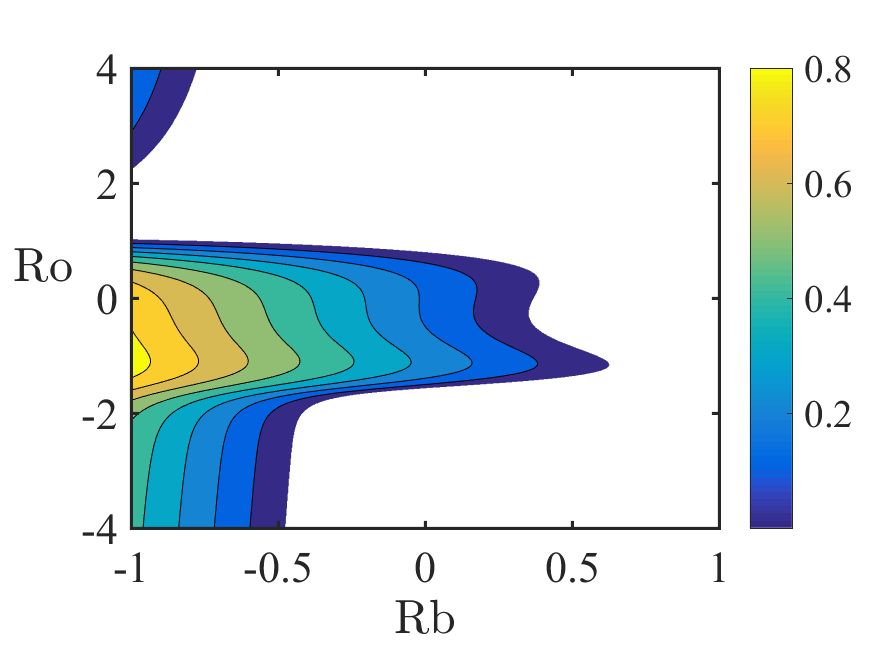}}
{\includegraphics*[scale=0.35]{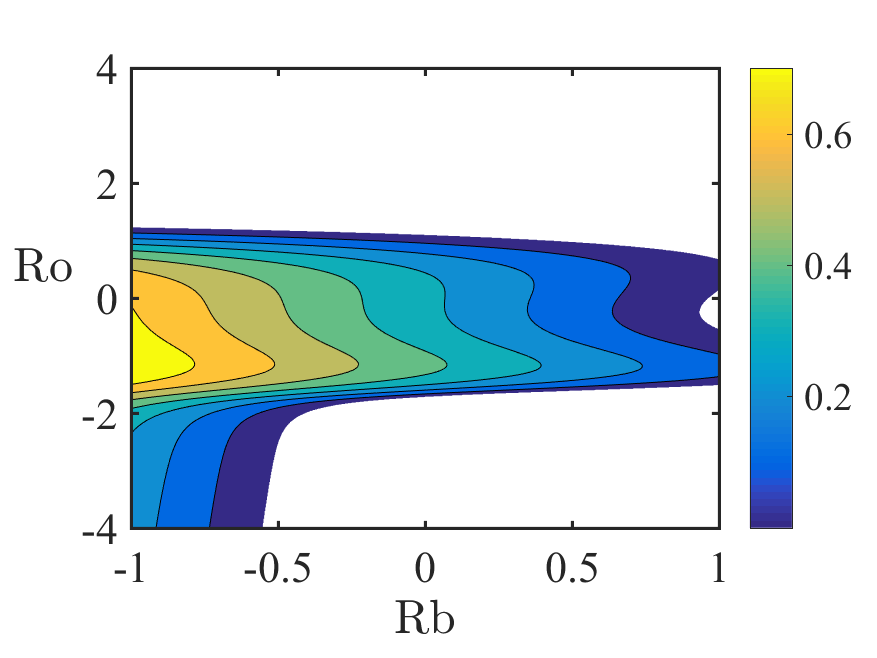}~
\includegraphics*[scale=0.35]{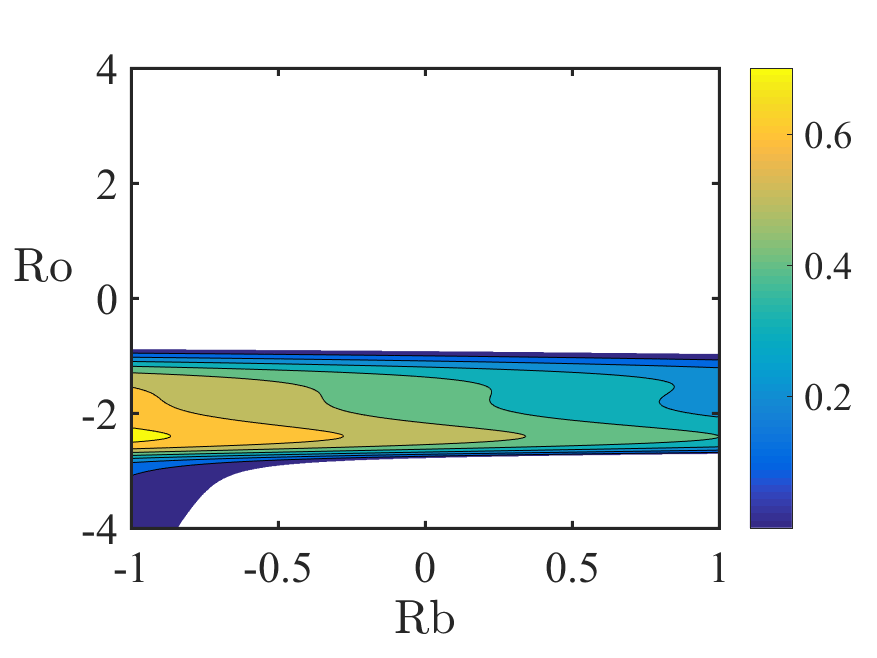}~
\includegraphics*[scale=0.35]{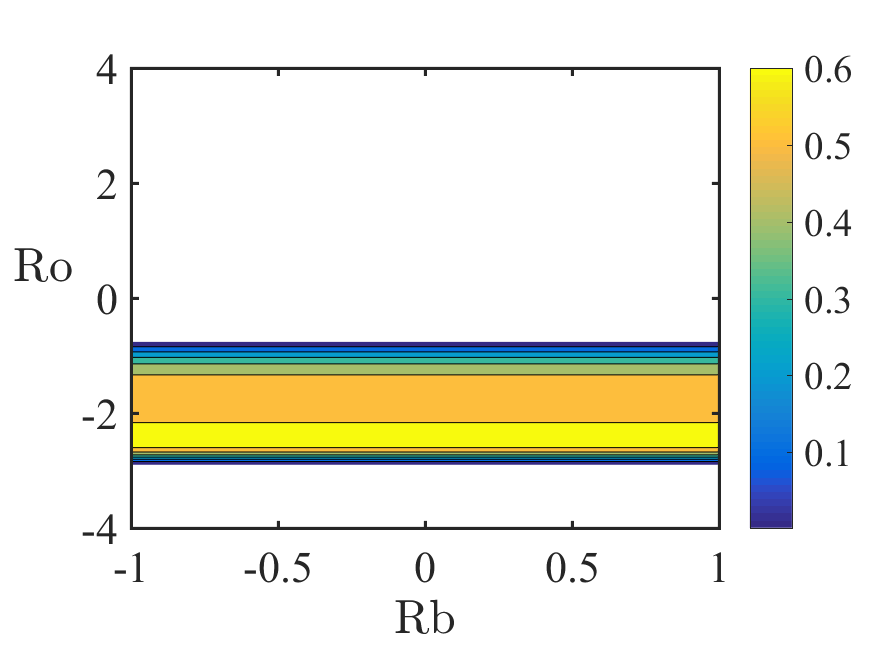}}
{\includegraphics*[scale=0.35]{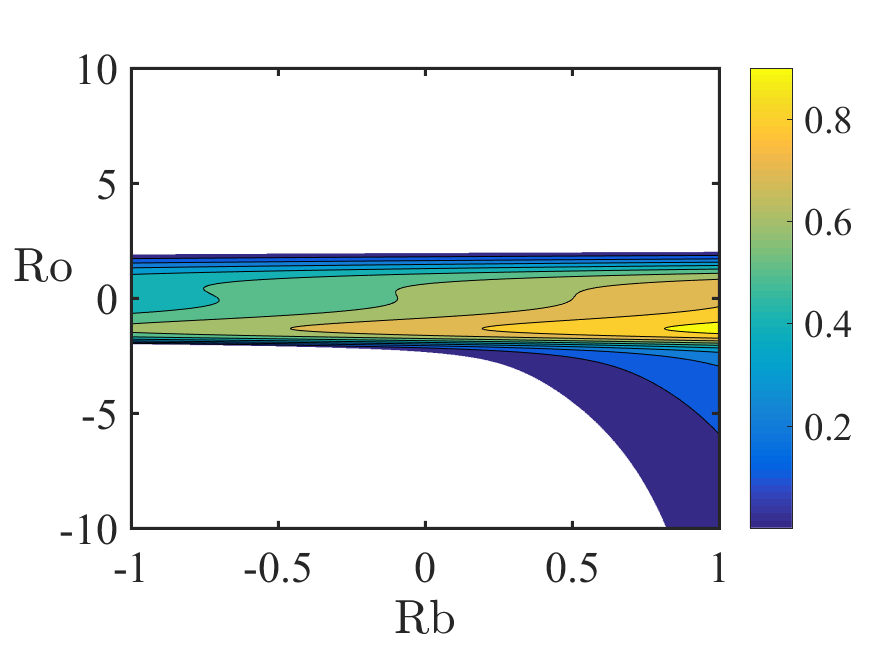}~
\includegraphics*[scale=0.35]{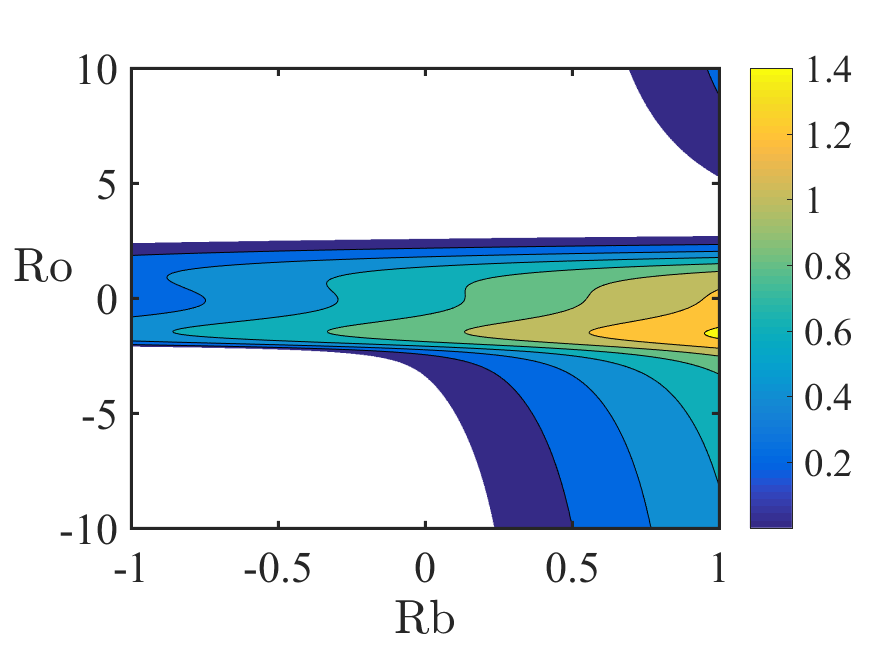}~
\includegraphics*[scale=0.35]{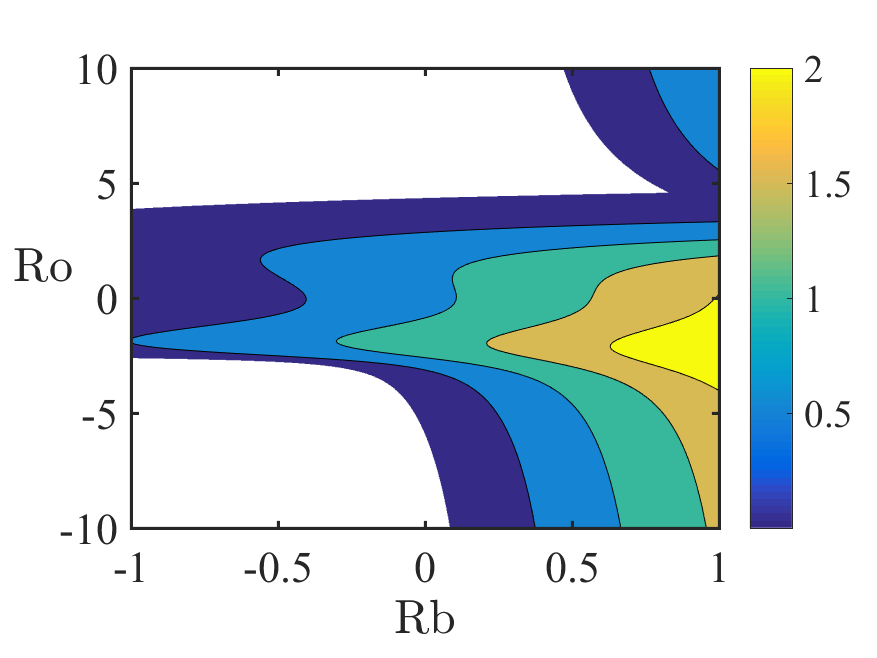}}
{\includegraphics*[scale=0.35]{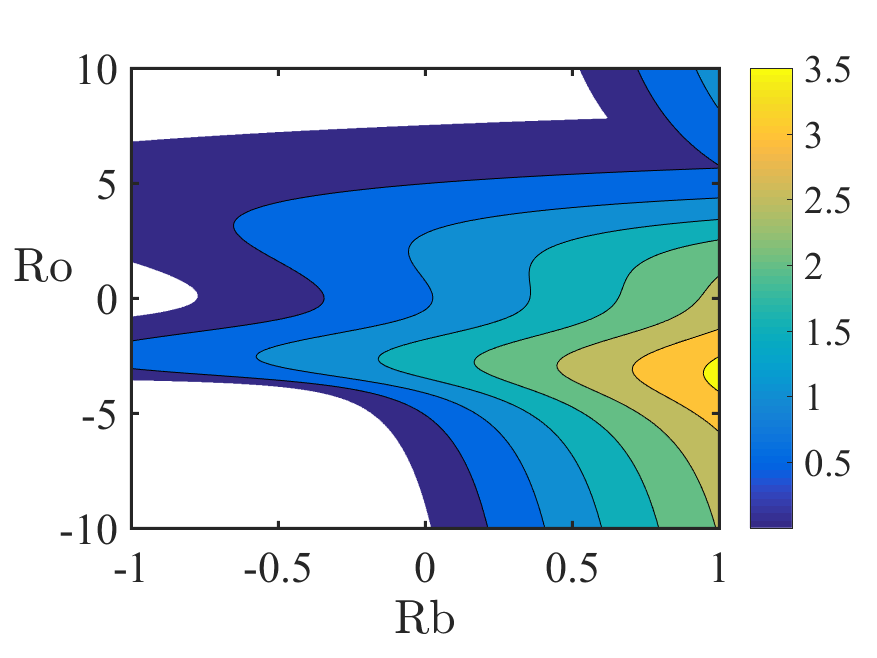}~
\includegraphics*[scale=0.35]{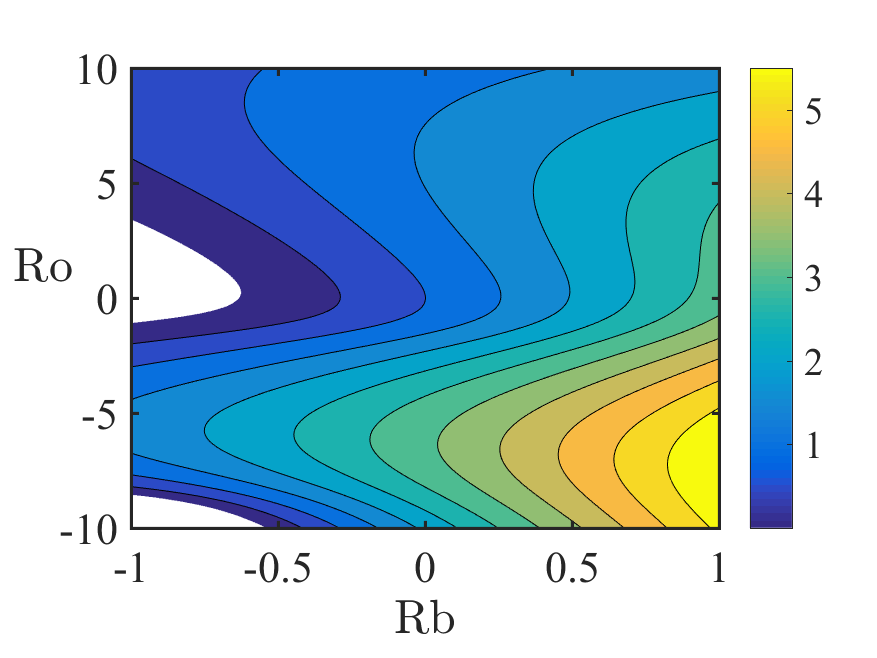}~
\includegraphics*[scale=0.35]{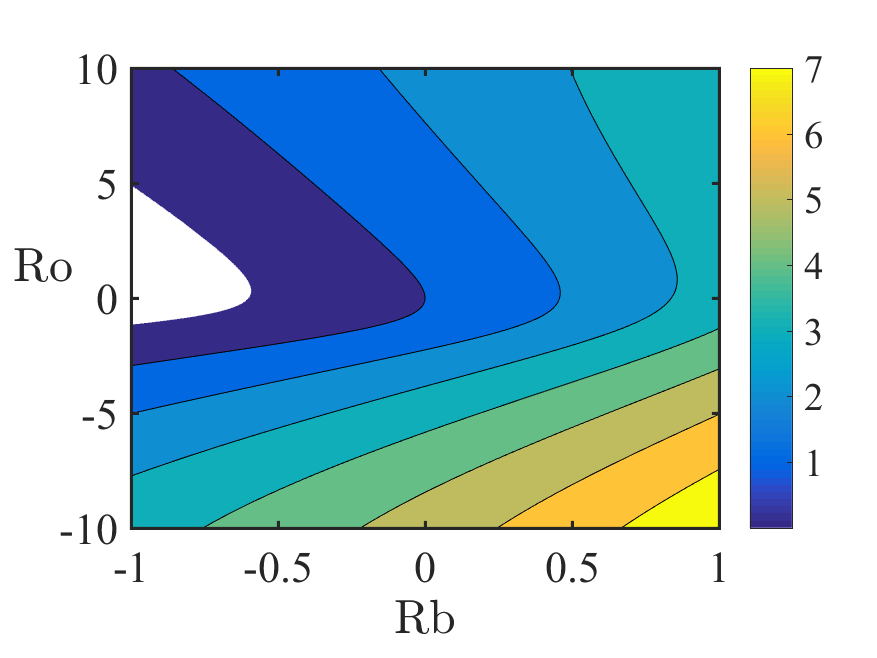}}
\caption{Growth rate calculated with the use of the Hain-L\"ust dispersion relation \rf{eqn:dispers-rel-inducless-azim} in projection to the Rossby plane ($\mathrm{Rb},\mathrm{Ro}$) for $\mathrm{Re}=10^4$, $\mathrm{Ha}_{\theta}=10^2$, $\widehat{q}=1$, $m=1$ and (from upper-left to lower-right panel): $\widehat{k}=0.01$, $0.4$, $0.7$, $0.8$, $0.9$, $1$, $1.1$, $1.3$, $1.8$, $2.5$, $5$ and $10$. The white domains represent stability.}
\label{fig:HL_Rossby_Plane}
\end{center}
\end{figure*}


\subsubsection{Evolution of AMRI region in the $(Ro,Rb)$-plane with $\widehat{k}$}
\label{sec:indefinite_Ro_Rb}

In order to understand how the instability region evolves from that described by \rf{eqn:AMRI-growthrate-k0-instab} at $\widehat{k}\rightarrow 0$ to \rf{eqn:AMRI-growthrate-kinfty-instab} at $\widehat{k}\rightarrow \infty$ we plot the growth rate of the dispersion relation \rf{eqn:dispers-rel-inducless-azim} in the projection to the $(Rb,Ro)$-plane, see FIG.~\ref{fig:HL_Rossby_Plane}.
The results are presented over a growing set of axial wavenumber $\widehat{k}$ for $Re=10^4$, $Ha_{\theta}=10^2$, $Pm=10^{-6}$, $\widehat{q}=1$, and $m=1$.

It is clearly seen that already for $\widehat{k}>1.8$ the neutral stability curve bounding the stability domain (shown in white in FIG.~\ref{fig:HL_Rossby_Plane}) is close to $Rb=-\frac{1}{8}\frac{(Ro+2)^2}{Ro+1}$ corresponding to the limit of $Pm \rightarrow 0$. Equivalently, the instability domain is close to \rf{eqn:AMRI-growthrate-kinfty-instab}.

At the lower values of $\widehat{k}$ the instability domain splits into two parts, one of which becomes dominant at $\widehat{k}=1.3$ stretching along the $Rb$-axis at $\widehat{k}=1$ and finally bifurcating into the instability domain corresponding to large negative values of $Rb$ and practically not depending on $Ro$, in agreement with the criterion \rf{eqn:AMRI-growthrate-k0-instab}.

Below we demonstrate a similar transition for the domain of Tayler instability.

   \begin{figure*}[tbhp]
  \begin{center}
{\includegraphics*[scale=0.4]{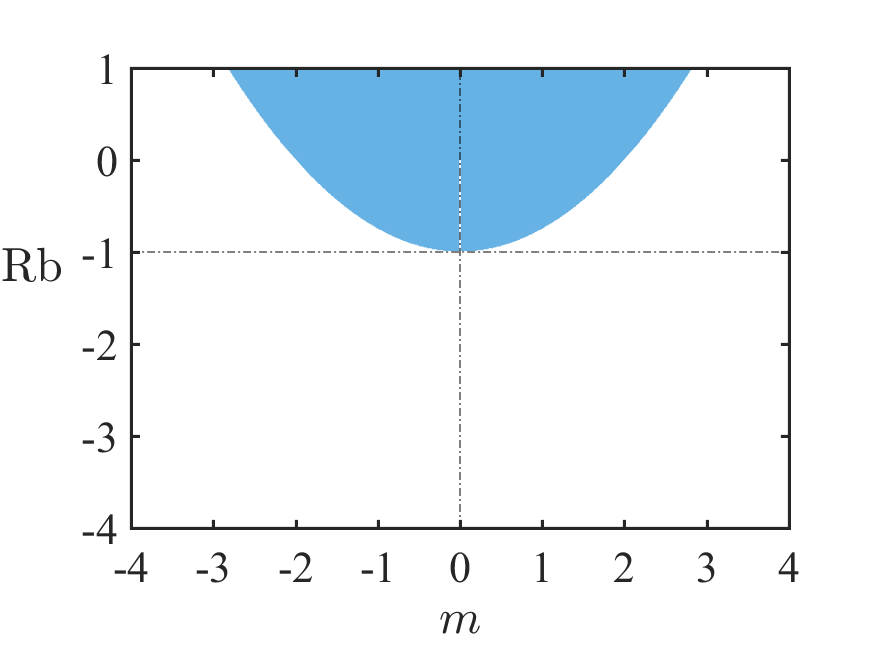}~
\includegraphics*[scale=0.4]{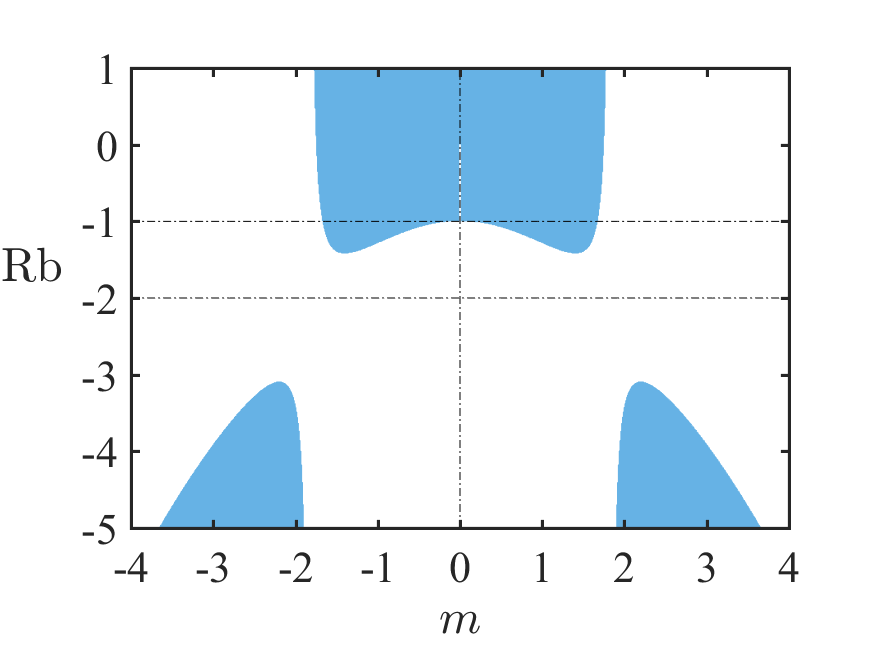}~
\includegraphics*[scale=0.4]{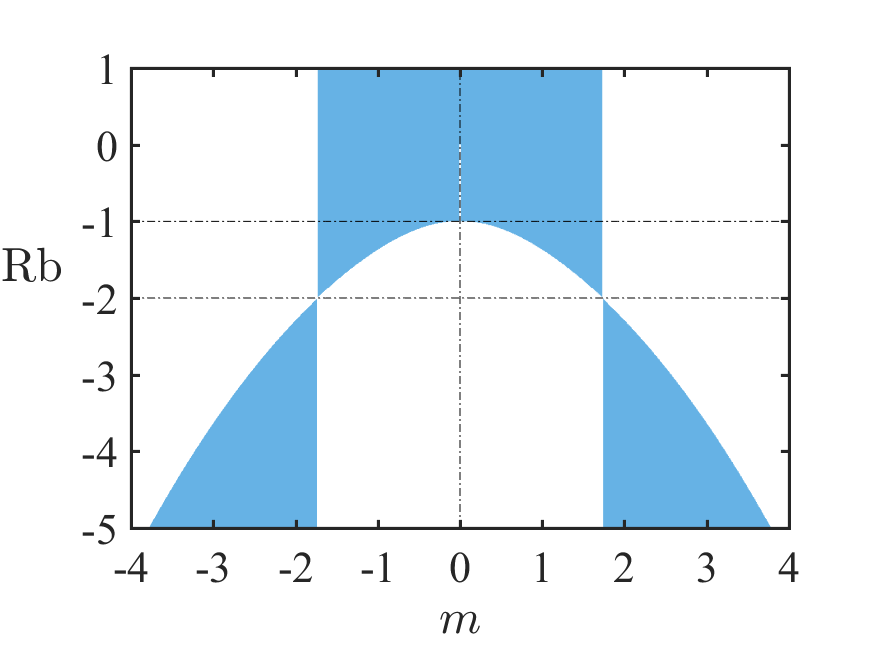}}
{\includegraphics*[scale=0.4]{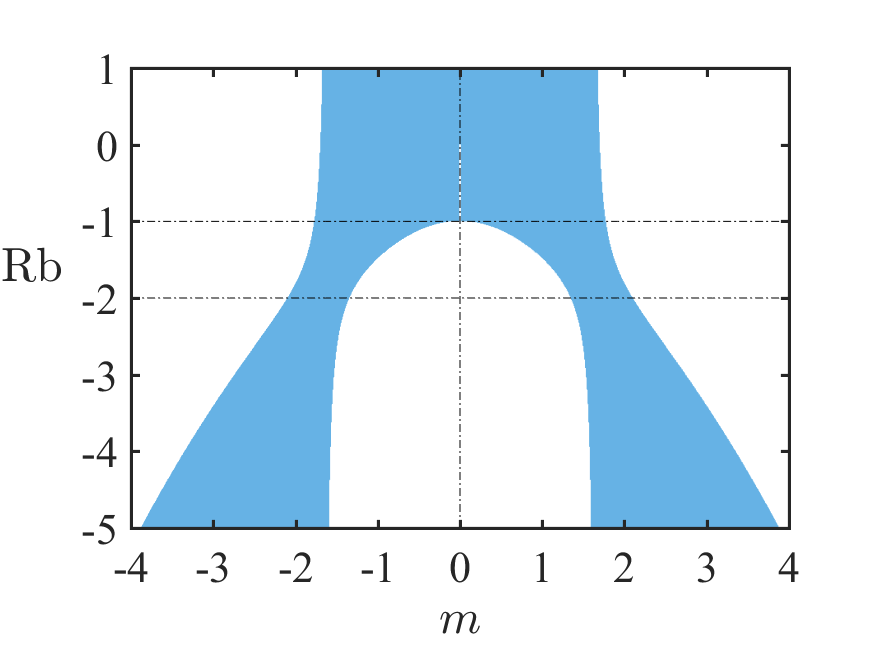}~
\includegraphics*[scale=0.4]{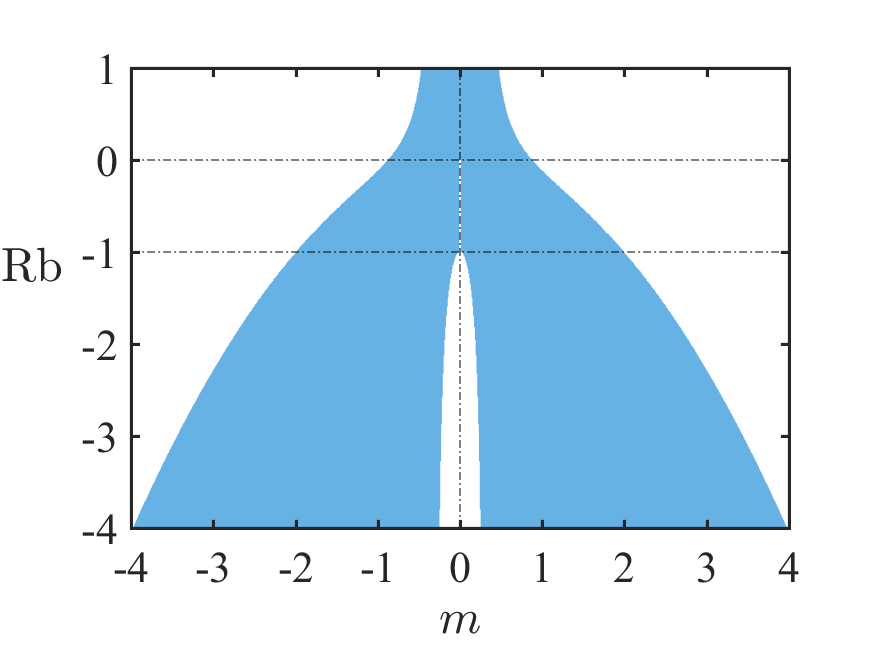}~
\includegraphics*[scale=0.4]{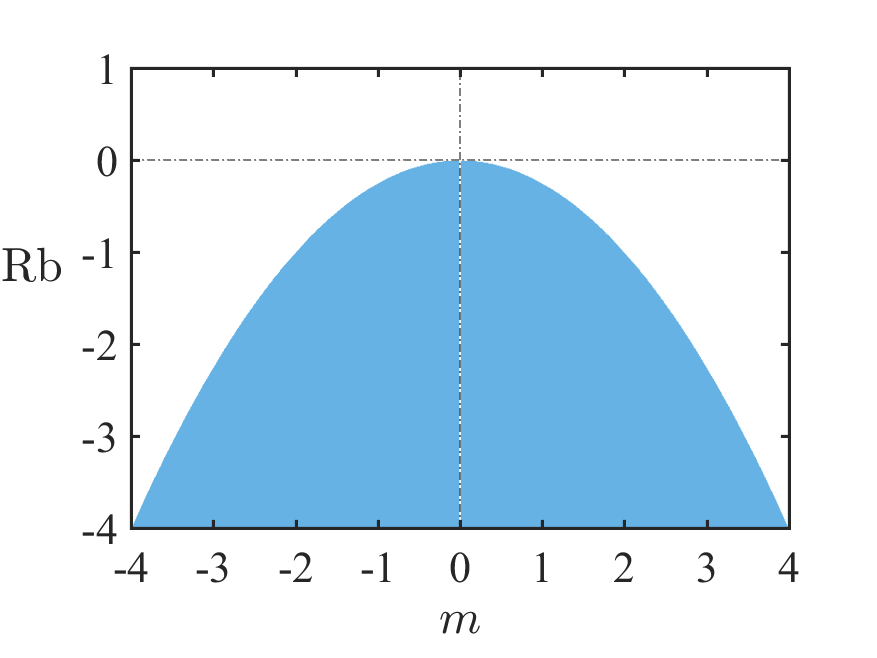}}
 \caption{ The regions of the Tayler instability (blue) with the boundary \rf{tig} for $\widehat{q}=0$ and (top row from left to right) $\widehat{k}=100$, $\widehat{k}=\sqrt{3}+0.1$, and $\widehat{k}=\sqrt{3}$ and (bottom row from left to right) $\widehat{k}=\sqrt{3}-0.1$, $\widehat{k}=0.3$, and $\widehat{k}\rightarrow0$.}
\label{fig:titransit}
\end{center}
\end{figure*}


\subsection{Tayler instability in the limit of $Pm \rightarrow 0$}
\label{sec:TI}
Tayler \cite{Tay73, RS2010} established that an ideal nonrotating perfectly conducting fluid in an azimuthal magnetic field is stable against nonaxisymmetric perturbations with the azimuthal wavenumber $m=1$ under the condition
\be{ti1}
\frac{d}{dr}(rB^2_{\theta}(r))<0.
\ee
Recalling the definition of the magnetic Rossby number \rf{eqn:def-Rossby-num} and taking into account that $B_{\theta}(r)=r\mu(r)$, the Tayler stability criterion for $m=1$ takes the form:
\be{ti2}
Rb<-\frac{3}{4},
\ee
which means that the azimuthal magnetic field $B_{\theta}(r)\sim r$ created by a current passing through a conducting fluid and corresponding to $Rb=0$ is unstable.

The work  \cite{RS2010} numerically predicted the Tayler instability (TI) caused by the field with $Rb=0$ to exist also in the limit of $Pm \rightarrow 0$, which allowed for its recent observation in the experiments with liquid metals  \cite{Tayler2012}.

Using the geometrical optics stability analysis Kirillov et al.  \cite{KSF14JFM} extended the criterion for the onset of the Tayler instability to the case of arbitrary $m\ge 1$
\be{ti3}
Rb>\frac{m^2}{4\alpha^2}-1,
\ee
where $\alpha=\widehat{k}^2/(\widehat{k}^2+\widehat{q}^2)$. When $m=\pm 1$ and $\alpha=1$, the criterion \rf{ti3} yields $Rb>-3/4$ for instability, which includes the case of $Rb=0$ observed in the experiment  \cite{Tayler2012}.

In order to explore the Tayler instability on the base of the dispersion relation (\ref{eqn:dispers-rel-inducless-azim}),
we assume $Re=0$ in it and take into account the relation
\be{43}
\frac{Re\sqrt{Pm}}{\Omega}=\frac{\mathit{Ha}_\theta}{\omega_{A_\theta}}.
\ee
This reduces (\ref{eqn:dispers-rel-inducless-azim}) to
\begin{eqnarray}
&& \Big[\Big(\frac{\lambda\mathit{Ha}_\theta}{\omega_{A\theta}}+\sqrt{Pm}\Big)\Big(\frac{\lambda\mathit{Ha}_\theta}
{\omega_{A\theta}}+\frac{1}{\sqrt{Pm}}\Big)+\mathit{Ha}_\theta^2 m^2\Big]^2
(\widehat{h}^2 + \widehat{q}^2) \notag\\
&& -4 \mathit{Ha}_\theta^4 \widehat{k}^2 m^2 -
 4 \mathit{Ha}_\theta^2 \Big((\widehat{k}^2- m^2)Rb+\frac{ m^2\widehat{k}^2 }{\widehat{h}^2}\Big)\\
 &&\times\Big[\Big(\frac{\lambda\mathit{Ha}_\theta}{\omega_{A\theta}}+\sqrt{Pm}\Big)\Big(\frac{\lambda\mathit{Ha}_\theta}{\omega_{A\theta}}+
 \frac{1}{\sqrt{Pm}}\Big)+\mathit{Ha}_\theta^2 m^2\Big]=0.\nn
\label{eqn:taylor-re=0-disp}
\end{eqnarray}

We consider the limit where $Pm$ is very small. Then the growth rate is of $O\left(\sqrt{Pm}\right)$ and we can renormalize the eigenvalue as
\begin{eqnarray}
&&\lambda=\lambda_0\sqrt{Pm}.
\label{eqn:taylor-re=0-disp1}
\end{eqnarray}
Then the leading-order terms of \rf{eqn:taylor-re=0-disp} are
\begin{eqnarray}
&& \bigg(1 + \frac{\lambda_0}{\omega_{A\theta}} \mathit{Ha}_\theta+\mathit{Ha}_\theta^2 m^2\bigg)^2 (\widehat{h}^2 + \widehat{q}^2) \notag\\
&&-4 \mathit{Ha}_\theta^4 \widehat{k}^2 m^2-4 \mathit{Ha}_\theta^2 \left[ \left(\widehat{k}^2-m^2\right)Rb+\frac{ m^2\widehat{k}^2 }{\widehat{h}^2}\right]\notag\\
&&\times\bigg(1 + \frac{\lambda_0}{\omega_{A\theta}}\mathit{Ha}_\theta + \mathit{Ha}_\theta^2 m^2\bigg)
=0.
\label{eqn:taylor-re=0-disp-pm=0}
\end{eqnarray}
For very large magnetic field  we have $\mathit{Ha}_\theta\gg 1$, and can further renormalize the eigenvalue as
\begin{eqnarray}
\lambda_0&=&\lambda_a \mathit{Ha}_\theta.
\label{eqn:taylor-gr0}
\end{eqnarray}
and solve (\ref{eqn:taylor-re=0-disp-pm=0}) for $\lambda_a$, to the leading order in ${\mathit{Ha}_\theta}^{-1}$, as
\ba{eqn:taylor-gr}
	\frac{\lambda_a}{\omega_{A\theta}}&=&-m^2+\frac{2}{\widehat{h}^2+\widehat{q}^2}\Bigg\{(\widehat{k}^2-m^2)Rb+\frac{m^2\widehat{k}^2}{\widehat{h}^2}
  \notag\\
	&\pm&
  \sqrt{\left[(\widehat{k}^2-m^2)Rb+\frac{m^2\widehat{k}^2}{\widehat{h}^2}\right]^2+m^2\widehat{k}^2(\widehat{h}^2+\widehat{q}^2)}\Bigg\}.\nn\\
\ea

In the limit of $\widehat{k} \rightarrow 0$, equation \rf{eqn:taylor-gr} yields
\be{k0}
\frac{\lambda_a}{\omega_{A\theta}}=\frac{\pm2m^2Rb-m^2(\widehat{q}^2+m^2+2Rb)}{\widehat{q}^2+m^2}.
\ee
One of the roots \rf{k0} is equal to $-m^2$, whereas another one becomes positive if
\be{ti0}
Rb<-\frac{1}{4}(m^2+\widehat{q}^2)
\ee
reproducing the first of the inequalities \rf{eqn:AMRI-growthrate-k0-instab}.

   \begin{figure*}[tbhp]
  \begin{center}
{\includegraphics*[scale=0.4]{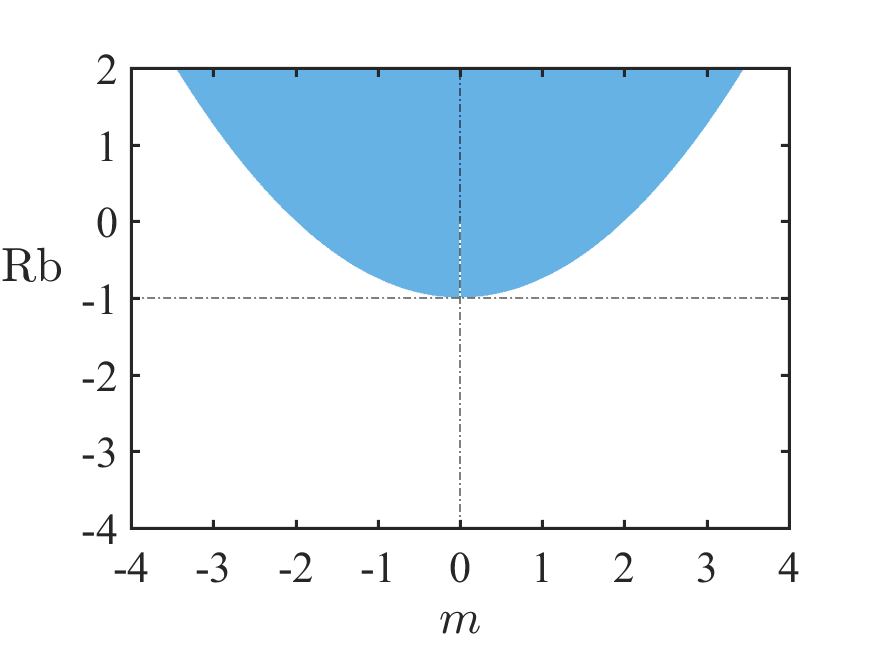}~
\includegraphics*[scale=0.4]{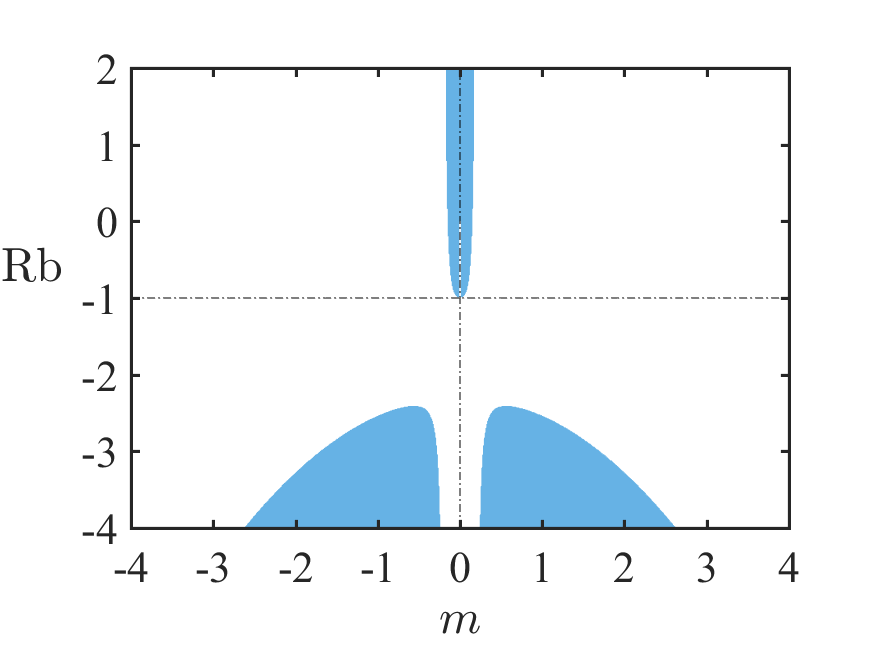}~
\includegraphics*[scale=0.4]{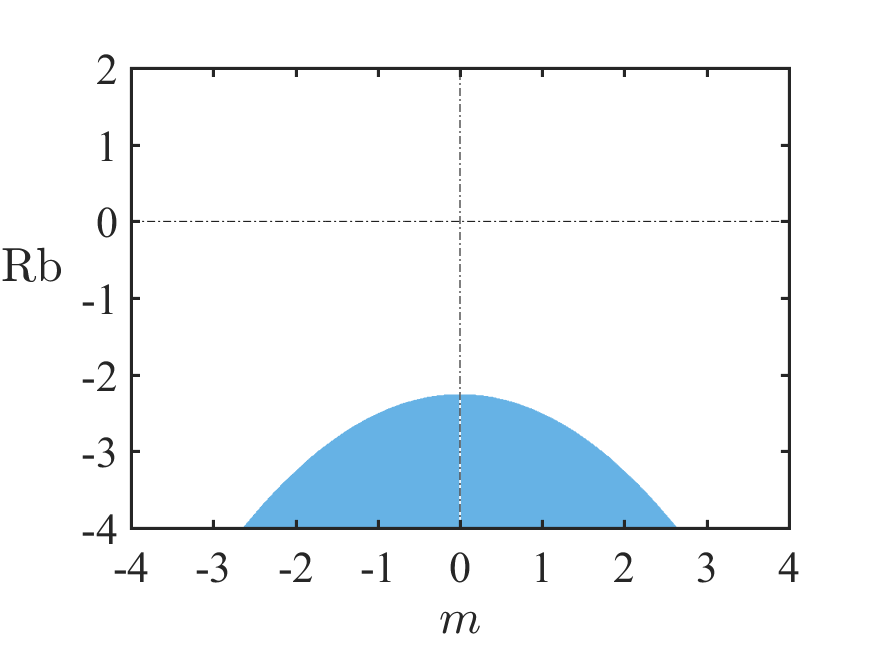}}
 \caption{ The regions of the Tayler instability (blue) with the boundary \rf{tig} for $\widehat{q}=3$ and (from left to right) $\widehat{k}=100$, $\widehat{k}=0.2$, and $\widehat{k}\rightarrow0$.}
\label{fig:titransit1}
\end{center}
\end{figure*}

In the limit of $\widehat{k} \rightarrow \infty$ and $\alpha \rightarrow 1$, equation \rf{eqn:taylor-gr} reduces to
\ba{kinf}
\frac{\lambda_a^{\pm}}{\omega_{A\theta}}&=&2Rb-m^2\pm2\sqrt{Rb^2+m^2}\nn\\
&=&(1+Rb)^2-\left(1\mp \sqrt{Rb^2+m^2}\right)^2.
\ea
The root $\lambda_a^{-}/\omega_{A\theta}$ in \rf{kinf} is always negative. The other,
$$
\frac{\lambda_a^{+}}{\omega_{A\theta}}=\left(Rb+\sqrt{Rb^2+m^2}\right)\left(2+Rb-\sqrt{Rb^2+m^2}\right),
$$
is a product of two expressions, the first of which is always positive whereas $2+Rb-\sqrt{Rb^2+m^2}$ is positive if
\be{ti1}
Rb>\frac{m^2}{4}-1
\ee
in accordance with \rf{ti3}, where $\alpha=1$. Therefore in the short axial wavelength approximation
we reproduce the result \cite{KSF14JFM}. Note that Ogilvie and Pringle \cite{OgiPri96} established criterion \rf{ti1} for the case of ideal MHD.

In general, setting the right hand side of \rf{eqn:taylor-gr} to zero, we get the critical $Rb$ at the neutral stability surface
\be{tig}
Rb=\frac{1}{4}\left\{\frac{\widehat{q}^2m^2}{\widehat{k}^2-m^2}+\frac{\widehat{k}^2(m^2-4)(\widehat{k}^2+2m^2)+m^6}{\widehat{k}^4-m^4}\right\}.
\ee
For $\widehat{k}=0$ the expression \rf{tig} yields the critical value of the criterion \rf{ti0} and for $\widehat{k}\rightarrow \infty$
the critical value of the criterion \rf{ti1}.

FIG.~\ref{fig:titransit} illustrates the transition from the criterion \rf{ti1} to the criterion \rf{ti0} as $\widehat{k}$ varies from $100$ to $0$ at the fixed $\widehat{q}=0$, based on the expression \rf{tig}. At the value
\be{bif}
\widehat{k} = \sqrt{3-\frac{1}{2}\widehat{q}^2}
\ee
(equal to $\sqrt{3}$ for $\widehat{q}=0$ in FIG.~\ref{fig:titransit}) there are two saddle points at
\be{saddle}
Rb = \frac{1}{8}\widehat{q}^2-2\quad {\rm and} \quad m = \pm \sqrt{3-\frac{1}{2}\widehat{q}^2},
\ee
corresponding to $Rb=-2$ and $m=\pm \sqrt{3}$ in FIG.~\ref{fig:titransit}.
The saddle points are formed by the straight lines $m = \pm \sqrt{3-\frac{1}{2}\widehat{q}^2}$ intersecting
with the curve
$$
Rb = \frac{2m^4+(18-\widehat{q}^2)m^2-4\widehat{q}^2+24}{4(\widehat{q}^2-2m^2-6)}.
$$

   \begin{figure*}[tbhp]
  \begin{center}
{\includegraphics*[scale=0.4]{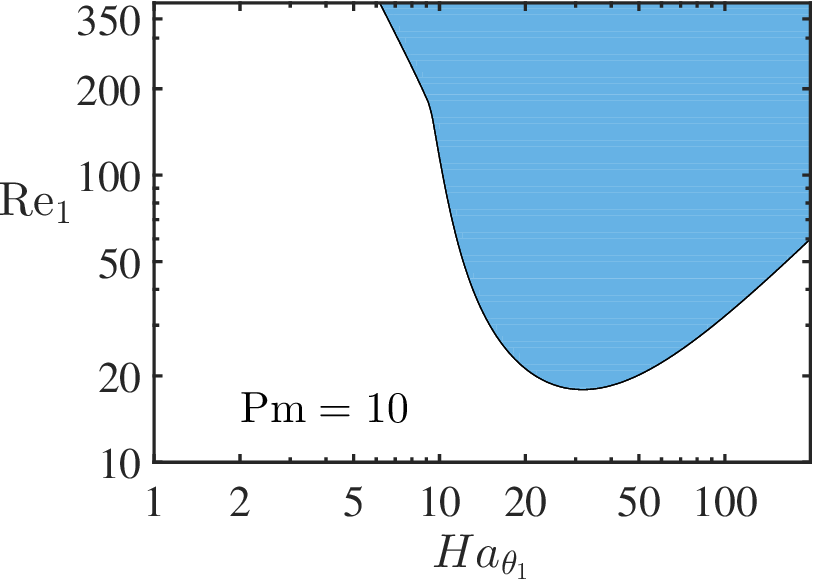}
\includegraphics*[scale=0.4]{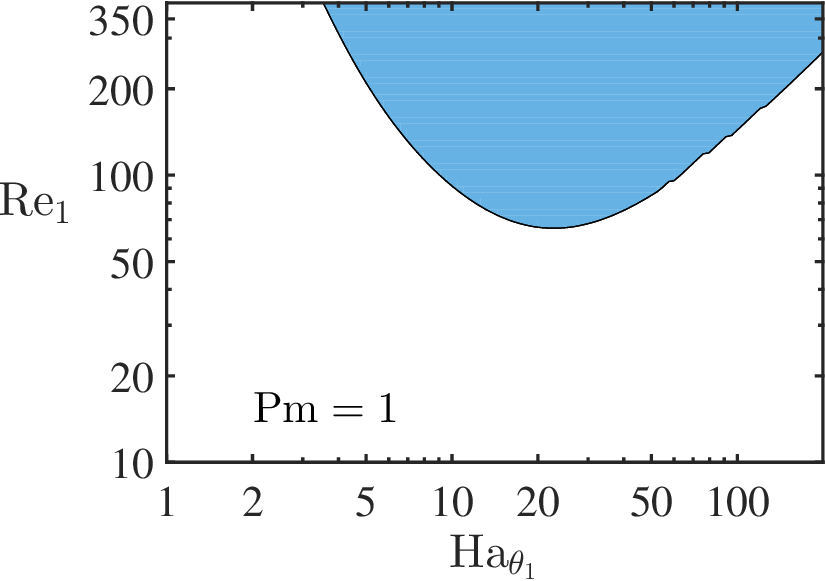}~
\includegraphics*[scale=0.4]{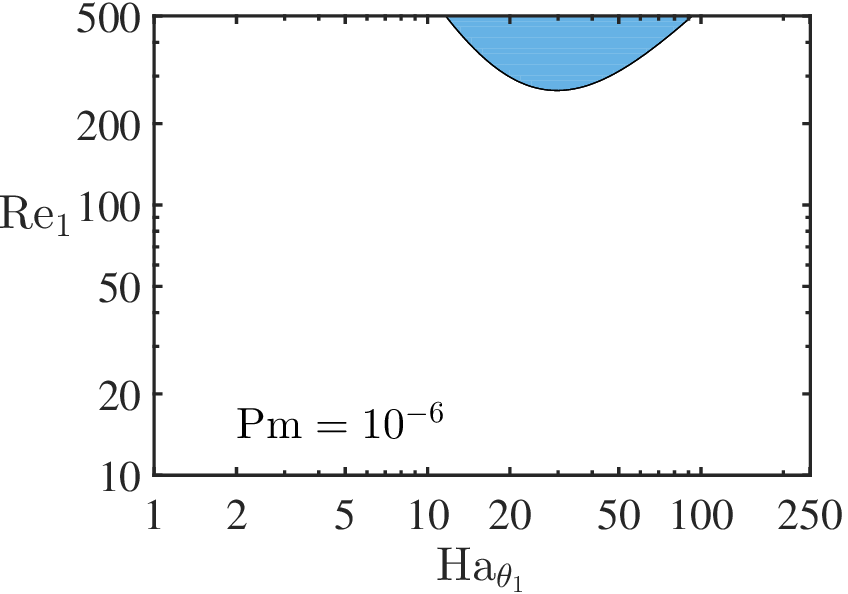}}
 \caption{ AMRI regions (above the neutral stability curves) in the $(\mathrm{Ha}_{\theta_1},\mathrm{Re}_1)$-plane for $\mathrm{Rb}=\mathrm{Ro}=-1$, $m=1$, $q=3r_0^{-1}$, $r=1.5r_0$ and (left to right) $\mathrm{Pm}=10$, $\mathrm{Pm}=1$, and $\mathrm{Pm}=10^{-6}$ found with the use of the growth rates maximized over $k$ of the roots of the dispersion relation \rf{eqn:dispers-rel-inducless-azim} with the parameters specified by \rf{eqn:new re and ha}.}
\label{fig:ro1rb1pm01}
\end{center}
\end{figure*}

Note that \rf{saddle} implies an upper bound on the value of $\widehat{q}$: $|\widehat{q}|<\sqrt{6}$. In these conditions
the bifurcation value \rf{bif} for the parameter $\widehat{k}$ sharply separates the cases of the short-axial-wavelength \rf{ti1} and long-axial-wavelength \rf{ti0} Tayler instability in the limit of vanishing $Pm$. However, in the case $|\widehat{q}|>\sqrt{6}$ the saddle point is absent and the transition scenario simplifies, see FIG.~\ref{fig:titransit1}

\section{AMRI and Tayler Instability at finite $Pm$}
\label{sec:Non-axisymmetric AMRI critical Re Ha}

The magnetorotational instability is, by definition, caused by the cooperative effect of rotating flow field and magnetic field. The cooperative action comes into play for a differential rotation. Assuming the expansion of the solution in terms of $Re$ as $\lambda^{\pm}/\omega_{A\theta}=a_0Re+a_1+a_2Re^{-1}+a_3Re^{-2}+\ldots$, we expand the dispersion relation (\ref{eqn:dispers-rel-inducless-azim}) with respect to $1/Re$ and solve the leading-order term to obtain $a_0$. We repeat the process to find the coefficient $a_1$ from the next-order term resulting in the following representation for the critical roots at large $Re$:{\color{black}
\begin{eqnarray}
\label{eqn:mri-re=infty-lambda1234}
\frac{\lambda_{1,2}}{\omega_{A\theta}}&=&-im\frac{Re \sqrt{Pm}}{\mathit{Ha}_\theta} - \frac{1}{\mathit{Ha}_\theta\sqrt{Pm}} \pm m\frac{\sqrt{-Ro(Ro+1)}}{Ro+1}\notag\\
&&+O\left(\frac{1}{Re}\right)\notag\\
\frac{\lambda_{3,4}}{\omega_{A\theta}}&=&\frac{-iRe \sqrt{Pm}}{\mathit{Ha}_\theta\widehat{h}^2(\widehat{q}^2+\widehat{h}^2)} \Big(m\big(\widehat{k}^2(\widehat{k}^2+\widehat{q}^2+2Ro-2)\notag\\
&&+m^4+m^2(2\widehat{k}^2+\widehat{q}^2+2Ro)\big)\pm 2i\sqrt{c_2}\Big)\notag\\
&&+c_3\notag\\
&&+O\left(\frac{1}{Re}\right),
\end{eqnarray}
where
\begin{eqnarray*}
c_1&=&\widehat{h}^2Ro-\widehat{k}^2\notag\\
c_2&=&-\mathit{Ha}^2_\theta Pm\left(\widehat{k}^2\widehat{h}^4(\widehat{h}^2+\widehat{q}^2)(1+Ro)+m^2c_1^2\right)\notag\\
c_3&=&-\frac{\sqrt{Pm}}{\mathit{Ha}_\theta}\notag\\
&&-\frac{(1-Pm)c_1Ro\left(-m^2c_1\mathit{Ha}_\theta\sqrt{Pm}\pm im\sqrt{c_2}\right)}{c_2\pm imc_1\sqrt{c_2}\mathit{Ha}_\theta\sqrt{Pm}}.
\label{eqn:c1c2c3}
\end{eqnarray*}
The growing wave $\Re(\lambda)>0$ corresponding to $\lambda_{1,2}$ for the particular case of Keplerian flow ($Ro=-3/4$) with $m=1$ is admitted for
\begin{eqnarray}
\mathit{Ha}_\theta>\frac{1}{\sqrt{3Pm}}.
\label{eqn:mri-insta1}
\end{eqnarray}
For $\lambda_{3,4}$, numerically we find that a growing wave is permitted for small $Pm$ and finite $\widehat{k}$. For example, for the Keplerian flow ($Ro=-3/4$) and $m=\widehat{k}=\mathit{Ha}_\theta=1$, $Pm=0.01$, $\widehat{q}=10$, the zeroth order growth rate $c_3\approx 1.68$. }

In the PROMISE laboratory facility  \cite{SGGGS14}, the experimental setup is a Taylor-Couette flow between two co-rotating cylinders of finite axial size. The inner cylinder is set with the radius $r_{\text{\rm in}}=40$mm and the outer cylinder is with $r_{\text{out}}=2r_{\rm in}=80$mm. The gap between the cylinders is $d=r_{\text{out}}-r_{\text{\rm in}}=r_{\rm in}$. By that reason, in this section we assume $r_{\text{\rm in}}=d=r_0$. Recalling (\ref{eqn:def-Rossby-num}), we can write
\be{omu}
\Omega(r_{\rm in})=\Omega(r)\left(\frac{r_0}{r}\right)^{2Ro},\quad \mu(r_{\rm in})=\mu(r)\left(\frac{r_0}{r}\right)^{2Rb}.
\ee
This allows us to redefine the Reynolds and Hartmann numbers as follows
\ba{eqn:new re and ha}
Re_1&=&\frac{\Omega(r_{\text{in}})d^2}{\nu}=Re|\bm{k}|^2r_0^2\left(\frac{r_0}{r}\right)^{2Ro},
              \notag\\	
\mathit{Ha}_{\theta_1}&=&\frac{\mu(r_{\text{in}})d^2}{\sqrt{\rho\mu_0\nu\eta}}=\mathit{Ha}_\theta|\bm{k}|^2r_0^2\left(\frac{r_0}{r}\right)^{2Rb},
\end{eqnarray}
where $|\bm{k}|^2=k^2+q^2+m^2/r^2$ and $Re$ and $\mathit{Ha}_\theta$ are given by (\ref{eqn:def-non-dim-num1}). The new Reynolds and Hartmann numbers \rf{eqn:new re and ha} match those of the numerical and experimental works  \cite{SGGGS14,RGSHS14, RudHol07}.

The critical Reynolds number at the onset of instability is crucial for the experimental realization of the MRI. The liquid metals used in the experiments have $Pm\sim10^{-6}$, and the standard MRI which scales with the magnetic Reynolds number and the Lundquist number corresponds to the Reynolds numbers of order $10^6$. Therefore it is hard to maintain the basic flow undisturbed before the onset of SMRI  \cite{B2011,JB2013}.

   \begin{figure*}[tbhp]
  \begin{center}
{\includegraphics*[scale=0.4]{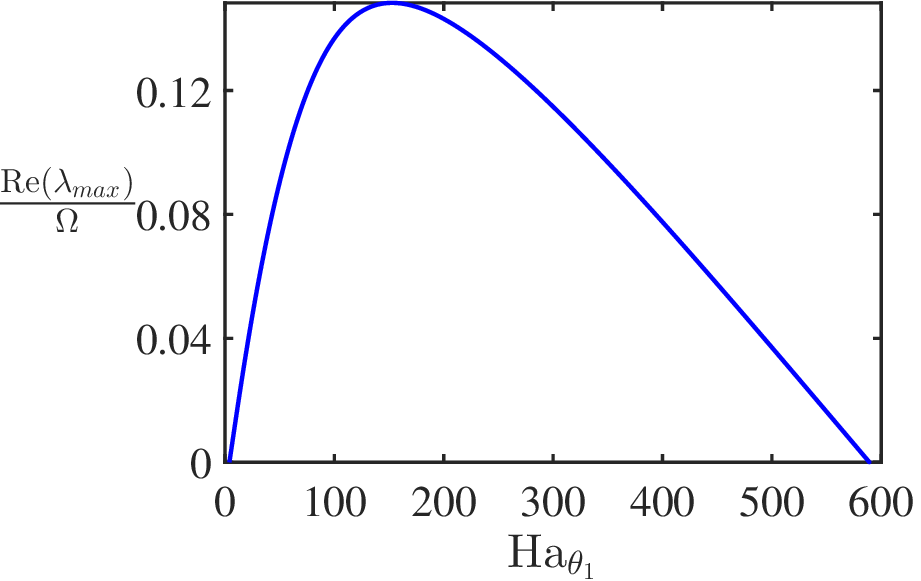}
\includegraphics*[scale=0.4]{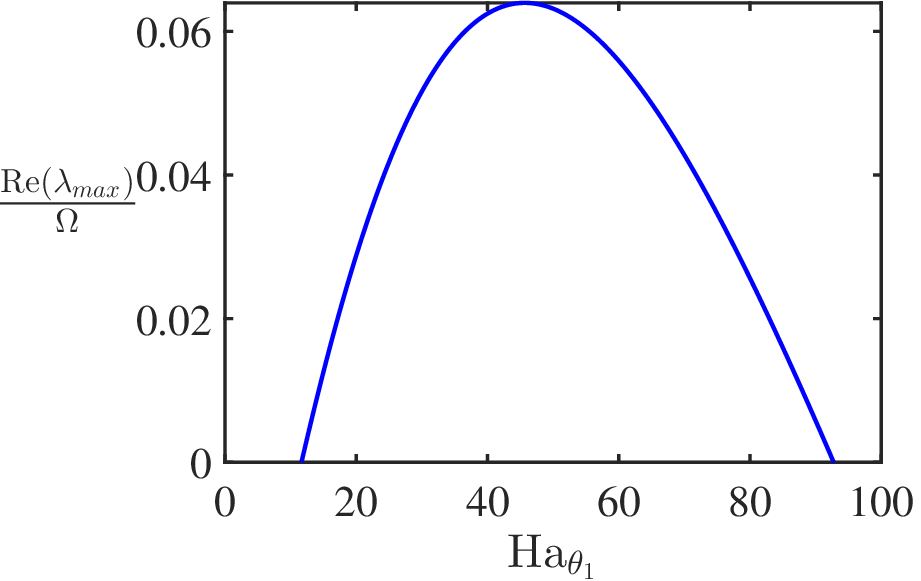}}
 \caption{The maximized over $k$ growth rate $\mathrm{Re}(\lambda_{max})$ in the units of $\Omega$ versus $\mathrm{Ha}_{\theta_1}$ according to equation \rf{eqn:dispers-rel-inducless-azim} with the parameters \rf{eqn:new re and ha} for the flow with $\mathrm{Ro}=-1$, $\mathrm{Rb}=-1$, $\mathrm{Pm}=10^{-6}$, $m=1$, $q=3r_0^{-1}$, and $r=1.5r_0$ when (left) $\mathrm{Re}_1=3000$ and (right) $\mathrm{Re}_1=500$.}
\label{fig:gr2}
\end{center}
\end{figure*}

The helical and the azimuthal MRI scale with the Reynolds and Harmann numbers and thus require moderate ranges of the Reynolds numbers compared to SMRI  \cite{HolRud05}.
By that reason both HMRI and AMRI were detected in the laboratory experiments \cite{SGGRSSH06,SGGHPRS09,SGGGS14,Stefani2019,PhysRep2018} for rotation which is a little bit shallower than the Rayleigh value $\Omega\sim r^{-1.9}$ and for the current-free azimuthal magnetic field corresponding to $Rb=-1$.
In \cite{KirSte13, KSF14} it was theoretically shown that the inductionless HMRI and AMRI for the Keplerian flow with $Ro=-3/4$ exist when the radial dependence of the azimuthal magnetic field is shallower than that of the current-free type: $Rb>-\frac{1}{8}\frac{(Ro+2)^2}{Ro+1}$. In section~\ref{sec:Non-axisymmetric perturbations}, we have verified this result for large axial wavenumbers, $k \gg 1$. The planned AMRI-TI experiment in the frame of the new DRESDYN facility \cite{Stefani2019,PhysRep2018} creates the azimuthal magnetic field both due to currents isolated of the liquid metal and passing directly through the metal thus allowing for variable $Rb$ including those satisfying the instability criterion \rf{eqn:HMRI-RbRo}.

On the other hand, in section~\ref{sec:Non-axisymmetric perturbations} we have found that for small axial wavenumbers, $k \ll 1$, the inductionless AMRI of the Keplerian flow may occur at $Rb<-1/4$, which includes the current-free azimuthal magnetic field with $Rb=-1$ used in the existing PROMISE experiment, see FIG.~\ref{fig:crossover3}. Using the redefined Reynolds and Hartmann numbers \rf{eqn:new re and ha} in this section we
compare our WKB-analysis with the results from the global analysis \cite{HolTee10, RudHol07} for arbitrary $Pm$ and discuss the implications for the experimental detection of the long-axial-wavelength instability. In view of the recent discovery of a long-wavelength linear instability of a hydrodynamical Taylor-Couette flow \cite{Deg17} this direction is worth pursuing.

\subsection{Case of $Ro=Rb=-1$ and $m=1$ with $q=3r_0^{-1}$}
\label{sec:Ro=-1, Rb=-1, m=1}
Since the Taylor-Couette experimental apparatus is radially bounded, we limit $q$ from below and choose e.g. $q=3r_0^{-1}$, which is reasonable when the radial velocity disturbance should be zero on the boundary and the width between the two cylinders is $r_0$.
In FIG.~\ref{fig:ro1rb1pm01} we present  the instability region in the $(\mathit{Ha}_{\theta_1},Re_1)$-plane.
To find it, we numerically calculate the maximum growth rate at every meshing point in the $(\mathit{Ha}_{\theta_1},Re_1)$-plane for a wide range of $k$. Zero growth rates correspond to the neutral stability curve. The calculation is performed locally at $r=1.5r_0$, the average of $r_{\rm{in}}=r_0$ and $r_{\rm{out}}=2r_0$. Notice that the Tayler instability is excluded in this parameter regime by \rf{ti0} and \rf{ti1}.
We can see that FIG.~\ref{fig:ro1rb1pm01} is similar to Figure 1 of Hollerbach et al. \cite{HolTee10} and Figure 1 of R\"udiger et al. \cite{RGSHS14}. The instability is invited when the Reynolds number is of the order $10^2$ when $Pm\ll 1$ and of the order $10$ when $Pm\approx 1,\ 10$. When $Pm=10^{-6}$, the critical Reynolds number is $Re_1\approx 265$ which is attained at $\mathit{Ha}_{\theta 1}\approx 30,\ k=3.4727r_0^{-1}$ and $q=3r_0^{-1}$. \textcolor{black}{Note, however, that in the works \cite{HolTee10,RGSHS14} the instability domains
have a finite size along the $Re$-axis which yields the existence of the second critical Reynolds number by exceeding which the AMRI vanishes. The neutral stability curves based on our local dispersion relation do not catch this upper critical Reynolds number.}

The left panel of FIG.~\ref{fig:gr2} shows that for $Re_1=3000$ and $Pm=10^{-6}$, the instability occurs when $\mathit{Ha}_{\theta_1}\in (4,590)$. The growth rate has its extremum $\Re{(\lambda_{max})}/\Omega_{in}\approx 0.1483$   at $\mathit{Ha}_{\theta_1}\approx 153$ with the extremizer $k\approx 7.43r_0^{-1}$. On the right panel of FIG.~\ref{fig:gr2} corresponding to $Re_1=500$, the instability occurs for $\mathit{Ha}_{\theta_1}\in (12,\ 93)$. The growth rate reaches its extremum $\Re{(\lambda_{max})}/\Omega_{in}\approx 0.06397)$ at $\mathit{Ha}_{\theta_1}\approx 46$ with the extremizer $k\approx 4.35r_0^{-1}$.
We see that in both cases no instability occurs when the magnetic field is sufficiently weak in agreement with the argument in Section \ref{sec:weak field}.

   \begin{figure*}[tbhp]
  \begin{center}
{\includegraphics*[scale=0.45]{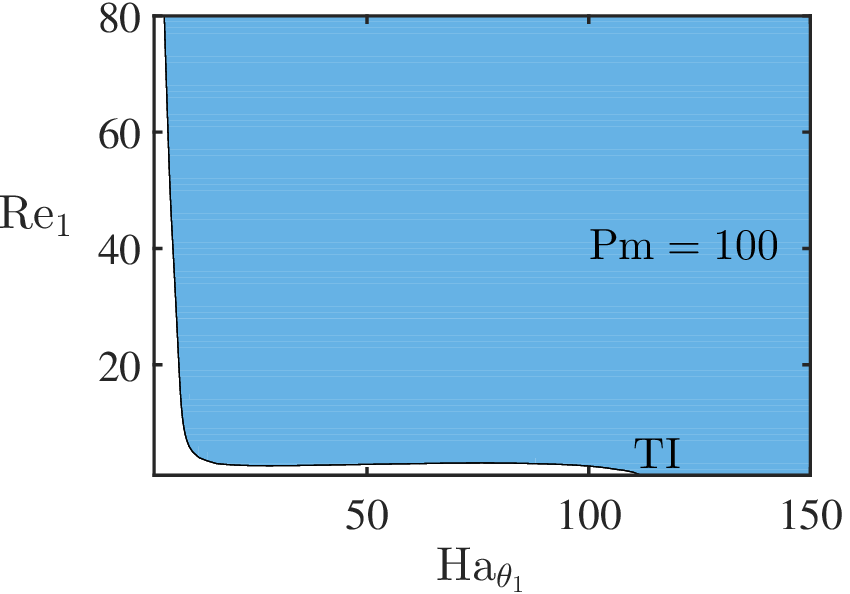}
\includegraphics*[scale=0.45]{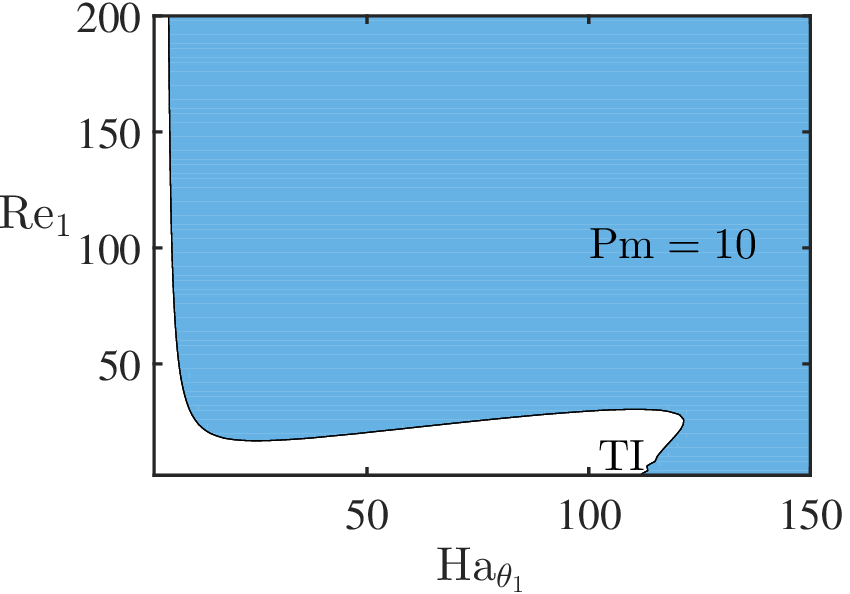}}
{\includegraphics*[scale=0.45]{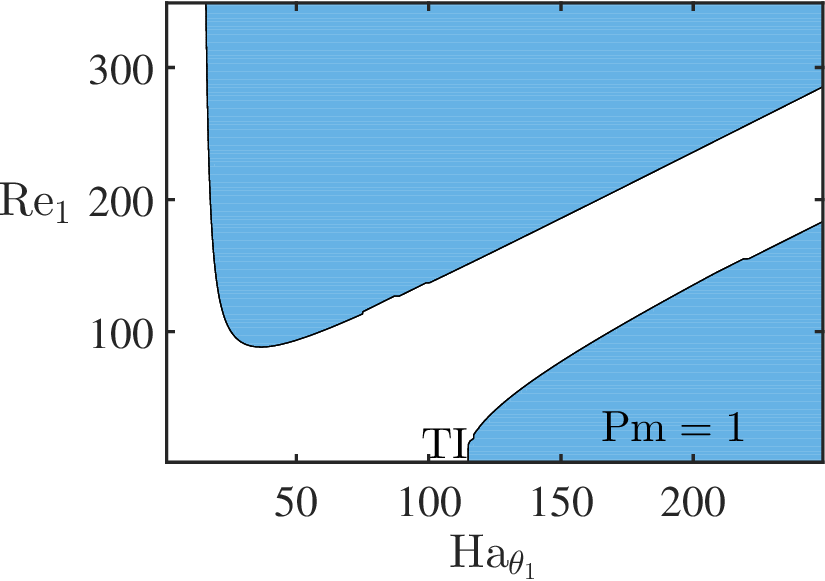}
\includegraphics*[scale=0.45]{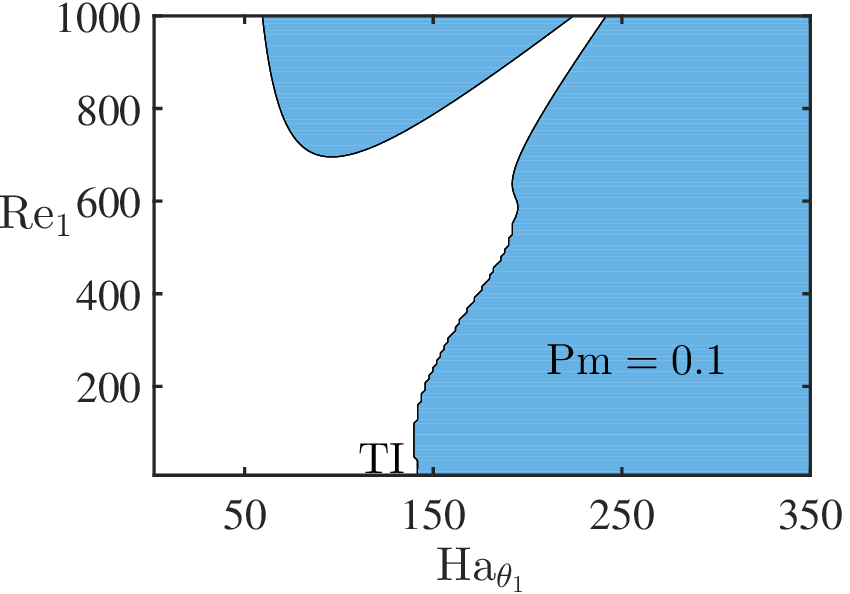}}
 \caption{ The instability region in the $(\mathrm{Re}_1,\mathrm{Ha}_{\theta_1})$-plane for $\mathrm{Ro}=-1/2$, $\mathrm{Rb}=-1/2$, $m=1$, $q=3r_0^{-1}$, $r=1.5r_0$, and $\mathrm{Pm}=100$, $\mathrm{Pm}=10$, $\mathrm{Pm}=1$, and $\mathrm{Pm}=0.1$. The instability domains represented in blue are found with the use of the growth rates maximized over $k$ of the roots of the equation \rf{eqn:dispers-rel-inducless-azim} with the parameters \rf{eqn:new re and ha}.}
\label{fig:four-insta-region}
\end{center}
\end{figure*}

\subsection{Case of $Ro=Rb=-1/2$ and $m=1$ with $q=3r_0^{-1}$}
\label{sec:Ro=-1/2, Rb=-1/2, m=1}
The magnetic Rossby number $Rb=-1/2$ and the azimuthal wavenumber $m=1$ lie inside the range \rf{ti1} and thus allow for the emergence of Tayler instability \cite{RudHol07,Tay73}.
FIG.~\ref{fig:four-insta-region} displays the variation of the instability regions in $(\mathit{Ha}_{\theta_1},Re_1)$ plane when the magnetic Prandtl number $Pm$ changes from $100$ to $0.1$. This result compares well with Figure 3 of R\"udiger et al. \cite{RudHol07}. We notice that there are two types of instabilities, with the lower part originating from the Tayler instability  occurring without rotation in the basic state, and with the upper part originating from the AMRI. As $Pm$ decreases, the critical Reynolds number becomes larger for the AMRI and the AMRI region shrinks to a seemingly separate upper region. The critical Hartmann number for the Tayler instability turns out to be insensitive to $Pm$.
FIG.~\ref{fig:four-insta-region} exhibits marked contrast with FIG.~\ref{fig:ro1rb1pm01} where TI is excluded by the criteria \rf{ti0} and \rf{ti1}.

\begin{figure}[tbhp]
  \begin{center}
\includegraphics*[scale=0.45]{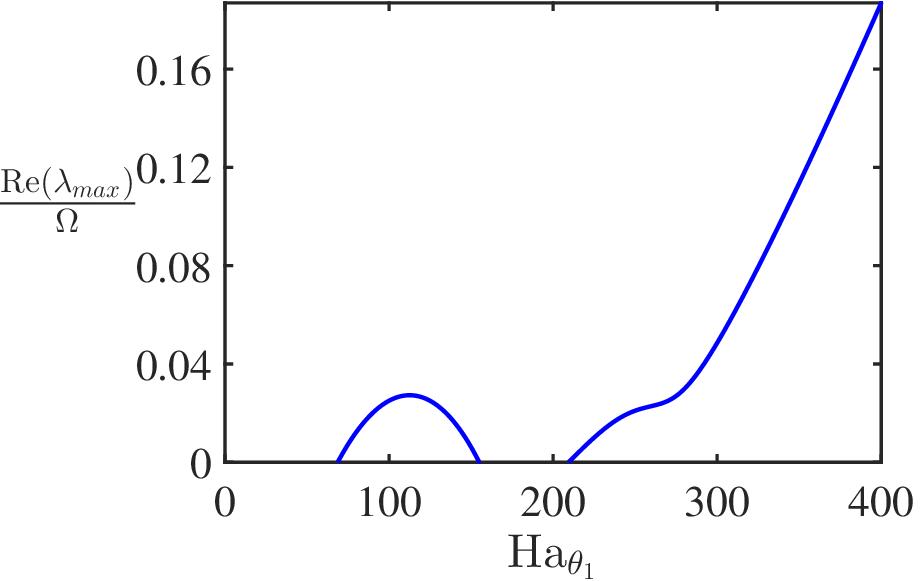}
 \caption{ The optimized over $k$ growth rate versus the Hartmann number $\mathrm{Ha}_{\theta_1}$ according to equation \rf{eqn:dispers-rel-inducless-azim} with the parameters \rf{eqn:new re and ha}. The parameters chosen are $\mathrm{Re}_1=800$, $\mathrm{Rb}=-1/2$, $\mathrm{Ro}=-1/2$, and $\mathrm{Pm}=0.1$ with $m=1$, $q=3r_0^{-1}$, and $r=1.5r_0$.}.
\label{fig:gr800}
\end{center}
\end{figure}

Closer to the experimental condition is the case of $Pm=0.1$ in FIG.~\ref{fig:four-insta-region}. Fixing $Pm=0.1$ and $Re_1=800$, we draw the optimized over $k$ growth rate as a function of $\mathit{Ha}_{\theta_1}$ in FIG.~\ref{fig:gr800}.
There are two instability intervals $\mathit{Ha}_{\theta_1}\in(69,\ 155)\cup (210, \infty)$. In the first one a local extremum is attained at $\mathit{Ha}_{\theta_1}\approx 112$ with the wavenumbers $k=2.7493r_0^{-1}$ and $q=3r_0^{-1}$. The growth rate increases monotonically with $\mathit{Ha}_{\theta_1}$ for $\mathit{Ha}_{\theta_1}>195$.

   \begin{figure*}[tbhp]
  \begin{center}
{\includegraphics*[scale=0.4]{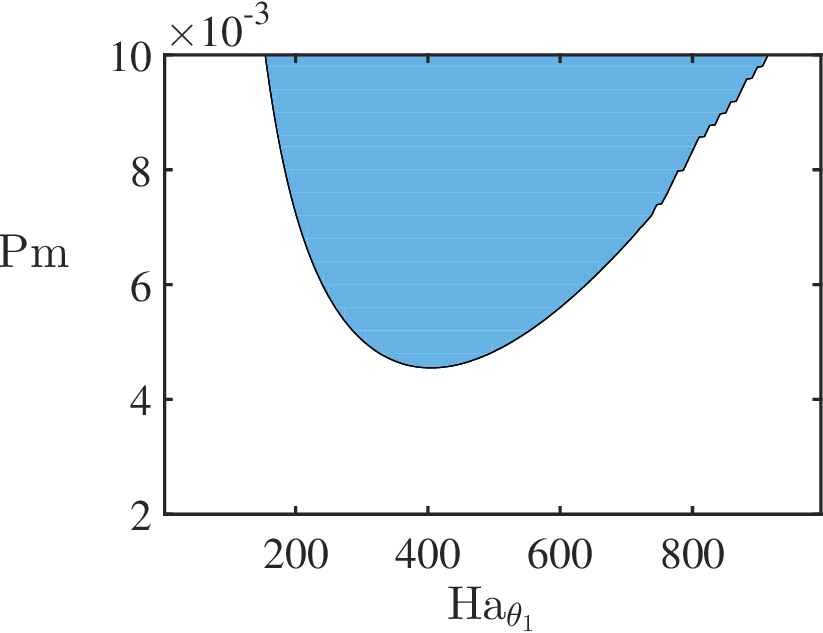}
\includegraphics*[scale=0.4]{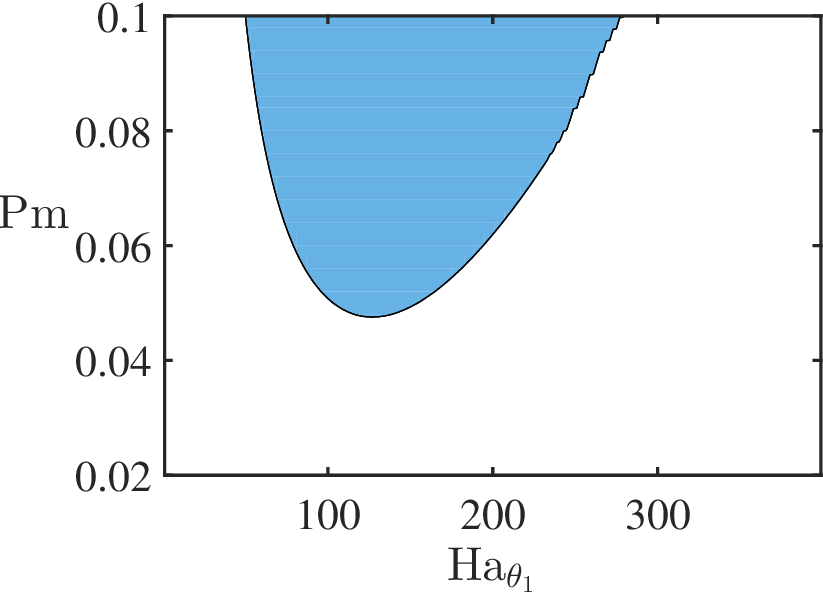}}
 \caption{ The region of AMRI (above the critical lines) in the $(\mathrm{Ha}_{\theta_1},\mathrm{Pm})$-plane when $\mathrm{Rb}=-1$, $\mathrm{Ro}=-3/4$ and (left) $\mathrm{Re}_1=10^4$ and (right) $\mathrm{Re}_1=10^3$ according to equation \rf{eqn:dispers-rel-inducless-azim} with the parameters \rf{eqn:new re and ha}. In the former case the instability occurs when $\mathrm{Pm}>0.0046$ with the smallest $\mathrm{Pm}$ corresponding to $\mathrm{Ha}_{\theta_1}\approx 400$ whereas in the latter when $\mathrm{Pm}>0.048$ with the lowest $\mathrm{Pm}$ corresponding to $\mathrm{Ha}_{\theta_1}\approx 127$.}
\label{fig:pm-re10000}
\end{center}
\end{figure*}

\subsection{Case of $Ro=-3/4, Rb=-1$ and $m=1$ with $q=3r_0^{-1}$}
\label{sec:Ro=-3/4, Rb=-1, m=1}

According to the instability condition \rf{eqn:HMRI-RbRo} the Keplerian rotation with $Ro=-3/4$ cannot be destabilized by the current-free
azimuthal magnetic field with $Rb=-1$ in the inductionless limit of $Pm=0$. Instead, the criterion \rf{eqn:HMRI-RbRo} suggests shallower radial profiles for the magnetic field with $Rb>-25/32$. Does this change for small but finite $Pm$? The work  \cite{KSF14JFM} predicted regions of HMRI existing at such values of the magnetic Prandtl number. What can we say about AMRI?

Here we demonstrate that, for $Ro=-3/4$ and $Rb=-1$, there is a minimum value of the magnetic Prandtl number $Pm$, below which the instability is ruled out. Let us choose $r=1.5r_0$ and search for the critical $Pm$ for instability. For the flow with $Re_1=10^4$, the left panel of FIG.~\ref{fig:pm-re10000} shows that the instability necessitates $Pm>0.0046$, with the critical value of $Pm$ corresponding to $\mathit{Ha}_{\theta_1}\approx 400$.  For $Re_1=10^3$, the critical value is raised to $Pm\approx 0.048$ which is attained at $\mathit{Ha}_{\theta_1}\approx 127$ as shown by the right panel of FIG.~\ref{fig:pm-re10000}.

\begin{figure}[tbhp]
  \begin{center}
{\includegraphics*[scale=0.4]{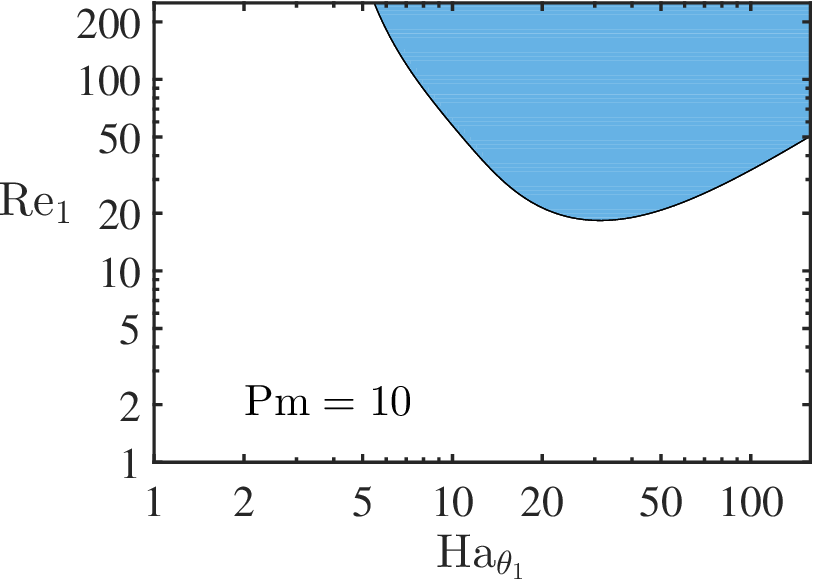}
\includegraphics*[scale=0.4]{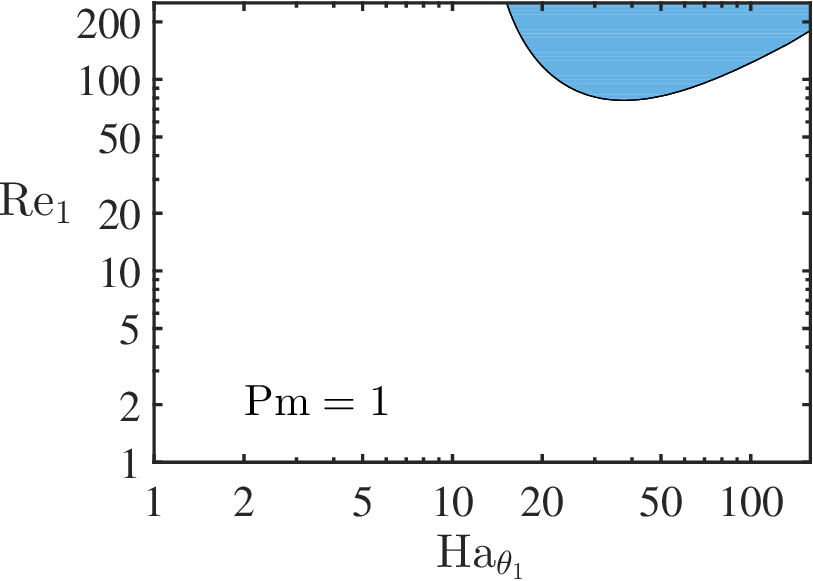}}
 \caption{ The region of AMRI in the $(\mathrm{Ha}_{\theta_1},\mathrm{Re}_1)$-plane when $\mathrm{Ro}=-3/4$ and $\mathrm{Rb}=-1$, $m=1$, $q=3r_0^{-1}$ and $r=1.5r_0$. The neutral stability curve is obtained by maximizing the growth rate over $k$ for $\mathrm{Pm}=10$ and $\mathrm{Pm}=1$ with the use of the equation \rf{eqn:dispers-rel-inducless-azim} with the parameters \rf{eqn:new re and ha}.}
\label{fig:region-kepler}
\end{center}
\end{figure}

As $Re_1$ is increased, the critical value of $Pm$ is decreased, which yields a larger strength of magnetic field according to \rf{eqn:mri-insta1}. Large Reynolds numbers mean turbulence in practice so that $Pm\gtrsim 10^{-3}$ is at least necessary for experimental realization of the AMRI. However the liquid eutectic alloy $GaInSn$ has $Pm=1.4\times 10^{-6}$ making the AMRI of a Keplerian flow virtually impossible for the experimental setup with the current-free azimuthal magnetic field \cite{SGGGS14}. Indeed, FIG.~\ref{fig:region-kepler} shows that as $Pm$ decreases, the instability region becomes smaller and smaller.

However, as FIG.~\ref{fig:crossover3} demonstrates, in the limit $(Pm\rightarrow 0)$, the $k\rightarrow 0$ mode has positive growth rate. To approach this instability, we set $r_{\text{out}}=r_0$ as the characteristic length but set $r$ to vary freely toward $r=0$. By setting $\widehat{k}=0$ in (\ref{eqn:AMRI-disp}), we find its roots in the following form
\begin{eqnarray}
\frac{\lambda_1}{\Omega}&=&-\sqrt{\frac{r_0}{r}}\frac{1}{1+(qr)^2}\frac{\mathit{Ha}_{\theta_1}^2}{Re_1}-\sqrt{\frac{r_0}{r}}\frac{1+(qr)^2}{Re_1}{\color{black}-i},\notag\\	
&& \notag\\	
\frac{\lambda_2}{\Omega}
 &=&\sqrt{\frac{r_0}{r}}\frac{3 - (qr)^2}{(1 +
   (qr)^2)^2 }\frac{\mathit{Ha}_{\theta_1}^2}{Re_1} - \sqrt{\frac{r_0}{r}}\frac{1 + (qr)^2}{Re_1}{-\color{black}i\frac{1+4q^2r^2}{4+4q^2r^2}}
. \notag\\&&
\end{eqnarray}

\begin{figure*}
\centering
{\includegraphics*[scale=0.4]{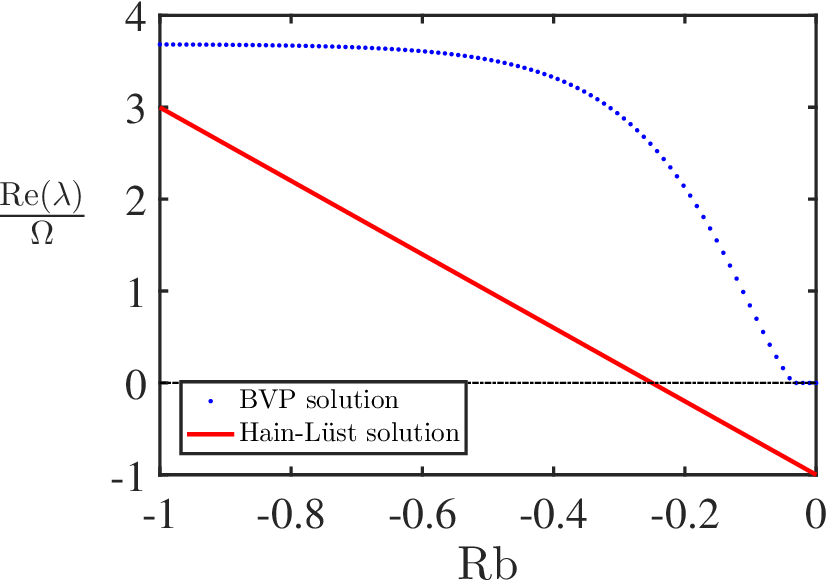}~
\includegraphics*[scale=0.4]{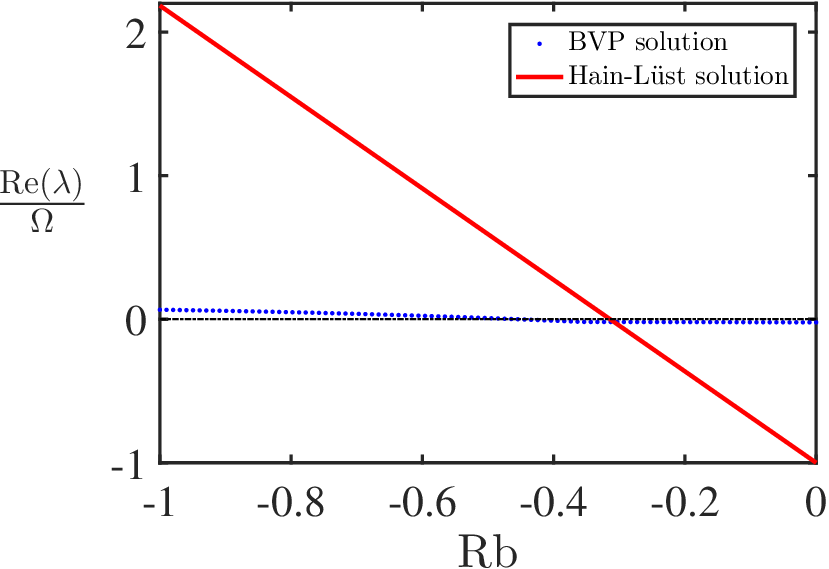}~
\includegraphics*[scale=0.4]{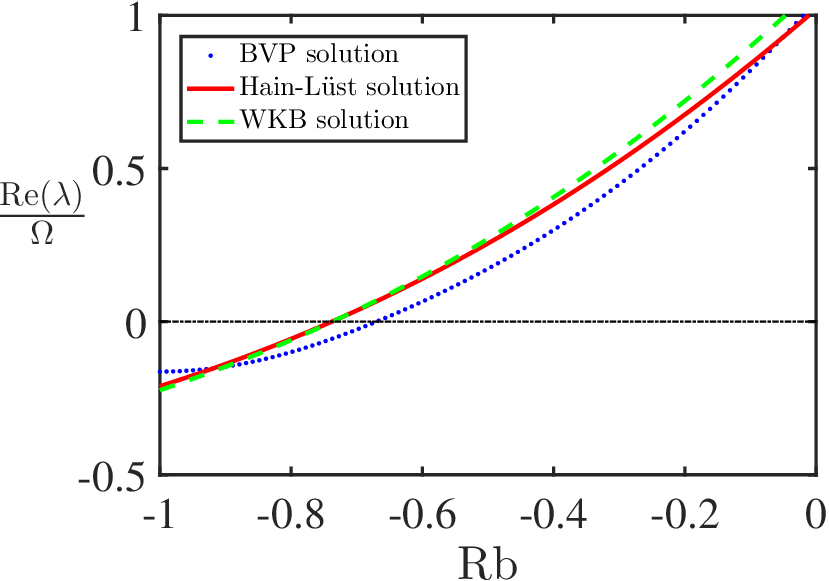}}
\caption{The growth rate $\mathrm{Re}(\lambda)$ in units of $\Omega_{in}$ over a range of magnetic Rossby number $\mathrm{Rb}$ for the Keplerian ($\mathrm{Ro}=-3/4$) flow with $\mathrm{Re}_{in}=10^4$, $\mathrm{Ha}_{\theta}=10^2$, $\mathrm{Pm}=0$, and $m=1$, in the case of the long axial wavelength ($\zeta=0.02$/$\zeta=0.336$ and $k=10^{-4}d^{-1}$, left/middle panels) and short axial wavelength ($\zeta=0.98$ and $k=3.5d^{-1}$, right panel). The dotted blue line comes from the dimensionless Taylor-Couette boundary value problem with the boundary conditions corresponding to the perfectly conducting walls. The red and dashed green lines correspond to the Hain-L\"ust dispersion relation \rf{eqn:AMRI-disp} and the WKB approximation \rf{WKB_sol}. The radial wavenumber is set to be $\widehat{q}=\zeta /(1-\zeta)$.}
\label{fig:HL_BVP_growth_rate}
\end{figure*}

The first root has $\Re(\lambda_1)<0$ and corresponds to a stable mode. The second one indicates that, for large values of $\mathit{Ha}_{\theta_1}$, the instability occurs when
\begin{eqnarray}
(qr)^2<3.
\label{eqn:induc-k=0,qr}
\end{eqnarray}
In addition, the radial wavenumber is bounded so as to satisfy the boundary conditions at the cylinders of $r=r_{\text{in}}$ and $r_{\text{out}}$, indicating
\begin{eqnarray}
q(r_{\text{out}}-r_{\text{in}})>\pi>3.
\label{eqn:induc-k=0,qr_2}
\end{eqnarray}
Combining (\ref{eqn:induc-k=0,qr}) and (\ref{eqn:induc-k=0,qr_2}), we obtain
\begin{eqnarray}
\frac{r}{r_{\text{out}}-r_{\text{in}}}<\frac{1}{\sqrt{3}}
\label{eqn:induc-k=0,r}
\end{eqnarray}
Because $r>r_{\text{in}}$, $r_{\text{in}}/(r_{\text{out}}-r_{\text{in}})<1/\sqrt{3}$, namely
\begin{eqnarray}
\frac{r_{\text{in}}}{r_{\text{out}}}<\frac{\sqrt{3}-1}{2}\approx0.366.
\label{eqn:induc-k=0,r1}
\end{eqnarray}
This crude argument suggests that the experimental set-ups with $r_{\text{in}}/r_{\text{out}}=1/2$ might need to be modified to have a wider gap in order to be able to capture the mode of $k=0$ for a Keplerian flow subject to the current-free magnetic field.

\section{Global stability analysis}
\label{sec:BVP}

In order to provide a numerical validation of the analytical results based on the Hain-L\"ust dispersion relation \rf{eqn:dispers-rel-inducless-azim}, in the following we consider the cylindrical Taylor-Couette flow as described in Section~\ref{sec:Non-axisymmetric AMRI critical Re Ha}. We will decompose the magnetic and velocity fields into toroidal and poloidal parts and after that reduce the original MHD system \rf{eqn:Euler-disturb}--\rf{eqn:div-b=0-disturb} to a one-dimensional boundary eigenvalue problem \cite{HolTee10,Hol15} by expanding the solution in the Heinrichs basis  \cite{Hein89,Deg11,Deg18}.

We assume a  finite radial gap $d:=|r_{out}-r_{in}|$ and a radius ratio $\zeta:=r_{in}/r_{out}$ that both define the geometry of the setup. Our numerical method is based on the pseudo-spectral expansion of the solution in terms of normal modes before collocating at the Chebyshev-Gauss nodes. The code we have developed has been benchmarked against several well-established results of similar stability analysis for either insulating or conducting boundary conditions with excellent agreement.

For the sake of clarity, we first render the problem \rf{eqn:Euler-disturb}--\rf{eqn:div-b=0-disturb} in a dimensionless form following the notations in Child et al.  \cite{Hol15} except for the background velocity field where we follow the work of Deguchi  \cite{Deg18}. This can be summarized by scaling length with $d$, time with the viscous time scale $d^2/\nu$, velocities with $\nu /d$, pressure with $\rho\nu^2/d^2$ and magnetic fields with $B_0$.

\subsection{Background fields and Rossby numbers}
\label{sec:Background fields and Rossby numbers}

The scaling introduced before leads to the following set of equations
\ba{eqn:dimensionless_MHD}
\frac{\partial\mathbf{\widehat{u}}}{\partial\widehat{t}} &=& - (\mathbf{\widehat{u}}\cdot\widehat{\nabla})\mathbf{\widehat{U}} - (\mathbf{\widehat{U}}\cdot\widehat{\nabla})\mathbf{\widehat{u}}  -\widehat{\nabla}\widehat{P} + \widehat{\nabla}^2\mathbf{\widehat{u}} \nn \\
&+& \frac{Ha_{\theta}^2}{Pm}\left[ (\mathbf{\widehat{b}}\cdot\widehat{\nabla})\mathbf{\widehat{B}} + (\mathbf{\widehat{B}}\cdot\widehat{\nabla})\mathbf{\widehat{b}}\right], \nn \\
\frac{\partial\mathbf{\widehat{b}}}{\partial\widehat{t}} &=& \widehat{\nabla}\times (\mathbf{\widehat{U}}\times\mathbf{\widehat{b}})+\widehat{\nabla}\times (\mathbf{\widehat{u}}\times\mathbf{\widehat{B}})+\frac{1}{Pm}\widehat{\nabla}^2\mathbf{\widehat{b}},\nn\\
\widehat{\nabla}\cdot\mathbf{\widehat{u}}&=&0, \nn \\
\widehat{\nabla}\cdot\mathbf{\widehat{b}}&=&0,
\ea
where $Ha_{\theta}=B_0d/\sqrt{\rho\mu_0\nu\eta}$ is the azimuthal Hartmann number and $Pm=\nu/\eta$ is the magnetic Prandtl number.

This dimensionless MHD system is therefore solved regarding to no-slip boundary conditions for the velocity field, which in the cylindrical coordinates $(r,\theta,z)$ lead to  \cite{Deg17}
\ba{eqn:no-slip}
	\widehat{U}_{\theta}(r_{in},\theta,z)=\frac{\Omega_{in}r_{in}d}{\nu}&=:&Re_{in}, \nn\\
    \widehat{U}_{\theta}(r_{out},\theta,z)=\kappa Re_{in}/\zeta&=:&Re_{out},
\ea
where $\widehat{U}_{\theta}$ is the azimuthal component of the fluid velocity, $\kappa:=\Omega_{out}/\Omega_{in}$ is the ratio between the angular velocities and the inner and outer radii can be defined according to the Taylor-Couette parameters as $r_{in}:=d\zeta/(1-\zeta)$ and $r_{out}:=d/(1-\zeta)$.

A fundamental solution for this system and the boundary conditions is the well-known Couette profile $\mathbf{\widehat{U}}=\widehat{r}\Omega(\widehat{r})\mathbf{e}_{\theta}$, given by
\ba{eqn:Couette-profile}
	\Omega(\widehat{r})&=&\frac{Re_{in}}{1+\zeta}\left[\left(\frac{\kappa}{\zeta}-\zeta\right) + \frac{\zeta (1-\kappa )}{(1-\zeta)^2} \frac{1}{\widehat{r}^2} \right],
\ea
where $\widehat{r}=rd^{-1}$ is the dimensionless radial coordinate.

The background magnetic field we consider here is purely azimuthal $\mathbf{\widehat{B}}=\widehat{B}_{\phi}(\widehat{r})\mathbf{e}_{\theta}$ and given by
\begin{equation}
\label{eqn:mag-field-profile}
	\widehat{B}_{\phi}(\widehat{r})=\frac{\zeta (\tau -\zeta)}{1-\zeta^2}\widehat{r}+\frac{1-\tau\zeta}{1-\zeta^2}\frac{1}{\widehat{r}},
\end{equation}
where $\tau:=B_{out}/B_{in}$ is the ratio between the outer and the inner azimuthal magnetic fields  \cite{Rud10}.

Using \rf{omu} and \rf{eqn:mag-field-profile} we can write
\ba{rorb}
	\kappa &=&\zeta^{-2 Ro},\nn\\
	\tau &=&\zeta^{-(2 Rb+1)}.
\ea
Then, the solid-body rotation ($Ro=0$) corresponds to $\kappa=1$ and the Keplerian flow ($Ro=-3/4$) to $\kappa = \zeta^{3/2}$.

In the following, we will specify the basic state of the magnetized flow via $Ro$, $Rb$, $\zeta$, $Re_{in}$, $Pm$ and $Ha_\theta$ defined earlier.

\begin{figure*}
\centering
\includegraphics*[scale=0.4]{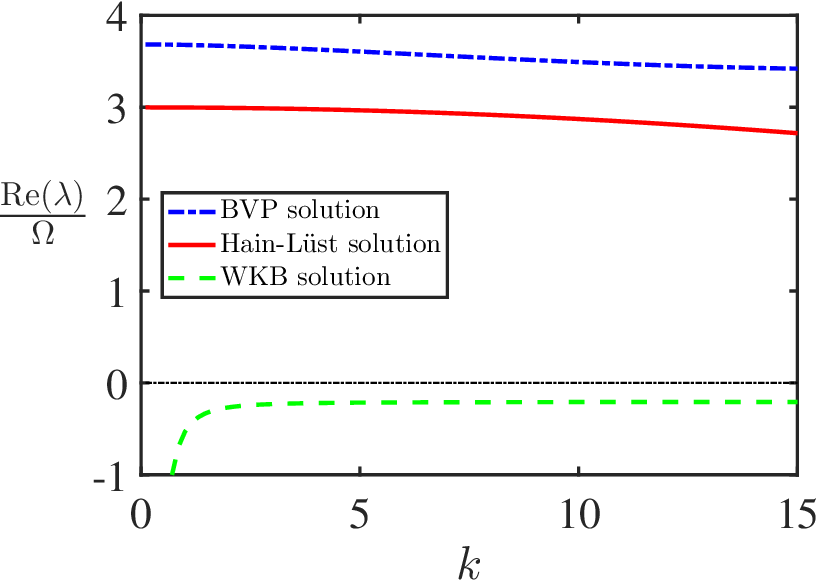}~
\includegraphics*[scale=0.4]{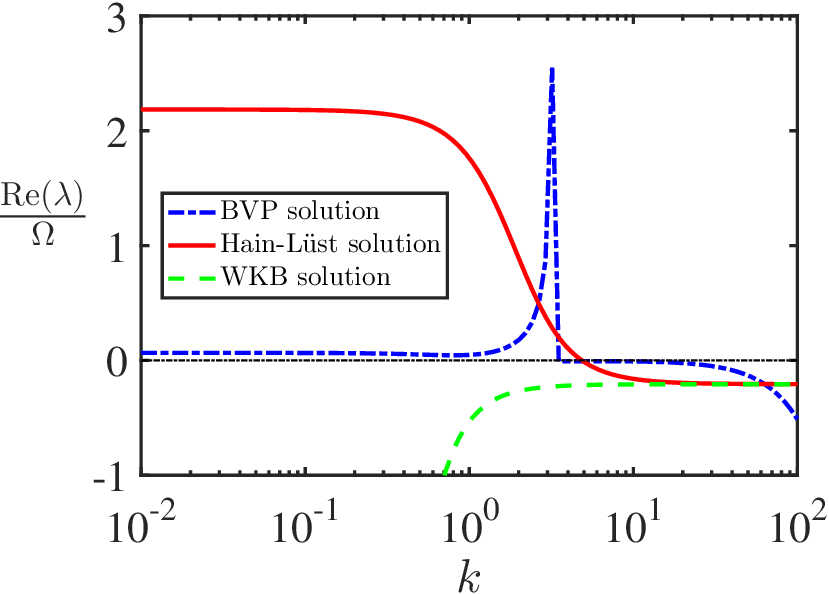}~
\includegraphics*[scale=0.4]{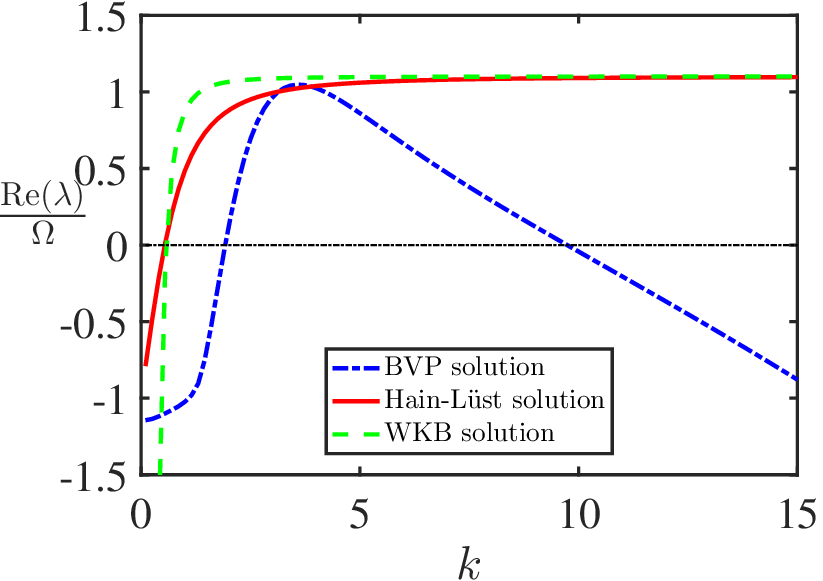}
\caption{The growth rate $\mathrm{Re}(\lambda)$ in the units of $\Omega_{in}$ from BVP (dash-dotted blue), WKB approximation (dashed green) and Hain-L\"{u}st (red) versus the axial wavenumber $k$ (in units of $d^{-1}$) for perfectly conducting boundaries. We set $\mathrm{Rb}=-1$ with $\zeta=0.02$ (left) and $\zeta=0.366$ (middle) and $\mathrm{Rb}=0$ with $\zeta=0.98$ (right). The parameter space is the same as in FIG.~\ref{fig:HL_BVP_growth_rate}.}
\label{fig:LWL_SWL_Stab_Curves}
\end{figure*}

\subsection{Pseudo-spectral expansion}
\label{sec:Pseudo-spectral expansion}

We seek for a solution to linearized MHD equations decomposed into toroidal and poloidal parts as follows
\begin{align}
\label{eqn:TP_Vel}
	\widetilde{\mathbf{u}}&= \nabla\times(\psi\,\mathbf{e}_r) + \nabla\times\nabla\times(\phi\,\mathbf{e}_r),\\
\label{eqn:TP_MF}
	\widetilde{\mathbf{b}}&= \nabla\times(\Psi\,\mathbf{e}_r) + \nabla\times\nabla\times(\Phi\,\mathbf{e}_r).
\end{align}
The disturbance fields ($\psi,\phi,\Psi,\Phi$) in (\ref{eqn:TP_Vel})-(\ref{eqn:TP_MF}) are expanded in terms of normal modes according to the pseudo-spectral Fourier method. In it, each variable is expressed with respect to Heinrichs basis  \cite{Hein89,Deg11,Deg18} for the radial direction and to Fourier basis for the axial and azimuthal directions.
Such expansion can be represented for an arbitrary field $\mathcal{L}$ as
\begin{equation}
\label{eqn:pseudo_spectral_exp}
	\mathcal{L}(x,t,\theta,z):=\sum_{n=0}^{\infty}\left[\mathcal{H}(x)T_n(x)\right]\exp{\left[\lambda t+i(m\theta+kz)\right]},
\end{equation}
where $T_n(x)$ is a Chebyshev polynomial, $\mathcal{H}(x)$ is the Heinrichs factor which depends on the boundary conditions considered, $\lambda$ is an eigenvalue, $(m,k)$ are the azimuthal and axial wavenumbers and $x$ is the length coordinate.

In order for the method to be computable, the infinite series are truncated at the $N$-th order and the mapping of the radial interval $[r_{in},r_{out}]$ to the Chebyshev interval $[-1,1]$ comes from the linear transformation  \cite{Deg11} $x=2(r-r_m)d^{-1}$ with $r_m=d(1+\zeta)/(2(1-\zeta))$ being the mean radius. Finally, the series are evaluated at the Chebyshev-Gauss collocation points $$x_i=\pm \cos\left(\pi\frac{i+1}{N+2}\right) ,\: i=0,\ldots ,N.$$

\begin{figure*}
\centering
{\includegraphics*[scale=0.4]{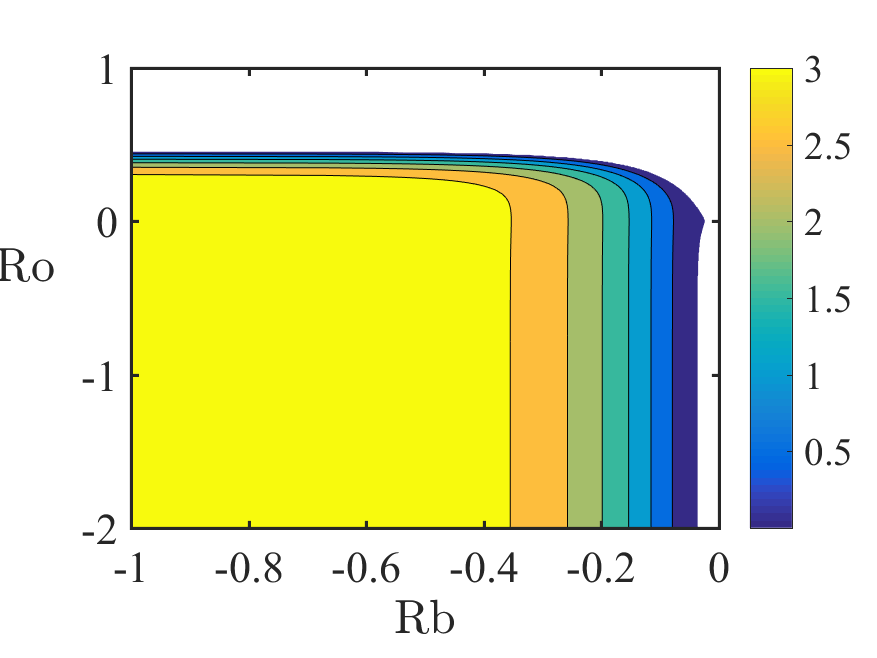}~
\includegraphics*[scale=0.4]{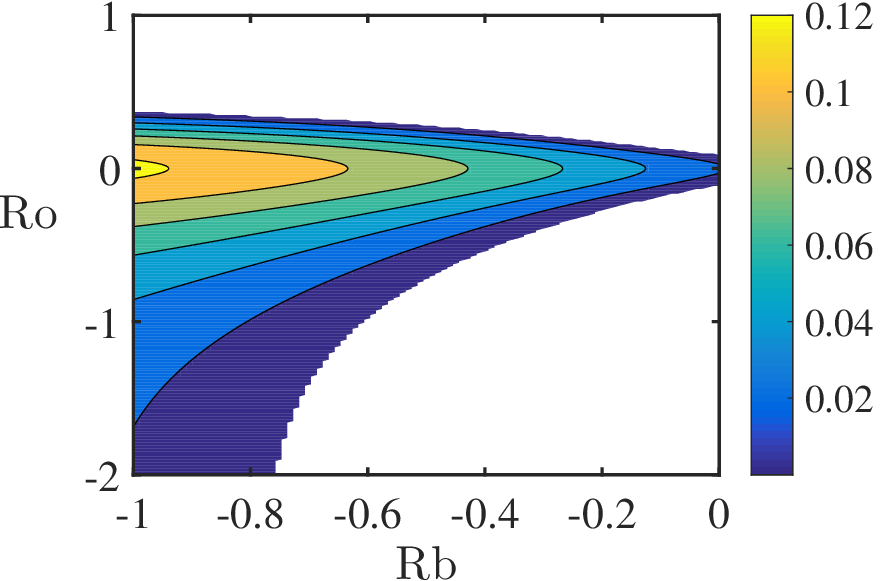}~
\includegraphics*[scale=0.4]{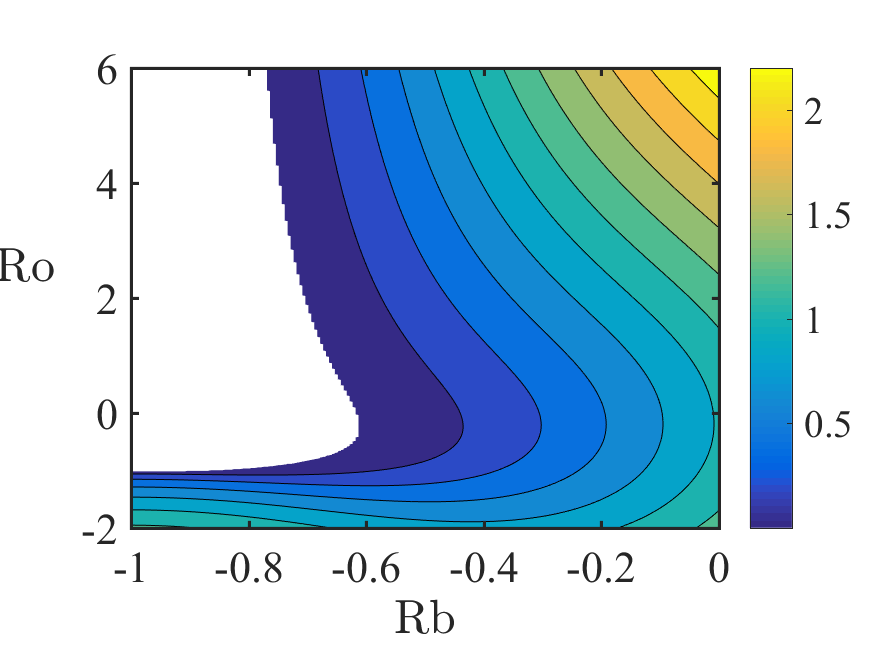}}
{\includegraphics*[scale=0.4]{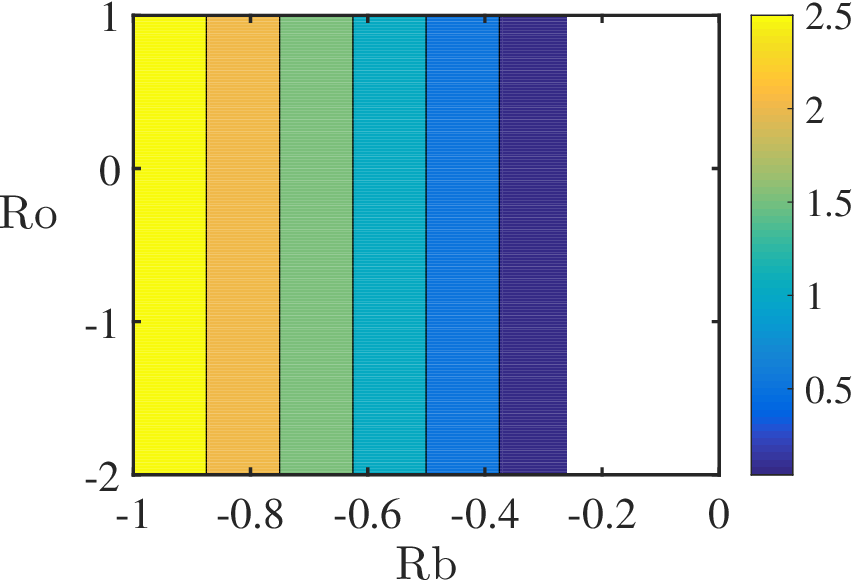}~
\includegraphics*[scale=0.4]{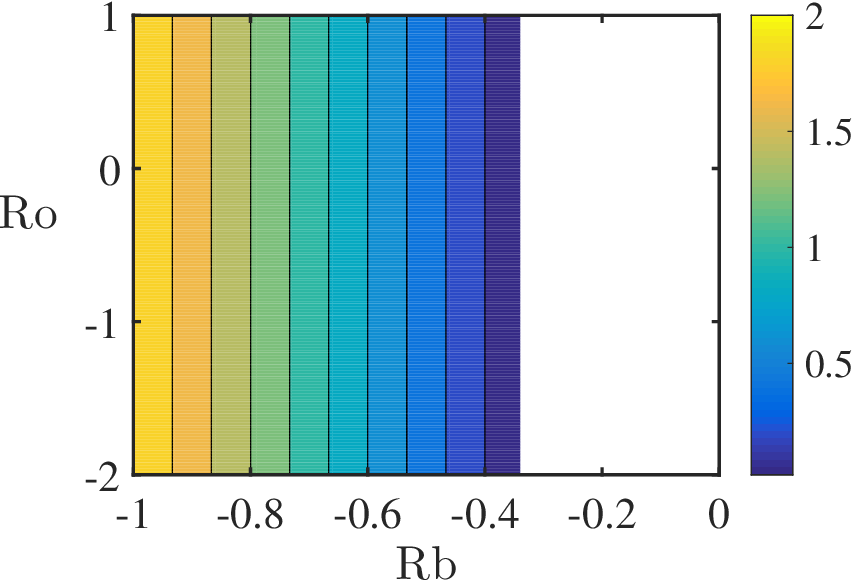}~
\includegraphics*[scale=0.4]{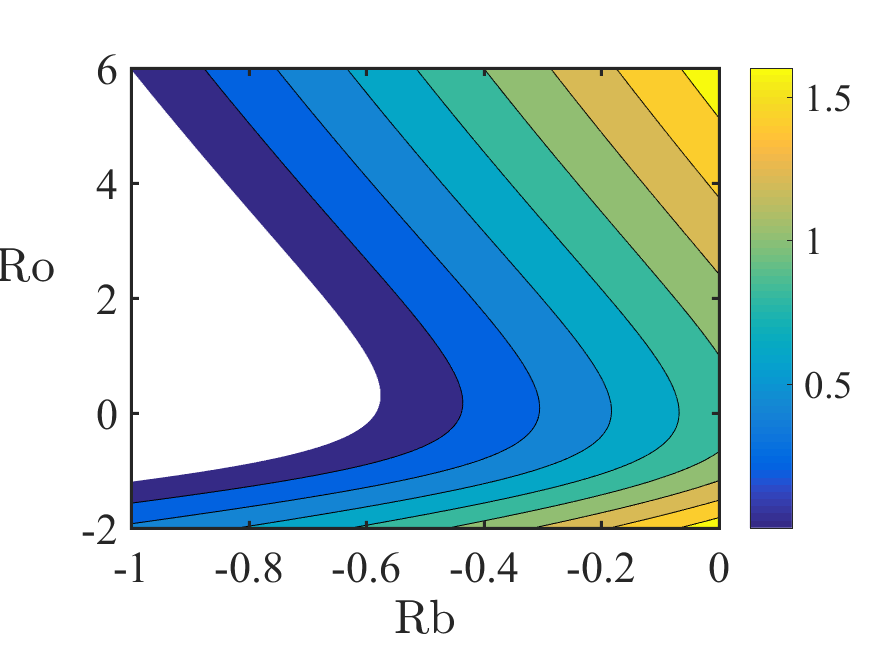}}
\caption{Growth rate magnitude in the Rossby plane ($\mathrm{Rb},\mathrm{Ro}$) from the boundary value problem with perfectly conducting boundaries for the upper panels and from Hain-L\"{u}st dispersion relation \rf{eqn:dispers-rel-inducless-azim} for the lower panels. The geometry correspond to $\zeta=0.02$, $k=10^{-4}d^{-1}$ (left column), $\zeta=0.366$, $k=10^{-4}d^{-1}$ (middle column) and $\zeta=0.98$, $k=3.5d^{-1}$ (right column). The parameter space is the same as in FIG.~\ref{fig:HL_BVP_growth_rate} and stability is represented in white.}
\label{fig:BVP_Rossby_Plane}
\end{figure*}

The decomposition \rf{eqn:pseudo_spectral_exp} allows us to express the differential operators as functions of the wavenumbers and parameters of the system. Details of this method and coefficients of the boundary value problem can be found, e.g., in Child et al.  \cite{Hol15} and Hollerbach et al.  \cite{HolTee10}.
The set of equations we obtain is solved regarding the boundary conditions considered, i.e., no-slip conditions for the velocity field and perfectly conducting for the magnetic field. Assuming the expansion in terms of normal modes for each variables, these conditions can be written in the following form  \cite{Gus15}
\begin{align}
\label{eqn:bc_bvp_pc}
	\psi=\phi=\partial_r\phi&=0,\\
	\Phi&=0,\\
	ik\partial_r\Psi+ikr^{-1}\Psi+imr^{-1}\partial_{rr}\Phi-imr^{-2}\partial_r\Phi&=0.
\end{align}


The system is therefore reduced to a generalised eigenvalue problem of the form $A\xi=\lambda B\xi$, where $\lambda$ is an eigenvalue and $\xi$ an eigenvector.

\subsection{Numerical results}
\label{sec:Numerical_Results}

For a fixed set of boundary conditions, the boundary value problem is solved and leads to the computation of the eigenvalues $\lambda$ of the magnetized Taylor-Couette flow. The global stability analysis is therefore conducted over similar sets of parameters from the previous sections of this paper in order to validate a large part of the results.

{\color{black} In FIG.~\ref{fig:HL_BVP_growth_rate} we compare growth rates given by the boundary value problem, the Hain-L\"ust dispersion relation \rf{eqn:dispers-rel-inducless-azim} and the original WKB approximation \cite{KSF14JFM}. The latter solution is given according to our notations as follows

\be{WKB_sol}
\frac{\Re{(\lambda)}}{\Omega_{in}} = N_{\theta}\left(2Rb - \frac{m}{\alpha_1}\right) - \frac{1}{Re} \pm \sqrt{2X+2\sqrt{X^2+Y^2}},
\ee

where $N_{\theta}=Ha_{\theta}^2/Re$, $\alpha_1=k/\sqrt{k^2+q^2}$ and

\ba{}
X &=& N_{\theta}^2\left(Rb^2 + \frac{m}{\alpha_1}\right) - (Ro + 1), \nn \\
Y &=& N_{\theta}(Ro+2)\frac{m}{\alpha_1}. \nn
\ea

The eigenvalues are both scaled with the inner angular velocity $\Omega_{in}$ and the stability analysis has been conducted for different but finite radial gaps over different values of $Rb$ in FIG.~\ref{fig:HL_BVP_growth_rate}. }In the interest of reaching the $k\to 0$ mode, we have fixed a wide radial gap in order to have a small radial wavenumber. We chose an arbitrary value for the gap between both cylinders $\zeta=0.02$ and according to this geometry, we manage to find a similar behavior for the growth rate of the long-wavelength domain but with nevertheless a discrepancy for the threshold of instability.
{\color{black}Not surprisingly, the WKB solution \rf{WKB_sol} in the long-wavelength approximation is diverging at such value of $k$ and therefore can not be represented in this plot. In a similar way, if we decrease the gap between both cylinders until the limit that equation \rf{eqn:induc-k=0,r1} predict, the numerical solution still behaves as the analytical result with nevertheless a worst accuracy in the magnitude. Nevertheless, while the WKB solution \rf{WKB_sol} is not able to catch such limit, it appears that our extented version of the Hain-L\"{u}st dispersion relation can. Regarding the short-wavelength domain, we used a narrow-gap $\zeta=0.98$ and a larger wavenumber $k$ and the growth rates from the BVP, the WKB solution and our dispersion relations are as expected in good agreement.

}The dependence of the growth rate on the axial wavenumber $k$ in our numerical scheme is represented in FIG.~\ref{fig:LWL_SWL_Stab_Curves} where we observe that for the long-wavelength domain the growth rate reaches its maximum for $k\to 0$ as expected. For the short-wavelength approximation, we are limited to a smaller interval of $k$ for which the growth rate is positive in the BVP solution but it still remains sufficient to produce smooth and correct comparisons with the analytic.
{\color{black}It is interesting to notice that in the middle panel for $\zeta=0.366$, the numeric is predicting a peak between both domains. A similar behavior has been observed in the second figure of Bodo et al. \cite{Bodo16} and is analyzed by the authors as a localized state of MRI.
As we can notice, the WKB and the Hain-L\"{u}st solutions are both asymptotically converging to the same value as $k$ is increasing (tending therefore to the short-axial-wavelength limit) only when the radial gap between both cylinder is not too large.

}
The last computation presents the stability domains in the $(Rb,Ro)$-plane as in the previous section \ref{sec:indefinite_Ro_Rb}. This case is presented in FIG.~\ref{fig:BVP_Rossby_Plane} where the growth rate magnitude from the BVP and from the dispersion relation \rf{eqn:dispers-rel-inducless-azim} is computed.
{\color{black}The left and middle column panels show the long-wavelength instability domains with $\zeta=0.02$, $\zeta=0.366$ and $k=10^{-4}d^{-1}$ and despite the difference in the neutral stability boundaries, the shape of both domains are in agreement with the analytical results of FIG.~\ref{fig:BVP_Rossby_Plane}.
When increasing the value of $k$ as in the right column of FIG.~\ref{fig:BVP_Rossby_Plane}, we also notice that the numerics is fitting well with our solution.
}

\section{Conclusion}
\label{sec:Discussions}

We have explored the AMRI and the Tayler instability of a rotating MHD flow, augmented by viscosity and electrical resistivity, with respect to the axisymmetric as well as non-axisymmetric perturbations. We have derived the extended Hain-L\"ust equation to include the viscosity and electrical resistivity. This is a second-order ordinary differential equation for the radial Lagrangian displacement.

We then applied the WKB approximation to it to derive a dispersion relation, valid in the regime of short wavelengths in the radial direction but allowing for arbitrary azimuthal and axial wavenumbers.

By that reason, the extended Hain-L\"ust dispersion relation contains the previously known dispersion relations derived by different methods, including the geometrical optics approximation.

On the other hand, the additional terms in it enable more accurate treatment of the non-axisymmetric perturbations with large axial wavelength.

While being in the limit of short axial wavelength we restored the well-known results of the inductionless approximation, including the necessary condition \rf{eqn:HMRI-RbRo} for both HMRI and AMRI, and the generalized Tayler instability condition \rf{ti1}, in the limit of long axial wavelength we discovered new instability that works both in the rotating and in the non-moving fluid.

We found a limitation on the radial wavelength providing an estimate for the gap in a Taylor-Couette setup which is necessary for detection of the new instability. Finally, we combined the numerical methods of Deguchi and Nagata  \cite{Deg11,Deg17} and Child et al.  \cite{Hol15} to validate the analytical findings, based on the Hain-L\"ust dispersion relation, using global stability analysis.

\begin{acknowledgements}
R.Z. was supported by a Ph.D.
Studentship from the China Scholarship Council. J.L. was supported by a Ph.D. Scholarship from Northumbria University. Y.F. was supported in part by a Grant-in-Aid for Scientific Research from the Japan Society for the Promotion of Science (Grant No. 19K03672).
\end{acknowledgements}


\appendix{}

\section{Derivation of equation (\ref{eqn:Hain-Lust-eq})}
\label{sec:Derivation of HL equation}

In this appendix we derive the extended Hain-L\"ust equation (\ref{eqn:Hain-Lust-eq}).
The lines 4--6 in \rf{eqn:M} allow us to express the magnetic field disturbance $\tilde{b_r}$, $\tilde{b_\theta}$, and $\tilde{b_z}$ in terms of the other variables. By eliminating the magnetic field disturbances, we can reduce \rf{eqn:M} to equations for
$\bm{\xi_1}=(\tilde{u_r},
\tilde{u_\theta},
\tilde{u_z},
\tilde{p})$
as
\begin{equation}
\sf{M_1}\bm {\xi_1=0},
\label{eqn:disturb-eq1}
\end{equation}
where, with use of (\ref{eqn:def-Lambda})
 \begin{eqnarray}
&\sf{M_1}=&\notag
\\
&\displaystyle
 \left(\begin{footnotesize}
\begin{array}{cccc}
 \displaystyle \Lambda+\frac{2\mu r}{\rho\mu_0\tilde{\lambda}_\eta}\left(\frac{iF}{\tilde{\lambda}_\eta}\Omega' -\mu'\right)
&\displaystyle-2\Omega+\frac{2iF\mu}{\rho\mu_0\tilde{\lambda}_\eta} &\displaystyle0 &\displaystyle\frac{1}{\rho}\frac{\dd}{\dd r}
        \\
 \displaystyle2\Omega+r\Omega'\left(1+\displaystyle\frac{F^2}{\rho\mu_0\tilde{\lambda}_\eta^2}\right)-
\frac{2iF\mu}{\rho\mu_0\tilde{\lambda}_\eta} &  \Lambda&\displaystyle0& \displaystyle\frac{1}{r\rho} im
          \\
 \displaystyle0&\displaystyle0 & \Lambda& \displaystyle\frac{1}{\rho} ik
                  \\
 \displaystyle\frac{1}{r}+\frac{\dd}{\dd r}&\displaystyle\frac{im}{r}&ik&0
\end{array}
\end{footnotesize}\right)&\notag\\
\label{eqn:M1}
\end{eqnarray}
and the prime denotes the derivative with respect to $r$.

We then combine all the equations into a single second-order differential equation for the radial component of the Lagrangian displacement field. As an intermediate step, we solve algebraic equations (\ref{eqn:disturb-eq1}) and express $(\tilde{u_r}, \tilde{u_\theta}, \tilde{u_z})$ in terms of $\tilde{p}$ as
\begin{eqnarray}
&&\tilde{u_r}=-\frac{\Lambda}{E \rho}
\frac{\dd \tilde{p}}{\dd r}+
   \frac{im}{E\rho r}\left(\frac{2iF\mu }{\tilde{\lambda}_\eta\rho\mu_0}-2\Omega \right)\tilde{p},
	     \notag\\
&& \tilde{u_\theta}=
\frac{1}{E\rho}
\left[2\Omega+r\Omega'\left(1+\displaystyle\frac{F^2}{\rho\mu_0\tilde{\lambda}_\eta^2}\right)-
\frac{2iF\mu}{\rho\mu_0\tilde{\lambda}_\eta}\right]
\frac{\dd \tilde{p}}{\dd r}
        \nonumber \\
&&\hspace*{10mm}
-
\frac{im}{Er\rho}\left[
 \Lambda+\frac{2\mu r}{\rho\mu_0\tilde{\lambda}_\eta}\left(\frac{iF}{\tilde{\lambda}_\eta}\Omega' -\mu'\right)\right]
\tilde{p}\notag,\\
\notag\\
&&\displaystyle\tilde{u_z}=-\frac{ik}{\rho \Lambda}\tilde{p},
\label{eqn:eq-tilde-u}
\end{eqnarray}
where
\begin{eqnarray}
E&=&\Lambda^2
   +\frac{2\Lambda\mu r}{\tilde{\lambda}_\eta\rho\mu_0}
 \left(\frac{i F}{\tilde{\lambda}_\eta}\Omega'
    -\mu'
\right)+2  \left(\Omega- \frac{i\mu F}{\tilde{\lambda}_\eta\rho\mu_0} \right)
        \nonumber \\
&&
\times\left[ 2\Omega+\left(1+\frac{F^2}{\tilde{\lambda}_\eta^2\rho\mu_0}
       \right)
r\Omega'
   -\frac{2i\mu F}{\tilde{\lambda}_\eta\rho\mu_0}
\right]
  .
\label{eqn:E-def}
\end{eqnarray}

Upon substitution from \rf{eqn:eq-tilde-u} for $\tilde{u_r}$, $\tilde{u_\theta}$, the continuity equation (\ref{eqn:solnoid-disturb}) produces a second-order differential equation for $\tilde{p}$
\begin{eqnarray}
&&\frac{\dd}{\dd r}\left(\frac{\Lambda}{\rho E}\frac{\dd \tilde{p}}{\dd r}\right)+\left[\frac{\Lambda}{rE\rho}
   -\frac{im}{E\rho}\left(1+\frac{F^2}{\rho\mu_0\tilde{\lambda}_\eta^2}\right)\Omega'\right]\frac{\dd \tilde{p}}{\dd r}
          \nonumber \\
&&
+\frac{2im}{Er^2\rho}\left(\Omega-\frac{iF\mu}{\rho\mu_0\tilde{\lambda}_\eta}\right)\tilde{p}+\frac{\dd}{\dd r}\left[\frac{2im}{Er\rho}\left(\Omega-\frac{iF\mu}{\rho\mu_0\tilde{\lambda}_\eta}\right)\right]\tilde{p}
           \nonumber \\
&&-\frac{2m^2}{Er^2\rho}\left(\frac{\Lambda}{2}+\frac{\mu r}{\rho\mu_0\tilde{\lambda}_\eta}\left(\frac{iF}{\tilde{\lambda}_\eta}\Omega' -\mu'\right)\right)\tilde{p}-\frac{k^2}{\Lambda\rho}\tilde{p}=0.
 \nonumber \\
&&
\label{eqn:p-disturb}
\end{eqnarray}

The first of equations \rf{eqn:eq-tilde-u} yields an expression for $\chi=-r\tilde{u_r}/\tilde{\lambda}_\eta$ in terms of $\tilde{p}$ and $\dd \tilde{p}/\dd r$
\begin{equation}
\chi=\frac{\Lambda r}{\tilde{\lambda}_\eta E \rho}
\frac{\dd \tilde{p}}{\dd r}+
   \frac{2im}{E\rho \tilde{\lambda}_\eta}\left(\Omega-\frac{iF\mu}{\rho\mu_0\tilde{\lambda}_\eta}\right)\tilde{p}.
\label{eqn:chi-def}
\end{equation}

In order to derive the equation for $\chi$, first we take the radial derivative of (\ref{eqn:chi-def}), and eliminate the second derivative of $\tilde{p}$, with the help of (\ref{eqn:p-disturb}), leaving
\begin{eqnarray}
\frac{\dd \chi}{\dd r}&=&\frac{im}{E\rho\tilde{\lambda}_\eta}\left[\left(1-\frac{\tilde{\lambda}_\nu}{\tilde{\lambda}_\eta}\right)r\Omega'+2\left(\Omega-\frac{iF\mu}{\rho\mu_0\tilde{\lambda}_\eta}\right)\right]\frac{\dd \tilde{p}}{\dd r}
          \nonumber \\
&&
+\frac{2m^2}{E\rho\tilde{\lambda}_\eta^2}\left(\Omega-\frac{iF\mu}{\rho\mu_0\tilde{\lambda}_\eta}\right)\Omega'\tilde{p}
           \nonumber \\
&&
+\frac{h^2r}{\Lambda E\rho\tilde{\lambda}_\eta}\left[\Lambda^2
   +\frac{2\mu r}{\tilde{\lambda}_\eta\rho\mu_0}
\Lambda \left(\frac{i F}{\tilde{\lambda}_\eta}\Omega'
    -\mu'
\right)\right]\tilde{p}
           \nonumber \\
&&
+\frac{2 k^2r}{\Lambda E\rho\tilde{\lambda}_\eta}\left[ 2\Omega+\left(1+\frac{F^2}{\tilde{\lambda}_\eta^2\rho\mu_0}
       \right)
r\Omega'
   -\frac{2i\mu F}{\tilde{\lambda}_\eta\rho\mu_0}
\right]
 \nonumber \\
&&
	\times\left(\Omega- \frac{i\mu F}{\tilde{\lambda}_\eta\rho\mu_0} \right)\tilde{p},
\label{eqn:chi-derivative}
\end{eqnarray}
where $h$ is defined by (\ref{eqn:def-h}).

A combination of (\ref{eqn:chi-def}) and (\ref{eqn:chi-derivative}) brings the expression for $\dd \tilde{p}/\dd r$ in terms of $\chi$ and $\dd \chi/\dd r$
\begin{eqnarray}
\frac{\dd  \tilde{p}}{\dd r}&=&-\frac{2i\rho m\tilde{\lambda}_\eta}{h^2r^2}\left(\Omega-\frac{iF\mu}{\rho\mu_0\tilde{\lambda}_\eta}\right)\frac{\dd \chi}{\dd r}+\frac{\rho\tilde{\lambda}_\eta E}{\Lambda r}\chi
     \nonumber \\
&&-\frac{2\rho m^2\tilde{\lambda}_\eta}{h^2r^3\Lambda}\Bigg(\Omega-\frac{iF\mu}{\rho\mu_0\tilde{\lambda}_\eta}\Bigg)\Bigg[\left(1-\frac{\tilde{\lambda}_\nu}{\tilde{\lambda}_\eta}\right)r\Omega'
     \nonumber \\
&&+2\left(\Omega-\frac{iF\mu}{\rho\mu_0\tilde{\lambda}_\eta}\right)\Bigg]\chi.
\label{eqn:p-derivative}
\end{eqnarray}
This helps us to rule out $\dd \tilde{p}/\dd r$ from (\ref{eqn:chi-derivative}) and obtain
\begin{eqnarray}
&&\Lambda r\frac{\dd \chi}{\dd r}-im\left[\left(1-\frac{\tilde{\lambda}_\nu}{\tilde{\lambda}_\eta}\right)r\Omega'+2\left(\Omega-\frac{iF\mu}{\rho\mu_0\tilde{\lambda}_\eta}\right)\right]\chi
 \nonumber \\
&&
=\frac{h^2r^2}{\rho\tilde{\lambda}_\eta}\tilde{p}.
\label{eqn:chi-p}
\end{eqnarray}

Multiplying both sides of (\ref{eqn:chi-p}) by $\rho\tilde{\lambda}_\eta/(h^2r^2)$, taking the derivative in $r$ and then substituting from (\ref{eqn:p-derivative}) for $\dd \tilde{p}/\dd r$  expressed in terms of $\chi$ and $\dd \chi/\dd r$, we eventually arrive at the extended Hain-L\"ust equation (\ref{eqn:Hain-Lust-eq})
\begin{eqnarray}
&&\frac{\dd }{\dd r}\left(f\frac{\dd \chi}{\dd r}\right)+s\frac{\dd\chi}{\dd r}-g\chi=0,
\label{eqn:Hain-Lust-eq2}
\end{eqnarray}
supplemented by (\ref{eqn:fsg}).


\section{Connection to the work \cite{KSF14}}

We write again the dispersion relation (\ref{eqn:dispers-rel0}), which we deduced from the extended Hain-L\"ust equation (\ref{eqn:Hain-Lust-eq})
\begin{eqnarray}
&& \tilde{\lambda}_\eta^2\Lambda^2
+4\alpha^2\left(\Omega\tilde{\lambda}_\eta-\frac{iF\mu}{\rho\mu_0}\right)
\notag\\
&&\times\left[\Omega Ro(\omega_\eta-\omega_\nu)
+\left(\Omega\tilde{\lambda}_\eta-\frac{iF\mu}{\rho\mu_0}\right)\right]
     \notag\\
&&
+ \frac{4\Lambda h^2\tilde{\lambda}_\eta}{h^2+q^2}
\Bigg[\left(\Omega^2Ro-\frac{\mu^2}{\rho\mu_0}Rb\right)\notag\\
&&+{\color{black}\frac{imr}{4} \frac{\dd}{\dd r}\left(\frac{2(\Omega\tilde{\lambda}_\eta-\frac{i\mu F}{\rho\mu_0})+(\omega_\eta-\omega_\nu)r\Omega'}{h^2r^2}\right)}\Bigg]
  =0,
	\notag\\
&&
\label{eqn:dispers-rel-again}
\end{eqnarray}
where $\Lambda=\tilde{\lambda}_\nu+\frac{F^2}{\tilde{\lambda}_{\eta}\rho\mu_0}$ and $\alpha^2=\frac{k^2}{h^2+q^2}$.
The dispersion relation in the work  \cite{KSF14} differs from \rf{eqn:dispers-rel-again} only by the term
\begin{eqnarray}
\frac{\Lambda h^2\tilde{\lambda}_\eta imr}{h^2+q^2}\frac{\dd}{\dd r}\left(\frac{{\color{black}2(\Omega\tilde{\lambda}_\eta-\frac{i\mu F}{\rho\mu_0})+(\omega_\eta-\omega_\nu)r\Omega'}}{h^2r^2}\right).
\label{eqn:derivative term}
\end{eqnarray}

We illustrate the difference by calculating the growth rates given by the two dispersion relations for $Ro=-1/2$, $Rb=-1/2$, $m=1$, and $Pm=1$. We define $\alpha_1$ by $\alpha_1^2=k^2/(k^2+q^2)$. Expanding the growth rates at large values of $\mathit{Ha}_\theta$ for \rf{eqn:dispers-rel-again}
with and without the term \rf{eqn:derivative term}, we find
\begin{eqnarray}
\frac{\Re{(\lambda)}}{\Omega}&=&a_H\mathit{Ha}_{\theta}+O(\mathit{Ha}^0_{\theta}),
\nonumber\\
\frac{\Re{(\lambda)}}{\Omega}&=&a_K\mathit{Ha}_{\theta}+O(\mathit{Ha}^0_{\theta}),
\label{eqn:linear dependence}
\end{eqnarray}
respectively, where
\ba{eqn:a}
a_H
&=&\frac{1}{Re\sqrt{(1 + (rk)^2) (\alpha_1^2 + (rk)^2)}}\nn\\
&&\times\Bigg\{(\alpha_1^2-1) (rk)^2 - (1 + \alpha_1^2) (rk)^4
 \nn\\
&&+
 \alpha_1\Big[
  4 (rk)^4 (1 + (rk)^2)^2 + {\color{black}\alpha_1^2 (1} + 8 (rk)^2
	\nonumber\\
&& + 10 (rk)^4 + (rk)^8)\Big]^{\frac{1}{2}}\Bigg\}^{\frac{1}{2}},
\nn\\
a_K
&=&\frac{1}{Re}\sqrt{
\sqrt{\alpha_1^2(4 + \alpha_1^2)}-(1 + \alpha_1^2)}.
\ea

The relation between $a_H$ and $a_K$ becomes clear, if we expand $a_H$ in power of $1/k$ at large values of $|k|$ leaving
\begin{eqnarray}
a_H&=&\frac{1}{Re}\sqrt{
\sqrt{\alpha_1^2(4 + \alpha_1^2)}-(1 + \alpha_1^2)}
\nonumber\\
&&+\frac{\Big((3 + \alpha_1^2) \sqrt{\alpha_1^2 (4 + \alpha_1^2)} -
   \alpha_1^2 (5 + \alpha_1^2) \Big) \sqrt{\alpha_1^2 }}{2 \sqrt{
 \alpha_1^2 + 4} \sqrt{\sqrt{\alpha_1^2 (4 + \alpha_1^2)}-1 - \alpha_1^2}Re }\frac{1}{k^2r^2}
 \nonumber\\
&&+O\left(\frac{1}{k^4}\right).
\label{eqn:dependence on k1}
\end{eqnarray}
We find that the leading order term is $a_K$.

\section{Connection to the work \cite{Bodo16}}\label{Connection to the work Bodo16}

\textcolor{black}{Bodo et al. \cite{Bodo16} consider \textit{compressible} MHD without viscosity and electrical resistivity in contrast to our setting, which is an incompressible MHD with viscosity and electrical resistivity. Here we demonstrate that in the limit of infinite speed of sound $(c_s\rightarrow\infty)$ the differential equation (35) of the work \cite{Bodo16} yields the same version of the Hain-L\"ust equation for the Lagrangian displacement as our Eq.~\rf{eqn:Hain-Lust-eq} does for $\nu=0$ and $\eta=0$.}

\textcolor{black}{Indeed, the differential equation (35) in \cite{Bodo16} is
\ba{eqn:Bodo-second-order}
&&\frac{\dd^2}{\dd r^2}(r\xi_r)+\frac{\dd}{\dd r}\ln\left(\frac{\Delta}{rC_2}\right)\frac{\dd}{\dd r}(r\xi_r)\nn\\
&&+\left[\frac{C_2C_3-C_1^2}{\Delta^2}-\frac{rC_2}{\Delta}\frac{\dd}{\dd r}\left(\frac{C_1}{rC_2}\right)\right](r\xi_r)
=0,
\ea}
\textcolor{black}{
which can be transformed to
\ba{eqn:Bodo-second-order1}
&&\frac{\dd}{\dd r}\left(\frac{\Delta}{rC_2}\frac{\dd}{\dd r}(r\xi_r)\right)\nn\\
&&+\bigg[\frac{C_2C_3-C_1^2}{r\Delta C_2}-\frac{\dd}{\dd r}\left(\frac{C_1}{rC_2}\right)\bigg](r\xi_r)=0.
\ea
}
\textcolor{black}{The coefficients $\Delta$, $C_1$, $C_2$, and $C_3$ are defined in \cite{Bodo16} by equations (27)-(30) that contain the sound speed $c_s$.
Retaining only the leading order terms in $c_s$ in the assumption that $c_s \gg 1$, we write these coefficients as
\ba{eqn:Bodo-Delta-C}
&&\Delta=c_s^2(\rho\tilde{\omega}^2-k_B^2)^2\nonumber\\
&&C_1=-\frac{2mc_s^2}{r^2}(k_BB_\phi+\rho v_\phi\tilde{\omega})(\rho\tilde{\omega}^2-k_B^2)\nonumber\\
&&C_2=-c_s^2\left(k^2+\frac{m^2}{r^2}\right)(\rho\tilde{\omega}^2-k_B^2)\nonumber\\
&&C_3=c_s^2(\rho\tilde{\omega}^2-k_B^2)^2\left(\rho\tilde{\omega}^2-k_B^2+r\frac{\dd}{\dd r}\left(\frac{B_\phi^2-\rho v_\phi^2}{r^2}\right)\right)\nonumber\\
&&-\frac{4c_s^2}{r^2}(\rho\tilde{\omega}^2-k_B^2)(kB_\phi+\rho  v_\phi\tilde{\omega})^2,
\ea}\textcolor{black}{
where we omit the subscript $_0$ that was used in \cite{Bodo16} to denote the equilibrium.}

\textcolor{black}{In \cite{Bodo16} $k_B=\frac{m}{r}B_\phi+kB_z$ and $\tilde{\omega}=\omega-\frac{m}{r}v_\phi-kv_z$.}

{\color{black}In our notation before the Appendix~\ref{Connection to the work Bodo16}, $\phi=\theta$ and $m,k$ are defined with opposite sign. Hence, comparing with our notation we have $k_B=-F$, $B_\phi=\mu r$, $v_\phi=r\Omega$, $v_z=0$ and $\omega=-i\lambda$. Now, using (\ref{eqn:Bodo-Delta-C}), we can write
\begin{eqnarray}
&&\frac{\Delta}{rC_2}=-\frac{\rho\tilde{\omega}^2-k_B^2}{r\left(k^2+\frac{m^2}{r^2}\right)},\nonumber\\
&&\frac{C_2C_3-C_1^2}{r\Delta C_2}-\frac{\dd}{\dd r}\left(\frac{C_1}{rC_2}\right)\nonumber\\
&&=\frac{1}{r}(\rho\tilde{\omega}^2-k_B^2)+\frac{\dd}{\dd r}(\mu^2-\rho\Omega^2)-\frac{4(k_B\mu+\rho\tilde{\omega}\Omega)^2}{r(\rho\tilde{\omega}^2-k_B^2)}\nonumber\\
&&+\frac{4m^2(k_B\mu+\rho\tilde{\omega}\Omega)^2}{r^3(k^2+\frac{m^2}{r^2})(\rho\tilde{\omega}^2-k_B^2)}-\frac{\dd}{\dd r}\left(\frac{2m(k_B\mu+\rho\tilde{\omega}\Omega)}{k^2r^2+m^2}\right).\nonumber\\
&&
\label{eqn:Bodo-second-order-terms}
\end{eqnarray}
}
\textcolor{black}{Substituting \rf{eqn:Bodo-second-order-terms} into \rf{eqn:Bodo-second-order1}, then dividing both sides of the resulting equation by a constant $\rho$, and noticing that in our notation $\chi=-ru_r/\tilde{\lambda}=-r\xi_r$, we arrive at the same Hain-L\"ust equation \rf{eqn:Hain-Lust-eq}
that is derived in our manuscript and in which one needs to set $\nu=0$ and $\eta=0$, see also \cite{ZouFuk14}.}

\textcolor{black}{We notice also that the dispersion relation (37) in \cite{Bodo16}, which is a sixth-degree polynomial derived in the assumption $c_s=0$, totally differs from our dispersion relation \rf{eqn:dispers-rel0} corresponding to the incompressible limit $c_s\rightarrow \infty$.}




\begin{thebibliography}{99}
\bibitem{Vel59}
E. Velikhov,
JETP (USSR) \textbf{36}, 1398 (1959).

\bibitem{Cha60}
S. Chandrasekhar,
Proc. Natl. Acad. Sci. \textbf{46}, 253 (1960).

\bibitem{BalHaw91}
A. Balbus and J. F. Hawley,
Astrophys. J. \textbf{376}, 214 (1991).

\bibitem{B2011}
S. A. Balbus, Nature, \textbf{470}(7335), 475 (2011).

\bibitem{JB2013}
H. Ji and S. Balbus, Phys. Today,
\textbf{66}, 27 (2013).

\bibitem{SGG2008}
F. Stefani, A. Gailitis and G. Gerbeth, Z. angew. Math. Mech., \textbf{88}, 930 (2008).

\bibitem{ELFB11}
F. Ebrahimi et al.,
Phys. Plasmas \textbf{18}, 062904 (2011).

\bibitem{PhysRep2018}
G. R\"udiger et al.
Physics Reports \textbf{741}, 1 (2018).

\bibitem{Stefani2019}
F. Stefani et al., Geophys. \& Astrophys. Fluid Dyn., \textbf{113}(1-2), 51 (2019).



\bibitem{KirSte10}
O. N. Kirillov and F. Stefani,
Astrophys. J. \textbf{712}, 52 (2010).

\bibitem{KirSte12}
O. N. Kirillov and F. Stefani,
Acta Appl. Math.
 \textbf{120}, 177 (2012).

\bibitem{KPS2011}
O. N. Kirillov, D. E. Pelinovsky and G. Schneider, Phys. Rev. E \textbf{84}, 065301(R) (2011).

\bibitem{WB2002}
A. P. Willis and C. F. Barenghi, Astron. \& Astroph. \textbf{388}, 688 (2002).

\bibitem{KS2011}
O. N. Kirillov and F. Stefani, Phys. Rev. E  \textbf{84}(3), 036304 (2011).

\bibitem{Deg18}
K. Deguchi, J. Fluid Mech. \textbf{865}, 492 (2019).

\bibitem{Des04}
S. J. Desch,
Astrophys. J. \textbf{608}, 509 (2004).


\bibitem{PapTer97}
J. C. B. Papaloizou and C. Terquem,
Mon. Not. R. Astron. Soc. \textbf{287}, 771 (1997).


\bibitem{BNST95}
A. Brandenburg et al.,
Astrophys. J. \textbf{446}, 741 (1995).

\bibitem{BalHaw92a}
A. Balbus and J. F. Hawley,
Astrophys. J. \textbf{400}, 610 (1992).

\bibitem{TerPap96}
C. Terquem and J. C. B. Papaloizou,
Mon. Not. R. Astron. Soc. \textbf{279}, 767 (1996).

\bibitem{FV95}
S. Friedlander and M. M. Vishik, Chaos \textbf{5},  416 (1995).

\bibitem{OgiPri96}
G. I. Ogilvie and J. E. Pringle,
Mon. Not. R. Astron. Soc. \textbf{279}, 152 (1996).

\bibitem{CurPud96}
C. Curry and R. E. Pudritz, Mon. Not. R. Astron. Soc. \textbf{281}, 119 (1996).

\bibitem{SalWar03}
R. Salmeron and M. Wardle,
Mon. Not. R. Astron. Soc. \textbf{345}, 992 (2003).

\bibitem{RGSHS14}
G. R\"udiger et al.,
Mon. Not. R. Astron. Soc. \textbf{438}, 271 (2014).

\bibitem{BalHen08}
A. Balbus and P. Henri,
Astrophys. J. \textbf{674}, 408 (2008).

\bibitem{Priede2007}
J. Priede, I. Grants and G. Gerbeth,
Phys. Rev. E \textbf{75}, 047303 (2007).

\bibitem{Pri11}
J. Priede, Phys. Rev. E \textbf{84}, 006314 (2011).

\bibitem{Priede2015}
J. Priede, Phys. Rev. E \textbf{91}, 033014 (2015).

\bibitem{KSF14JFM}
O. N. Kirillov, F. Stefani and Y. Fukumoto,
J. Fluid Mech. \textbf{760}, 591 (2014).

\bibitem{HolRud05}
R. Hollerbach and G. R\"udiger,
Phys. Rev. Lett. \textbf{95}, 124501 (2005).

\bibitem{Rud10}
G. R\"udiger et al.,
Phys. Rev. E \textbf{82}, 1 (2010).

\bibitem{SGGRSSH06}
F. Stefani et al.,
Phys. Rev. Lett. \textbf{97}, 184502 (2006).

\bibitem{SGGHPRS09}
F. Stefani et al., Phys. Rev. E \textbf{80}, 066303 (2009).

\bibitem{HolTee10}
R. Hollerbach, V. Teeluck and G. R\"udiger,
Phys. Rev. Lett. \textbf{104}, 044502 (2010).

\bibitem{SGGGS14}
M. Seilmayer et al.,
Phys. Rev. Lett. \textbf{113}, 024505 (2014).

\bibitem{KirSte13}
O. N. Kirillov and F. Stefani,
Phys. Rev. Lett. \textbf{111}, 061103 (2013).

\bibitem{LGHJ06}
W. Liu et al., Phys. Rev. E \textbf{74}, 056302 (2006).

\bibitem{Tay73}
R. J. Tayler, Mon. Not. R. Astron. Soc. \textbf{161}, 365 (1973).

\bibitem{Tay80}
R. J. Tayler, Mon. Not. R. Astron. Soc. \textbf{191}, 151 (1980).

\bibitem{RS2010}
G. R\"udiger and M. Schultz, Astron. Nachr. \textbf{331}, 121 (2010).

\bibitem{Tayler2012}
M. Seilmayer et al., Phys. Rev. Lett. \textbf{108}, 244501 (2012).

\bibitem{KSF14}
O. N. Kirillov, F. Stefani and Y. Fukumoto,
Fluid Dyn. Res. \textbf{46}, 031403 (2014).

\bibitem{Hol15}
A. Child, E. Kersal\'e and R. Hollerbach, Phys. Rev. E. \textbf{92}, 033011 (2015).

\bibitem{KSF12}
O. N. Kirillov, F. Stefani and Y. Fukumoto,
Astrophys. J. \textbf{712}, 52 (2012).


\bibitem{SB2014}
J. Squire and A. Bhattacharjee, Phys. Rev. Lett. \textbf{113}, 025006 (2014).

\bibitem{Kir15}
O. Kirillov and F. Stefani, Phys. Rev. E \textbf{92}, 051001 (2015).

\bibitem{Kir17}
O. N. Kirillov,
Proc. R. Soc. A \textbf{473}(2205), 20170344 (2017).

\bibitem{HL1958}
K. Hain and R. L\"ust, Z. Naturforsch. \textbf{13a}, 936 (1958).

\bibitem{HMS2017}
S. Hassi, M. M\"oller and H. de Snoo,
Math. Nachr. \textbf{291}(4), 652 (2017).

\bibitem{FR1960}
E. Frieman and M. Rotenberg, Rev. Mod. Phys. \textbf{32}, 898 (1960).

\bibitem{GP04}
J. P. Goedbloed and S. Poedts,
\textit{Principles of Magnetohydrodynamics}
(Cambridge University Press. Cambridge, 2004).

\bibitem{GKP10}
J. P. Goedbloed, R. Keppens and S. Poedts,
\textit{Advanced Magnetohydrodynamics}
(Cambridge University Press. Cambridge, 2010).


\bibitem{GKP2019}
J. P. Goedbloed, R. Keppens, S. Poedts,
\textit{Magnetohydrodynamics of laboratory and astrophysical plasmas}
(Cambridge University Press. Cambridge, 2019).

\bibitem{ZouFuk14}
R. Zou and Y. Fukumoto, Prog. Theor. Exp. Phys. 113J01, (2014).











\bibitem{RudHol07}
G. R\"udiger et al.,
 Mon. Not. R. Astron. Soc. \textbf{377}, 1481 (2007).

\bibitem{H2000}
R. Hollerbach, Int. J. Num. Methods in Fluids \textbf{32}(7), 773 (2000).

\bibitem{Deg11}
K. Deguchi and M. Nagata, J. Fluid Mech. \textbf{678}, 156 (2011).

\bibitem{Deg17}
K. Deguchi, Phys. Rev. E \textbf{95}, 021102 (2017).

\bibitem{VMI1999}
V. A. Vladimirov, H. K. Moffatt and K. I. Il'in,
J. Fluid Mech. \textbf{390}, 127 (1999).












\bibitem{Hein89}
W. Heinrichs, Math. Comput. \textbf{53}, 187 (1989).






\bibitem{Gus15}
A. Guseva et al., New J. Phys. \textbf{53}, 093018 (2015).


\bibitem{Bodo16}
G. Bodo et al.,
Mon. Not. R. Astron. Soc. \textbf{462}, 3031 (2016).






\end{thebibliography}
\end{document}